\documentclass[fleqn,usenatbib]{mnras}

\usepackage{newtxtext,newtxmath}

\usepackage[T1]{fontenc}
\usepackage{ae,aecompl}

\usepackage{pifont} 

\usepackage{graphicx}	
\usepackage{amsmath}	
\usepackage{amssymb}	
\usepackage[normalem]{ulem}
\usepackage{enumitem}

\title[Retrograde day-side flow in WASP-43b]{Equatorial retrograde flow in WASP-43b elicited by deep wind jets?}

\author[L. Carone et al.]
{Ludmila Carone,$^{1}$\thanks{E-mail: carone@mpia.de }
Robin Baeyens,$^{2}$
Paul Molli\`ere,$^{1}$
Patrick Barth,$^{1,3}$
Allona Vazan,$^{4,5}$\newauthor
Leen Decin,$^{2}$
Paula Sarkis,$^{1,6}$
Olivia Venot$^{7}$
and Thomas Henning$^{1}$
\\
$^{1}$ Max-Planck-Institut f\"ur Astronomie, K\"onigstuhl 17, 69117 Heidelberg, Germany\\
$^{2}$ Instituut voor Sterrenkunde, KU Leuven, Celestijnenlaan 200D, B-3001 Leuven, Belgium\\
$^{3}$ Centre for Exoplanet Science, University of St Andrews, North Haugh, St Andrews, KY169SS, UK\\
$^{4}$ Racah Institute of Physics, The Hebrew University,  Jerusalem 91904, Israel \\
$^{5}$
Institute for Computational Science, Center for Theoretical Astrophysics and Cosmology, University of Z{\"u}rich, 8057 Z{\"u}rich, Switzerland\\
$^{6}$ Physikalisches Institut, Universit\"at Bern, Gesellschaftsstrasse 6, 3012 Bern, Switzerland\\
$^{7}$ Laboratoire Interuniversitaire des Syst\`{e}mes Atmosph\'{e}riques (LISA), UMR CNRS 7583, Universit\'{e} Paris-Est-Cr\'eteil, Universit\'e de Paris,\\ Institut Pierre Simon Laplace, Cr\'{e}teil, France
}

\date{Accepted XXX. Received YYY; in original form ZZZ}

\pubyear{2019}
\begin{document}
\label{firstpage}
\pagerange{\pageref{firstpage}--\pageref{lastpage}}
\maketitle

\begin{abstract}
 We present WASP-43b climate simulations with deep wind jets (down to 700~bar) that are linked to retrograde (westward) flow at the equatorial day side for $p<0.1$~bar. Retrograde flow inhibits efficient eastward heat transport and naturally explains the small hotspot shift and large day-night-side gradient of WASP-43b ($P_{\text{orb}}=P_{\text{rot}}=0.8135$~days) observed with Spitzer. We find that deep wind jets are mainly associated with very fast rotations ($P_{\text{rot}}=P_{\text{orb}}\leq 1.5$~days)  which correspond to the Rhines length smaller than $2$ planetary radii. We also diagnose wave activity that likely gives rise to deviations from superrotation. Further, we show that we can achieve full steady state  in our climate simulations by imposing a deep forcing regime for $p>10$~bar: convergence time scale $\tau_{\text{conv}}=10^6-10^8$~s  to a common adiabat, as well as linear drag at depth ($p\geq 200$~bar), which mimics to first order magnetic drag.  Lower boundary stability and the deep forcing assumptions were also tested with climate simulations for HD~209458b ($P_{\text{orb}}=P_{\text{rot}}=3.5$~days). HD~209458b simulations always show shallow wind jets (never deeper than 100~bar) and unperturbed superrotation. If we impose a fast rotation ($P_{\text{orb}}=P_{\text{rot}}=0.8135$~days), also the HD~209458b-like simulation shows equatorial retrograde flow at the day side. We conclude that the placement of the lower boundary at $p=200$~bar is justified for slow rotators like HD~209458b, but we suggest that it has to be placed deeper for fast-rotating, dense hot Jupiters ($P_{\text{orb}}\leq 1.5$~days) like WASP-43b. Our study highlights that the deep atmosphere may have a strong influence on the observable atmospheric flow in some hot Jupiters.
\end{abstract}

\begin{keywords}
hydrodynamics -- planets and satellites: atmospheres -- planets and satellites: gaseous planets
\end{keywords}



\section{Introduction}
\label{sec:Intro}

A fast (1-7~km/s), equatorial eastward wind jet is consistently produced in 3D climate simulations of tidally locked hot Jupiters  \citep[e.g.][]{Showman2002,Showman2009,Dobbs2010,Tsai2014,Kataria2015,Amundsen2016,Zhang2017,Mendonca2018,Parmentier2018}. This superrotating flow leads to an eastward hot spot shift with respect to the substellar point  \citep{Knutson2007} and efficient day-to-night-side heat transport.

Several planets, however, may show deviations from equatorial superrotation: CoRoT-2b has a westward shifted hot spot  \citep{Dang2018} and  the optical peak offset in HAT-P-7b oscillates west- and eastward around the substellar point  \citep{Armstrong2016}. Several mechanisms have been proposed to explain differences between observations and predictions with cloud-free 3D GCMs that exhibit very strong superrotation with an eastward hot spot shift: the neglected influence of clouds \citep{Parmentier2016,Helling2016,Mendonca2018,Mendonca2018b} and a higher atmospheric metallicity \citep{Kataria2015,Drummond2018} may reduce the speed of the equatorial jet. Magnetic fields \citep{Rogers2014a,Kataria2015,Arcangeli2019,Hindle2019} are also proposed to reduce eastward equatorial wind jets in part of the atmosphere. Also non-synchronous planetary rotation can in some cases lead to retrograde instead of prograde flow along the equator\citep{Rauscher2014}. Also, \cite{Armstrong2016,Dang2018} state that cloud coverage variability could explain the anomalous HAT-P-7b and CoRoT-2b observations.

Another planet that has started a discussion about abnormal flow properties, cloud properties and deviations from equilibrium chemistry is WASP-43b \citep{Stevenson2017,Kataria2015,Mendonca2018,Mendonca2018b}. WASP-43b is one of the closest-orbiting hot Jupiters (see Table~\ref{table_params}) that transits its host star every 0.8315~days \citep{Hellier2011}. It has a moderately hot effective temperature given its proximity to its host star ($T_{\rm eff,Pl}\approx 1450$~K) because it orbits a cool K dwarf star. Furthermore, WASP-43b is unusually dense with a radius of 1.04 $R_{\rm J}$ and a mass of $2.05 M_{\rm J}$. Observations in the infrared \citep{Stevenson2014,Stevenson2017,Keating2017} suggest that the eastward hot spot shift is unusually small and that the day-to-night-side temperature contrast is unusually large, compared to planets of similar effective temperature like HD~209458b \citep{Chen2014,Stevenson2017,Komacek2017,Keating2017,Keating2019}. The Spitzer observations of \citet{Stevenson2017} have come under scrutiny \citep{Mendonca2018,Morello2019}, but after reanalysis the day-night temperature contrast still remains high \citep{Keating2019} and the hot spot shift remains small \citep{Mendonca2018}.

The WASP-43b observations have been attempted to be explained by magnetic drag and higher solar metallicity \citep{Kataria2015} and by night-side clouds with disequilibrium chemistry \citep{Mendonca2018,Mendonca2018b}. It is noteworthy, however, that HD~209458b, which is of similar effective temperature than WASP-43b, does not appear to exhibit this large temperature gradient and small hot spot gradient. Furthermore, the formation of clouds at high altitudes are more favored for planets of low surface gravity such as HD~209458b compared to high surface gravity planets as WASP-43b is \citep{Stevenson2016}. A comparison of the bond albedo estimates for HD~209458b and WASP-43b shows instead a higher albedo for the former compared to the latter (Table~1, \citet{Keating2019}). Some observations of WASP-43b in transmission appear to also favor a cloud-free atmosphere \citep{Kreidberg2014,Weaver2019} as expected for hot Jupiters with high surface gravity \citep{Stevenson2016}, while others favor thick clouds \citep{Chen2014}. Even when taking into account clouds, a direct comparison between between WASP-43b and HD~209458b shows that there are interesting differences between these despite both planets having similar thermal properties.  We also note that CoRoT-2b, the planet with an observed westward hot spot shift, has a night-side temperature several 100K colder than other hot Jupiters in the same effective temperature regime \citep{Dang2018,Keating2019}, which is difficult to explain with night-side clouds. In any case, none of the given possible explanations are sufficiently well understood to fully account for `outliers' in the hot Jupiter population in terms of hot spot shift and day-to-night-side temperature differences that are displayed by HaT-P-7b, CoRoT-2b and WASP-43b.

Here, we tackle an alternative scenario to understand why WASP-43b is different compared to HD~209458b by investigating for both planets climate simulations that take into account deeper climate layers than previously considered. We place the lower boundary at $p_{\rm boundary}= 700$~bar and stabilize the model at depth via friction, which we choose as a first order representation of magnetic drag. This drag should couple predominantly with very deep wind jets, as observed in Jupiter \citep{Kaspi2018} and also proposed in hot Jupiters \citep{Rogers2014b}. This lower boundary prescription was primarily selected as a means of stabilizing the lower boundary.

We report here that with this prescription, deep wind jets appear in our 3D climate model of WASP-43b. They appear to be linked with retrograde wind flow at the equatorial day side for $p\geq 100$~mbar, embedded in strong equatorial superrotation at other longitudes. These retrograde winds are at the same time absent in HD~209458b that continues to exhibit unperturbed superrotation with efficient horizontal heat transport. Even with deeper layers, we reproduce for the latter planet  results consistent with those from previous `shallower' 3D climate models \citep{Showman2008,Rauscher2012}.

We postulate that deep wind jets that trigger retrograde flow may be another possible mechanism to explain why we observe anomalies in the wind flow of some hot Jupiters like WASP-43b but not in others like HD~209459b. In this work, we also investigate why the wind jets in WASP-43b extend much deeper into the interior than in HD~209458b and why these deep wind jets are associated with retrograde wind flow along the equator.

In Section~\ref{sec: methods} we present the 3D atmosphere model, the thermal forcing and lower boundary prescriptions, where we focus mainly on results from the nominal (\textit{full\_stab}) set-up, see Table~\ref{table_sims}. A more in-depth investigation of all employed methods to stabilize the lower boundary can be found in Appendices~\ref{sec: lower_stab} and \ref{sec: angular_mom}. In Section~\ref{sec: sims} we present simulation results for WASP-43b and HD~209458b, an in-depth comparison of eddy and actual wind flow and we also present simulations for different rotation periods. Furthermore, we compare predictions based on our nominal WASP-43b simulation with HST/WFC3 and Spitzer observations. We summarize our results in Section~\ref{sec: results} and discuss the crucial differences between our WASP-43b and HD~209458b simulations. We further show that our results are complementary to and consistent with previous climate studies in the fast rotating hot Jupiter and tidally locked Earth regime, where waves at depth can arise to shape the wind flow. We stress again in Section~\ref{sec: Conclusions} the importance of the lower boundary and wind flows in the deep atmosphere ($p>10$~bar) for fast rotating planets $P_{\text{orb}}<1.5$~days. The deep layers may give rise to instabilities, which can have a significant effect on the observable day-to-night-side redistribution. We present possibilities for further investigations in Section~\ref{sec: Outlook}.

\section{Methods}
\label{sec: methods}
\subsection{3D atmosphere model}
\label{sec:atmo}
We employ the dynamical core of \textit{MITgcm} \citep{Adcroft2004}, where we solve the three-dimensional hydrostatic primitive equations (HPE) on a cubed-sphere grid  \citep{Showman2009}. The ideal gas law is used as equation of state in all our simulations. We use like \citep{Showman2009} a horizontal fourth-order Shapiro filter with $\tau_{\text{shap}}=25$~s to smooth horizontal grid-scale noise. We assume 40 vertical layers in logarithmic steps from 200 bar to 0.1~mbar, resolving three levels per pressure scale height. We further place additional layers with 100 bar spacing below 200~bar to ensure that we resolve deep vertical momentum transport. In total, we thus have 45 (53) vertical layers between $p_{\rm bottom}=$700~bar ($p_{\rm bottom}=$1500~bar, for testing purposes, see Section~\ref{sec: complete_stab}) and 0.1~mbar. We use a time step of $\Delta t=25$~seconds and a cubed-sphere (C32) horizontal resolution, which corresponds to a resolution in longitude and latitude of $128\times 64$ or approximately $2.8^{\circ}\times 2.8^{\circ}$. Horizontal resolution and vertical resolution for $p < 200$~bar is chosen in accordance with the \textit{SPARC/MITgcm} simulation of WASP-43b \citep{Kataria2015}. Our dynamical model set-up deviates from the \textit{SPARC/MITgcm} set-up by extending the vertical grid downward and by employing additional lower boundary stabilization measures (see Section~\ref{sec: low_boundary}).

\begin{table*}
	\caption[table]{Overview of the simulation parameters used in our nominal simulations of WASP-43b and HD~209458b. An overview of the different simulations and their adjusted parameters is given in Table~\ref{table_sims}.}
	\label{table_params}
	\centering
	\begin{tabular}{l c c}
    \noalign{\smallskip}
	\hline \hline
	\noalign{\smallskip}
	\textbf{Quantity} & &  \textbf{Value} \\
	\noalign{\smallskip}
	\hline
	\noalign{\smallskip}
	Horizontal resolution  &  &   128 $\times$ 64 \\
    Vertical resolution  &  & 45 \\
    Time step, $\Delta t$ [s]  & &  25 \\
    Outer boundary pressure, $p_{\rm top}$ [bar]  & &   10$^{-4}$ \\
    Inner boundary pressure, $p_{\rm bottom}$ [bar]  & &   700 \\
    Specific heat capacity (at constant pressure), $c_p$ [J kg$^{-1}$ K$^{-1}$]  & &   $1.199 \cdot 10^4$ \\
    Mean molecular weight, $\mu_{\rm MMW}$ [g mole$^{-1}$]  & &   2.325 \\
    Specific gas constant $R$ [J kg$^{-1}$ K$^{-1}$] &  & 3576.1 \\
    Sponge layer Rayleigh friction, $k_{\rm top}$ [days$^{-1}$]  &  &  0.056 \\
	\noalign{\smallskip}
	\hline
    \noalign{\smallskip}
    \textbf{Quantity}  &  \textbf{WASP-43b} & \textbf{HD~209458b (benchmark)} \\
    \noalign{\smallskip}
    \hline
    \noalign{\smallskip}
    Intrinsic temperature, $T_{\rm int}$ [K]  & 170  &  400 \\
    Rotation period, $P_{\rm rot}$ [days$^{-1}$] &  0.8135  &  3.47 \\
    Gravity, $g$ [m s$^{-2}$ ]  &  46.9  &  9.3 \\
    Mass, $M_{\rm p}$ [$M_{\rm J}$]  & $2.05$  &  $0.69$ \\
    Radius, $R_{\rm p}$ [$R_{\rm J}$]  &  $1.04$  &  $1.38$ \\
    \noalign{\smallskip}
    \hline
	\end{tabular}
\end{table*}%

To avoid unphysical gravity wave reflection at the upper boundary, we treat the uppermost layer as a `sponge layer'. We impose a Rayleigh friction term on the horizontal velocities $\mathbf{v}$ in the topmost layer, which is similar to the `sponge layer' set-up used in other climate models  \citep{Zalucha2013,Carone2014,Carone2016,Jablo2011}:
\begin{equation}
\mathbf{F}_v=-k_{\rm top}\mathbf{v},
\end{equation}
where $k_{\rm top}$=1/18 days$^{-1}$. In this work, we focus mainly on the interplay between the very deep layers ($p<10$~bar) and the observable photosphere ($1<p<0.01$~bar). It is thus beyond the scope of this work to discuss in detail how our `hard' sponge layer, that only affects the uppermost atmosphere layers ($p\leq 10^{-4}$~bar) is different from the `soft' set-up used by \citet{Mendonca2018b}, where the authors are forcing the upper-most layers to the zonal mean of the wind flow. We note, however, that \citet{Deitrick2019} discuss the sponge layer used by \citet{Mendonca2018b} in more details. See also Section~\ref{sec: sponge layer}, where we performed a first comparison between simulations with our nominal sponge layer set-up, with a set-up similar to the one used by \citet{Mendonca2018b} and to simulations without any sponge layer.

For the gas properties, we adopt values calculated by the radiative transfer model \textit{petitCODE}  \citep{Molliere2015,Molliere2017} that is also used for thermal forcing (see next subsection). For a hydrogen-dominated atmosphere with solar metallicity  \citep{asplund2009} and $T_{\rm eff,Pl}=1450$~K, the equilibrium chemistry in \textit{petitCODE}  \citep{Molliere2017} yields at $p=1$ bar a heat capacity $c_p=1.199\times 10^4$J~kg$^{-1}$~K$^{-1}$ and mean molecular weight $\mu_{\rm MMW}=2.325$~g~mole$^{-1}$. An overview of the model parameters used in our simulations is given in Table~\ref{table_params}.

\subsection{Thermal forcing}

\begin{figure*}
	\includegraphics[width=1.9\columnwidth]{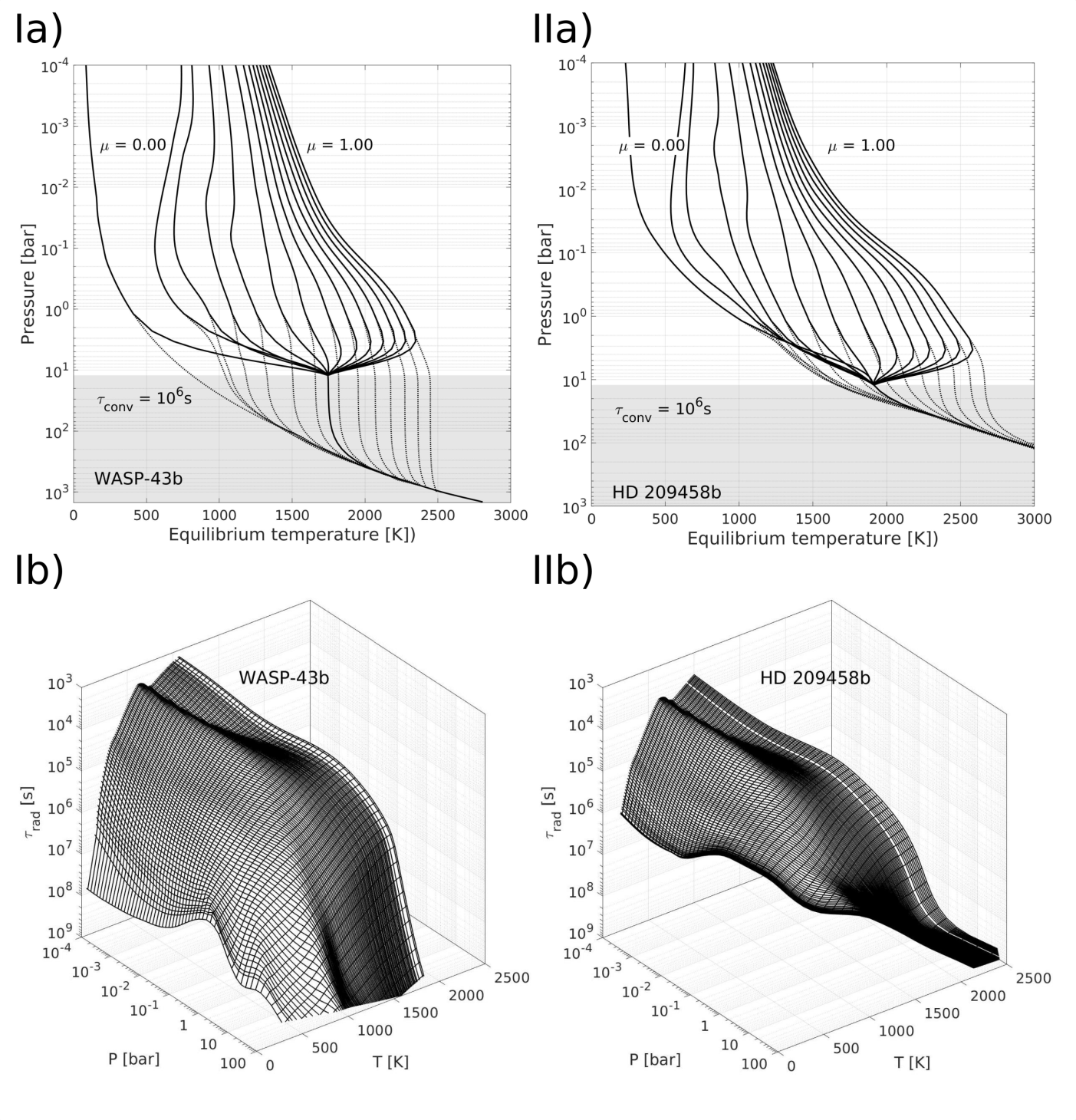}
    \caption{Equilibrium temperature \textit{($T_{\rm eq}$, top row)} and radiative timescales \textit{($\tau_{\rm rad}$, bottom row)} for WASP-43b (panels Ia, Ib) and HD~209458b (panels IIa, IIb) computed using \textit{petitCODE}. In the upper panels, the dotted lines denote the equilibrium temperature profiles for different angles of incidence $\mu$, which converge to a common deep adiabat. Here, WASP-43 converges to a common adiabat at higher pressures compared to HD~209458b. As one additional stabilization measure, we also converged the profiles between 1~bar and 10~bar to the planetary average temperature (solid black lines). In the nominal simulation, the (p-T) profiles following the solid black lines are used.}
    \label{fig:rad}
\end{figure*}

Thermal forcing in the model is provided via the Newtonian cooling mechanism. It has been shown to be suitable to qualitatively investigate flow dynamics in tidally locked hot Jupiters \citep{Menou2009,Komacek2016,Showman2002,Showman2008,Mayne2014} and rocky exoplanets \citep{Carone2015}:

\begin{equation}
    \frac{\partial T} {\partial t} = \frac{T_{\rm eq}-T} {\tau_{\rm rad}(p,T)} \label{eq:Newton},
\end{equation}

where $T_{\rm eq}(\theta, \phi, p)$ is the radiative-convective equilibrium temperature for different latitudes $\theta$, longitudes $\phi$ and pressure levels $p$ and $\tau_{\rm rad}(p,T)$ is the radiative time scale.

We place the substellar point at latitude and longitude $(0^{\circ},0^{\circ})$ and calculate the equilibrium temperature $T_{\rm eq}(\theta, \phi, p)$ for different irradiation incidence angles $\zeta$. These are connected to latitude $\theta$ and longitude $\phi$ on the planet via
\begin{eqnarray}
&&\cos\zeta=0 \textrm{ for } |\phi|\geq 90^{\circ} \nonumber\\
&&\cos\zeta=\cos\theta\cdot \cos\phi \textrm{ for }|\phi|< 90^{\circ},
\end{eqnarray}
assuming a planetary obliquity of $0^{\circ}$. We will use henceforth $\mu=\cos \zeta$ for the cosine of the incidence angle.

We follow a well-tested `recipe' for simplified thermal forcing in hot Jupiters \citep{Showman2008}. We calculate ($T_{\rm eq}$--$p$) profiles for concentric rings on the planet around the substellar point for $\mu=$1, 0.9, 0.8, 0.7, 0.6, 0.5, 0.4, 0.3, 0.2, 0.1, 0.05, 0.02, 0.01 and 0. The night side with $\mu=0$ corresponds to a 1D radiative-equilibrium profile without stellar irradiation. This selection of angles of incidence yields rings with a latitudinal width of 9$^\circ$ on the day side. Near the planetary terminator for $0 < \mu < 0.1$, extra sampling is used to account for the big temperature contrast between the equilibrium temperature profile corresponding to $\mu = 0.1$ and the much colder night side (see Figure~\ref{fig:rad}). The equilibrium temperatures corresponding to each vertical column are set to the ($T_{\rm eq}$--$p$) profile associated with the $\mu$-value of the column, rounded up to the next sampled $\mu$-value. We have tested different samplings for the radiation incidence angles and have found no significant changes in the atmospheric circulation for our planets. The radiative time scales $\tau_{\rm rad}$ are calculated by adding a thermal perturbation $\Delta T=10 K$ for each equilibrium temperature profile $T_{\rm eq}$ at a given pressure level $p$  and calculating the time it takes for the perturbed air parcel to return to radiative equilibrium within the radiative transfer part of \textit{petitCODE} \citep{Molliere2015,Molliere2017}. The \textit{petitCODE} is a state-of-the-art, versatile 1D code for exoplanet atmosphere modeling. Furthermore, \textit{petitCODE} is benchmarked with other state-of-the-art multi-wavelength radiative transfer codes  \citep{Baudino2017}. We thus create a grid of radiative time scales $\tau_{\rm rad}$ for a given $(p,T)$ combination (Figure~\ref{fig:rad}):
\begin{equation}
\tau_{\rm rad}(p,T)=\Delta T \frac{c_p\rho}{\Delta F/\Delta z}\label{eq:tau_rad},
\end{equation}
where $\rho$ is the atmospheric density assuming an ideal gas and $\Delta F/\Delta z$ is the ratio between the net vertical flux in the perturbed atmosphere layer and the vertical extent of that layer in meter. Figure~\ref{fig:rad} shows $T_{\rm eq}$ and $\tau_{\rm rad}$ for WASP-43b and HD~209458b.

Deep in the planet, all ($T_{\rm eq}$--$p$) profiles (Figure~\ref{fig:rad}) are converged to a common temperature adiabat as is assumed in most 3D climate models \citep{Showman2008,Mayne2014,Amundsen2016}. The location of the convective layer is calculated assuming planetary averaged energy flux from the interior. The intrinsic temperature $T_{\rm int}$ is the temperature associated with the intrinsic flux ($F_{\rm int} = 4 \pi R_p^2 \sigma T_{\rm int}^4$) of the planet, i.e.~not including irradiation \citep{Barman2005}. The intrinsic or internal temperature is derived using state-of-the-art interior models \citep{Mordasini2012,Vazan2013}. The evolution model presented in \citet{Mordasini2012} is now also coupled to the non-gray atmospheric model \textit{petitCODE} and accounts for extra energy dissipation deep in the interior of the planet (Sarkis et al. in prep). We report the intrinsic temperature that reproduces the mass and radius of each planet (Table~\ref{table_params}).

3D  General Circulation Models (GCMs) with simplified thermal forcing are one possible intermediate step between 3D GCMs with full coupling between radiation and dynamics like those used by \citep{Showman2009,Amundsen2016}, and shallow water models, i.e. atmosphere models with one atmosphere layer comprising vertically averaged flow \citep{ShowmanBook}. Fully coupled GCMs have the highest accuracy in stellar radiation and flow coupling and thus the highest predictive power. They are, however, computationally much more expensive and their complexity makes it more difficult to test underlying modeling assumptions compared to GCMs with simplified thermal forcing. The latter are thus better suited to run simulations for various scenarios, to understand large scale flow and circulation properties in 3D climate models under different conditions \citep{Tsai2014,Mayne2014,Komacek2016,Liu2013,Carone2015,Carone2016,Mayne2017,Hammond2017}. Such models have been proven to be very useful: superrotation in hot Jupiters was first inferred by \citet{Showman2002} in a 3D GCM with Newtonian cooling. Recently, \citet{Showman2019} used Newtonian cooling to establish a clean, simple environment to diagnose flow dynamics in brown dwarfs, Jupiter and Saturn-like planets. Shallow water models present an even simpler model framework and represent 3D flow patterns in an atmosphere depth-dependent (2D) formalism \citep{Showman2010,Showman2011,Penn2017}. There are other useful radiative forcing parametrizations such as those using the dual-band radiative scheme, which can also explore a large parameter space and basic assumptions (see e.g. the model used by \cite{Komacek2017}). Generally, a hierarchy of models with various levels of complexity has proven to be extremely beneficial to understand complex flow patterns in full 3D climate simulations. Here, we establish a clean, simple environment to understand possible dynamical feedback between the lower boundary and observational flow via Newtonian cooling.

\subsection{Lower boundary}
\label{sec: low_boundary}
It is known within the 3D climate modelling community for hot Jupiters that flow near the lower boundary is challenging for the numerical stability of the simulations. Possible instabilities where documented also in GCMs using a different dynamical core than the one used here \citep{Menou2009,Rauscher2010,Mayne2017,Cho2015}. Flow at the bottom of the no-slip friction-less lower boundary can lead to meandering of wind jets at depth and to crashes of the simulation \citep{Menou2009,Rauscher2010}. In the past, several measures have been employed in different models to tackle model convergence problems related to deep flow:~e.g.~it was pointed out that one can circumvent problems induced by deep flow by using drag at depth \citep{Liu2013}. Other modellers converged the temperature at $p=10$~bar \citep{Mayne2014,Rauscher2010}.

In the following, we present the set-up of our nominal WASP-43b simulation, and a benchmark simulation for HD~209458b. We carefully checked that these treatments did not lead to spurious wind flow and that they reduced fluctuations at depth by running simulations with different combinations of stabilization measures. A selection of test simulations that we performed to validate different lower boundary set-ups can be found in the Appendix (Section~\ref{sec: lower_stab}). We find that once the flow at the lower boundary is stabilized, we get the same qualitative wind flow structure for all test simulations.

The measures described here also allow us to evaluate if there is impact of deep circulation on the observable planetary atmosphere in the WASP-43b-model within reasonable simulation times, including the possible depth of wind jets  (Figure~\ref{fig2m}) and their influence on the observable wind flow (Figure~\ref{fig6m}). Furthermore, we always checked these measures within the HD~209458b benchmark simulation and made sure that all stabilization measures yield results consistent with previous work \citep{Mayne2014,Showman2008}.

\subsubsection{Temperature convergence for $p\geq 10$~bar}
As one possible method of stabilization, we choose that all prescribed equilibrium temperature-pressure profiles ($T_{\rm eq}$--$p$) converge towards the planetary average below $p=10$~bar. This approach was already previously successfully used to remove lower boundary instabilities in a 3D GCM with simplified forcing for hot Jupiters  \citep{Menou2009,Mayne2014}. We choose to interpolate with a spline fit between 1 and 10~bar. It was also shown in several 3D GCMs and in 2D planet atmosphere models that temperatures in these hot Jupiter models converge to the same temperature for $p\geq 10$~bar  \citep{Tremblin2017,Amundsen2016,Kataria2015}. To make certain that the rather steep interpolation at 10~bar does not cause problems in itself for WASP-43b simulations, we performed several test simulations without any stabilization measure (see Section~\ref{sec: lower_stab}). We also testes a simulation, where we selectively switched on other stabilization measures and switched off temperature convergence at 10~bar (Figure~\ref{fig5mx}). All these simulations showed qualitatively a similar picture, once we stabilized against fluctuations at the lower boundary.  Thus, we are confident that the temperature treatment, while crude, is not in itself the cause for instabilities at the lower boundary nor is it the cause for retrograde equatorial flow over the day side in our WASP-43b simulations.

\subsubsection{Deep evolution time scale $\tau_{\rm conv}$ for $p\geq 10$~bar:}
Radiative time scales increase rapidly from $10^6-10^8$~s at $p=10$~bar to up to $10^{12}$~s in deeper atmosphere layers. Thus, so far, many GCMs for hot Jupiters have left the atmospheric layers below 10~bar unconverged, arguing that the observable atmosphere ($p<1$~bar) has already reached steady state and that still ongoing thermal evolution of the deeper atmosphere appears to have a negligible influence on the observable atmosphere  \citep{Amundsen2016}. We test here if there is dynamical feedback between deeper layers and the observable atmosphere by accelerating the evolution of the `radiatively inactive' atmospheric layers $p>10$~bar. We replace in these deep layers the long radiative time scales in the Newtonian cooling prescription with a shorter convergence time scale $\tau_{\rm conv}=10^6$~s. This approach is similar to the work of \citet{Mayne2014, Liu2013}, who have used this measure to likewise reach a full steady state from top to bottom of their hot Jupiter 3D climate models.

\subsubsection{Deep magnetic drag}
 We found that shear flow instabilities at the lower boundary can give rise to problematic behaviour that can affect the entire simulated atmospheric flow (see Section~\ref{sec: shear}). We found that we can reach complete steady state and at the same time avoid shear flow instabilities by applying Rayleigh friction which dissipates horizontal winds $\mathbf{v}$ at the lower boundary via (see Section~\ref{sec: complete_stab}):
\begin{equation}
\mathcal{\mathbf{F}}_v=-\frac{1}{\tau_{\rm fric}}\mathbf{v},
\end{equation}
where a time scale of $\tau_{\rm fric}$ is applied between the lowest boundary layer $p_{\rm boundary,bottom}$ and $p_{\rm boundary,top}$. The friction time scale $\tau_{\rm fric}$ decreases linearly between the maximum value $\tau_{\rm bottom,fric}$ to zero. The prescription for $\tau_{\rm fric}$ is:
\begin{equation}
\tau_{\rm fric}=\tau_{\rm bottom,fric} \max\left(0,\frac{p - p_{\rm boundary,top}}{p_{\rm boundary,bottom} - p_{\rm boundary,top}}\right).
\end{equation}
When we adopt deep friction with parameters $p_{\rm boundary, top}=490$~bar, $p_{\rm boundary, bottom}=700$~bar and $\tau_{\rm bottom,fric}=1$~days, this prevents shear flow instabilities and unphysical, mainly numerically driven changes in the flow pattern of the simulated observable atmosphere ($p >$1~bar). In this set-up, drag only acts on the wind flow in the deepest atmospheric layers and does not affect the observable atmosphere directly. The moderate friction time scale of $\tau_{\rm bottom,fric}=1$~days was found to be a good value for regulating the dissipation of fast wind jets at depth without yielding other numerical problems. The result is a very stable climate simulation framework that preserves the general flow properties. A similar mechanism was used in work by \citet{Liu2013} to numerically stabilize hot Jupiter climate studies.

Although the main motivation to apply deep drag was to stabilize the flow at the lower boundary, there is also a compelling physical reason for a deep atmospheric drag force: Recent Juno observations similarly indicate a truncation of deep wind jets in the interior of our Solar System Jupiter by magnetic fields  \citep{Kaspi2018}. We thus justify that our deep  drag formalism acts as a first-order parametrization of this effect. It only acts at depth ($p>100$~bar) in the atmosphere and it is thus different from the parametrized magnetic field coupling used by  \citet{Kataria2015,Parmentier2018} that is applied to the observable atmosphere ($p<1$~bar) as well.

We find that the stability of the lower boundary is in our set-up for a 3D climate model of extreme importance to yield reliable, numerically stable results for the fast-rotating, dense hot Jupiter WASP-43b. In particular, we find that our WASP-43b simulations exhibit different wind flows depending on if shear flow instabilities at depth occur or not (Section~\ref{sec: shear}, Figures~\ref{fig3m}, \ref{fig6m} and \ref{fig5m} top). When we stabilize the lower boundary such that instabilities at depth are suppressed, the wind flow in the observable atmosphere is maintained and stabilized. Thus, we choose this set-up as the nominal WASP-43b simulation (Figure~\ref{fig5m}, bottom).

\section{Results}
\label{sec: sims}
In the following, we show our main results: 3D climate simulations for WASP-43b and HD~209458b, taking into account possible dynamical feedback from atmospheric layers deeper than 100~bar. We further investigate how the planetary rotation and interior structure affect the dynamical feedback from deep atmospheric layers. We investigate under which circumstances we find the following horizontal wind flow patterns, focusing mainly on the equatorial region. First, superrotation is in hot Jupiters characterized as a fast ($u \gg 1$~km/s) eastward wind flow along the equator, which circumnavigates the whole planet. (Equatorial) prograde superrotation is thus usually present at all longitudes. Second, we find strong  ($u \ll -1$~km/s) retrograde, that is, westward flow along the equator, which is confined at the equatorial day side (longitudes: $-90^\circ$~to~$90^\circ$). We show that retrograde flow is part of the well-known Matsuno-Gill flow pattern \citep{Matsuno1966,Gill1980} that emerges in our simulations with deep wind jets.

Furthermore, we show how our WASP-43b simulations and the physical effects therein may aid our understanding of current and future observations.

\subsection{WASP-43b simulation}
\label{sec: WASP-43b}

We report that the wind jets in our WASP-43b simulation can reach pressure depths of at least 700~bar (Figure~\ref{fig1}, panel Ia). Wind jets in the hot Jupiter benchmark planet HD~2094589b are much shallower and taper off at $p\approx 100$~bar (Figure~\ref{fig1}, panel IIa). The wind jets shown in these simulations are without `deep magnetic drag' (see Section~\ref{sec: low_boundary}) to demonstrate how deep the wind jets can descend into the interior for WASP-43b. See also Figure~\ref{fig2m}.

We further report that the horizontal flow in our WASP-43b simulation deviates from that found in WASP-43b simulations by previous climate studies \citep{Mendonca2018,Kataria2015}.  We find a westward flow along the equator at the day side in part of the atmosphere as soon as we allow the model to develop deep wind jets. This retrograde flow is found in our WASP-43b simulations at the upper thermal photosphere ($p\leq 80$~mbar) and is accompanied by large day-to-night-side temperature differences ($\Delta T \approx 1200$~K at $p=12$~mbar) (Figure~\ref{fig1}, panel Ib). Among the simulation setups that we tested, the only cases in which WASP-43b develops an unimpeded equatorial superrotating wind jet are consistently linked to instabilities at the lower boundary (as demonstrated in Sections~\ref{sec: shear} and~\ref{sec: complete_stab}) fast ($u \ll -1$~km/s), which we deem to be unphysical.

At the equatorial day side, an eddy-mean-flow analysis (see Section~\ref{sec: perturb}) shows a very strong tendency for an equatorial westward (retrograde) flow for $p<0.1$~bar in our WASP-43b simulation (Figure~\ref{fig1}, panel Ic). Regions of strong vertical transport of zonal momentum are identified in a similar fashion in our simulations by analysing deviations from the zonally or longitudinally averaged product of the upward velocity $w$ and eastward velocity $u$. After performing the latter analysis, we report for WASP-43b also a strong upward transport of horizontal zonal momentum between 20 and 100~bar (Figure \ref{fig1}, panel Id, yellow region).\footnote{It should be noted that the vertical axis is in pressure and that pressure decreases with height. Therefore, an upward tendency is a negative tendency in pressure vertical coordinates. Analogously, positive zonal momentum is eastward, negative is westward. Upward transport of westward momentum results thus in a net-positive tendency, as observed here.}

We hence conclude that deep circulation provides a zonal momentum reservoir at depth $p>10$~bar, which is sufficient to disrupt the equatorial eastward jet (superrotation) in the case of WASP-43b, by enforcing retrograde flow on the day side via upward transport of zonal momentum.

\subsection{HD~209458b simulation}
\label{sec: HD20}
We test our model framework also with the benchmark planet HD~209458b  \citep{Showman2008, Showman2009, Heng2011, Mayne2014, Amundsen2016}. Thus, we want to identify why HD~209458b is observed to have a significantly smaller day-to-night-side temperature difference and a larger westward hot spot shift compared to WASP-43b  despite having similar effective temperatures ($T_{\rm eff,Pl}\approx 1450$~K). We postulate that these differences may also be due to differences in dynamics, which may led to weakly efficient horizontal heat transfer for WASP-43b and strongly efficient horizontal heat transfer for HD~209458b in one and the same 3D climate frame work. Cloud effects also need to be taken into account for a comprehensive comparison of heat circulation, but in this work we focus first on the basic flow properties, which sets the stage in temperature and vertical mixing for cloud formation.

Another notable difference between WASP-43b and HD~209458b is the planets' interior structure. HD~209458b is inflated, i.e.~this planet has a larger radius and consequently a lower density than expected from planetary evolution models. To explain the puffiness of such hot Jupiters, an inflation mechanism is assumed to inject energy into the interior of this planet, e.g.~via flow interactions with the planetary magnetic field or Ohmic dissipation \citep{Batygin2010, Thorngren2018}. In contrast to that, WASP-43b is not inflated. In fact, WASP-43b is roughly eight times denser than HD~209458b.

The higher density of WASP-43b compared to HD~209458b has two consequences. One consequence is that the radiative time scales of WASP-43b are larger by a factor of eight compared to HD~209458b for the same pressure and temperature range (Figure~\ref{fig:rad}), due to the difference in density (see Equation~(\ref{eq:tau_rad}) and Table~\ref{table_params}). Thus, even though the two planets have similar effective temperatures, the thermal forcing is not exactly the same. However, the difference is less than one full order of magnitude. The other consequence is that the fully convective layer in HD~209458b is located at lower pressure levels (higher up in the atmosphere) compared to WASP-43b due to the much higher intrinsic or internal temperature required to explain the inflated radius of the former (see Section~\ref{sec: low_boundary} and Figure~\ref{fig:rad}).

\begin{figure*}%
  \centering%
  \includegraphics[width=1.5\columnwidth]{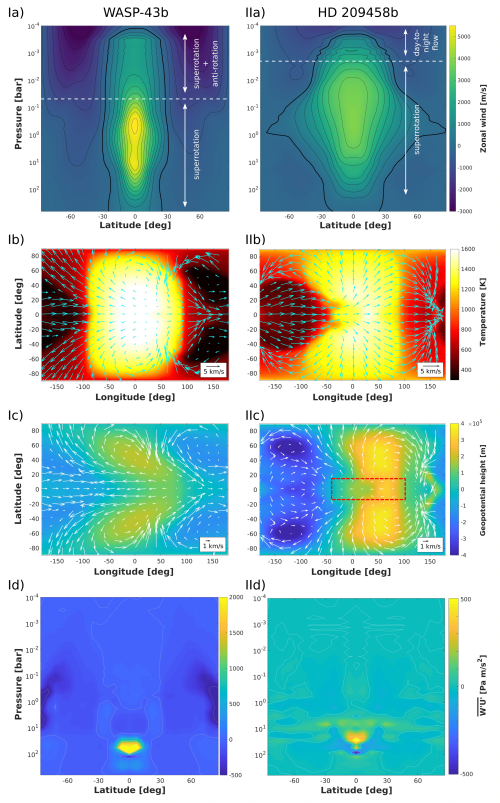}%
  \caption{Basic climate properties of the WASP-43b (I) and HD~209458b (II) simulations. The top row (\textit{a}) shows longitudinally averaged zonal (eastward) wind speed versus vertical pressure coordinates with regions of a dominant horizontal wind flow type (arrows). Panels Ia and IIa show simulations without lower boundary friction (`deep magnetic drag') to elucidate the possible depth of wind jets. Contours are shown every 500 m/s, the contour of 0 m/s is bold. The second row (\textit{b}) shows temperature and horizontal wind velocity at $p = 12$~mbar in a longitude-latitude map, where the substellar point is at (0$^\circ$,0$^\circ$). The third row (\textit{c}) shows the eddy geopotential height and eddy wind velocities which display deviations from the (zonal) mean climate state (see Section~\ref{sec: perturb}). The red dashed box highlights the retrograde eddy wind flow at the equatorial day side in HD~209458b. The bottom row (\textit{d}) shows the longitudinally averaged vertical transport of the zonal momentum $\overline{w'u'}$. Note that the color bars are scaled differently between panels Id and IId.}%
  \label{fig1}%
\end{figure*}

We investigate in the following if deep circulation can also be present in HD~209458b, and if differences in the depth of wind jets between WASP-43b and HD~209458b can explain why the former is less efficient in heat transfer than the latter. Extending the lower boundary downwards, we find that the HD~209458b simulation with $p_{\text{boundary}}=700$~bar develops full equatorial superrotation for $p\geq 10^{-3}$~bar, that is, in the planetary photosphere (Figure~\ref{fig1}, panel IIa,b). The simulated superrotating flow is very similar to the superrotation reported in other 3D GCMs \citep{Mayne2017,Mayne2014,Showman2008}.

\begin{figure}%
  \centering%
  \includegraphics[width=\columnwidth]{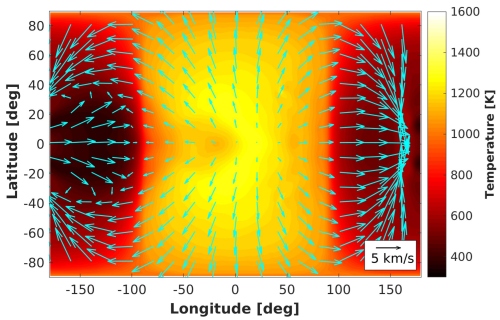}%
  \caption{Temperature map and flow field for our nominal HD~209458b simulation at a pressure level of 0.6~mbar. The arrows, displaying the speed and direction of the horizontal wind flow, can be seen to point radially from the substellar point to the night side, indicating a direct day-to-night-side flow characterized by short radiative time scales.}%
  \label{fig:day-to-night}%
\end{figure}

In even higher atmospheric levels (for $p < 10^{-3}$~bar), we find that direct day-to-night-side flow starts to become dominant (Figure~\ref{fig:day-to-night}). Direct day-to-night-side flow is also observed in \citet{Showman2008} with their climate model with $p_{\text{boundary}}=200$~bar and simplified forcing (their Figure 5) and in \citet{Rauscher2014} with their climate model, using double-gray radiative transfer (their Figure 2 for synchronous rotation). Direct flow has also been confirmed to be dominant for $p<10^{-3}$~bar in the hot Jupiter HD~189733b \citep{Flowers2018, Brogi2016}. Here, it should be noted that the emergence of direct versus jet-dominated wind flow at the very upper atmosphere mainly depends on the relation between the dynamical and radiative responses in the atmosphere \citep{ZhangShowman2017, Perez-Becker2013}. If radiative time scales are much shorter than the time scales associated with wave propagation, then the formation of jets is suppressed in favor of thermally direct, radial day-to-night-side flow. The radiative time scales for the planets we investigate here are similar within one order of magnitude ($\tau_{\rm rad}=10^3-10^4$~s for $p<10^{-3}$~bar, Figure~\ref{fig:rad}). Clearly, for our nominal HD~209458b simulation ($P_{\text{orb}}=3.5$~days) the dynamical response time is larger than $10^4$~s to allow direct flow to emerge at the very upper atmosphere. In the WASP-43b simulation ($P_{\text{orb}}=0.8135$~days), on the other hand, direct flow does not emerge at the top despite similar radiative time scales, because the wave response is apparently faster. We show later that direct flow does emerge at the top for a WASP-43b-like simulation with much slower rotation ($P_{\text{orb}}=3.5$~days), elucidating that faster rotation decreases the wave response time.

Our main focus in this work is, however, on possible feedback between much deeper layers and observable jet-dominated atmospheric layers. For our HD~209458b simulations, we find that the horizontal flow patterns in simulations with a very deep boundary are the same compared to another simulation, where we put the lower boundary at $p_{\text{boundary}}=200$~bar and leave the lower boundary unconverged\footnote{Not shown here.}.

The case of HD~209458b demonstrates that we indeed recover full equatorial superrotation -- even when we extend the lower boundary downwards and apply all our stabilization measures for the deep ($p>10$~bar) atmosphere. Therefore, we have ensured that our lower boundary stabilization measures do not in itself affect the wind flow patterns in the observable atmosphere in a significant way, by hindering the formation of a superrotating jet stream. Based on the HD~2094598b simulation alone, we would reach the same conclusion as \citet{Amundsen2016}: that the state of the deep atmosphere can be neglected for comparison with observations, since it does not impact the observable atmosphere. The opposite is true for WASP-43b. For this hot Jupiter, the deep atmosphere is vital for a full understanding of the horizontal wind flow.

\subsection{Comparison eddy wind flow, actual wind flow and link to deep zonal momentum in HD~209458b and WASP-43b}
\label{sec: eddy_comp}

 For a first diagnosis, we compare the momentum transport and wind flow in the WASP-43b and HD~209458b simulations, we separate the wind flow field into its mean (zonal average) and eddy (deviations from the zonal average) components (see Appendix~\ref{sec: perturb}).

\begin{figure*}
\includegraphics[width=0.6\textwidth]{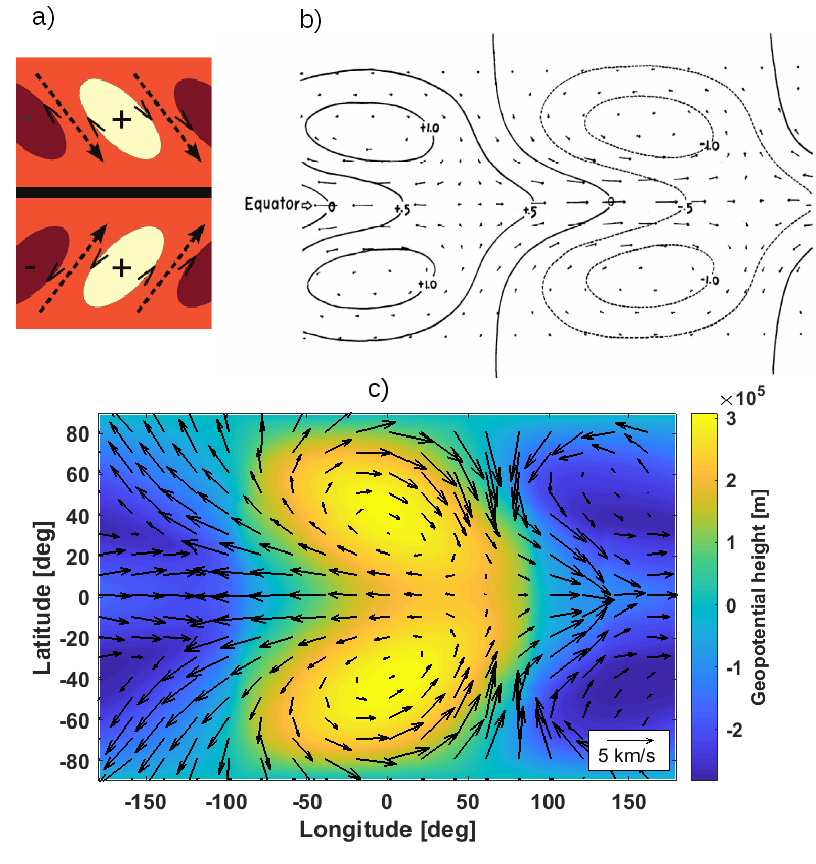}
\caption{a) Schematic of the tilt of the Rossby wave gyres from \citet{Showman2011} that generate superrotation. b) \textit{Matsuno-Gill} flow patterns from \citet{Matsuno1966}.
c) The eddy geopotential wind flow $\mathbf{(u',v')}$ (arrows) and geopotential height anomaly $z'$ (colour) in our (nominal) WASP-43b simulation at $p=12$~mbar, which is repeated here for a better comparison with theoretical concepts.}
\label{fig: WASP43b_eddy}
\end{figure*}

In the WASP-43b simulation, the eddy horizontal wind flow (Figure~\ref{fig: WASP43b_eddy}) reveals two theoretical mechanisms that drive the wind flow in tidally locked planets: a) a tilt of Rossby wave gyres, which was identified in \citet{Showman2011} as the underlying reason for equatorial superrotation and b) the Matsuno-Gill flow pattern from \citet{Matsuno1966}. The tilt of the eddy Rossby wave gyres are stronger in WASP-43b compared to HD~209458b due to the faster rotation of the former ($P_{\text{rot}}=0.8135$~days compared to $P_{\text{rot}}=3.5$~days, respectively, compare also Figure~\ref{fig1} panel Ic and II c). The connection between rotation and tilt of Rossby gyres was also outlined for tidally locked Exo-Earths in \citet{Carone2015}.

We note here that superrotation emerging from the tilt in Rossby wave gyres and the Matsuno-Gill wave response are closely related but separate mechanisms. The former arises out of the latter, as can be clearly seen in Figure~10 of \citet{Showman2011}. There, the formation of a Matsuno-Gill flow pattern, of which tilting Rossby wave gyres are part, is explicitly identified during spin-up of a typical hot Jupiter simulation before the formation of equatorial superrotation. The zonal jet needs time to develop out of the momentum transport due to the shear between the tilted Rossby gyres and the Kelvin wave, and thus typically supersedes the original Matsuno-Gill flow in the emergent wind flow later in the simulation. Furthermore, as already pointed out in \citet{Showman2010}, whereas superrotation requires the formation of the Matsuno-Gill wave response, the inverse is not necessarily true.

We further note that in the WASP-43b simulation the westward eddy flow at the equatorial day side (latitude: $0^\circ$ and longitudes $-50^\circ$ to $50^\circ$ (Figures~\ref{fig1} panel Ic and \ref{fig: WASP43b_eddy} c) is reflected by actual westward flow at the equatorial day side. There, we find it to be dominant between the morning terminator and substellar point (longitude $-90^\circ$ to $0^\circ$, Figure~\ref{fig1} panel Ib).

In the HD~209458b simulation (Figure~\ref{fig1} panel Ic), we find strong similarities in the eddy horizontal wind flow compared to the WASP-43b simulation (Figure~\ref{fig1} panel IIc) despite displaying very different (actual) flow patterns  (Figure~\ref{fig1} panel Ib and II b). We also find here a westward eddy wind flow along the equator (albeit much weaker), exactly at the same longitudes where we similarly find a westward eddy wind in our WASP-43b simulation (longitude $-50^\circ$ to $50^\circ$, marked by a red rectangle). In the actual flow along the equator, the wind flow is always eastward (superrotating) along the equator.

Upon careful inspection of the emerging total flow (Figure~\ref{fig1} panel Ib) it becomes clear that superrotation is not uniformily strong across all latitudes: between morning terminator and substellar point (longitude $-90^\circ$ to $0^\circ$), there is a weakening of the eastward wind strength compared to the wind strength at the evening terminator (longitude $90^\circ$). Thus, we conclude that although equatorial superrotation is dominant in our HD~209458b simulations, equatorial retrograde appears to be `lurking' in the background.

For WASP-43b, we find further that the tendency for equatorial retrograde flow is linked to vertical zonal momentum transport from depth with maxima at around 70-100~bar. When we examine the vertical momentum transport in our nominal HD~209458b simulation, we also find (weak) zonal momentum transport and at shallower depth (between 10 and 40~bar) compared to our WASP-43b nominal simulation  (Figure \ref{fig1}, panel IId, yellow region). We note here that the vertical eddy momentum flux for the HD~209458b simulation is qualitatively similar to the same property investigated by \citet{Showman2015} for hot and slowly rotating (8.8~days) planets (their Figure 7b). The relative weakness of vertical zonal momentum transport is in line with our interpretation that retrograde flow can be also elicited in the HD~209458b simulation, but is too weak to emerge into the foreground in the actual wind flow. The wind flow is instead dominated by the other mechanism: strong equatorial superrotation. In the WASP-43b simulation, contrarily, equatorial retrograde flow emerges as the dominant wind flow at the day side for $p<0.1$~bar (compare Figure~\ref{fig1} panel II d with Id, respectively).

The underlying physical reason for the vertical momentum transport and retrograde equatorial flow is strongly connected to the orbital period of the hot Jupiter that has to be shorter than 1.5~days, as will be shown in the next section and discussed in more details in Sections~\ref{sec: emergence_westward}, \ref{sec: Rhines}, \ref{sec: EP}.

\subsection{Climate evolution with different orbital periods}
\label{sec: Rotation}

We performed several additional simulations to pinpoint the parameters influencing the retrograde wind flow.
\begin{figure*}%
	\centering%
    \includegraphics[width=1.95\columnwidth]{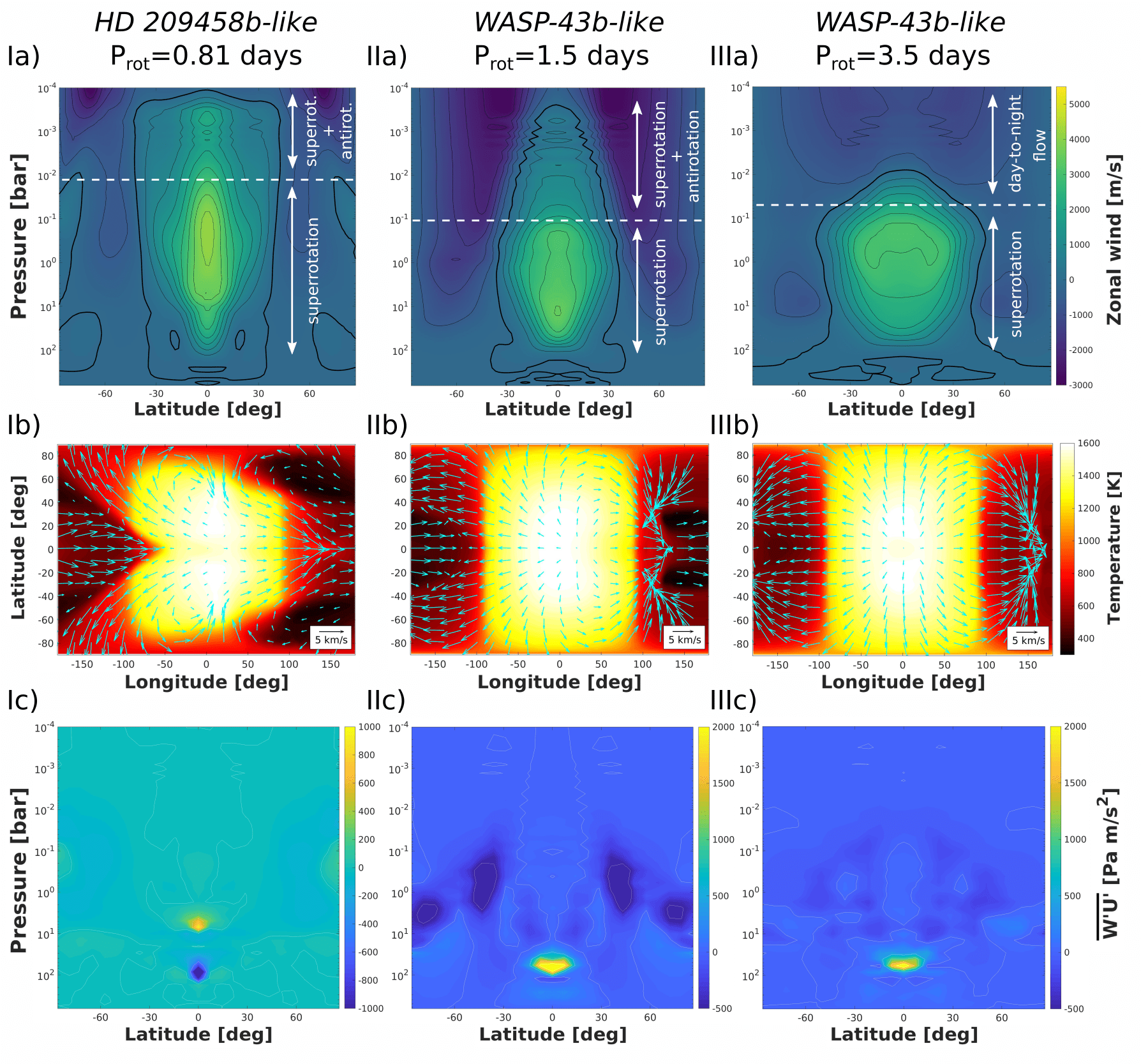}%
	\caption{Climate simulations for different orbital (rotation) periods: an HD~209458-like planet with short period ($P_{\rm rot} = P_{\rm orb} = P_{\text{WASP-43b}} = 0.8135$ days), a WASP-43b-like planet with an intermediate rotation period ($P_{\rm rot} = P_{\rm orb} = 1.5$ days), and a WASP-43b-like planet with long period ($P_{\rm rot} = P_{\text{HD~209458b}} = 3.5$ days). The top row (\textit{a}) shows the longitudinally averaged zonal (eastward) wind speed versus vertical pressure coordinate. Contours are displayed for 500 m/s intervals with the 0 m/s contour shown bold. The middle row (\textit{b}) shows maps of the temperature and horizontal wind at $p=12$ mbar. The bottom row (\textit{c}) shows the longitudinally averaged vertical transport of the zonal momentum $\overline{w'u'}$. Note that the colors are scaled differently between I\textit{c}, II\textit{c} and III\textit{c}.}%
    \label{fig2}%
\end{figure*}

We report that a fast HD~2094589b-like case, where the rotation is set to the rotation rate of WASP-43b ($P_{\rm orb}=P_{\rm rot}=0.8135$~days), develops narrower wind jets below $p=10$~bar that are not as strong as similar jets in the WASP-43b simulation (Compare Figure~\ref{fig1} panel Ia with Figure~\ref{fig2} panel Ia). Vertical momentum transport is stronger compared to the nominal HD~209458b simulation, but weaker compared to WASP-43b by a factor of 2 (compare Figure~\ref{fig2}, panel Ia versus Figure~\ref{fig1}, panel Ia and Figure~\ref{fig2}, panel Ic versus Figure~\ref{fig1}, panel Id, respectively). As a result, a retrograde flow pattern also emerges in the fast HD~209458b-like simulation, but at relatively high altitudes ($p\leq 10^{-2}$~bar). The retrograde flow pattern further appears weaker compared to the WASP-43b nominal simulation with the same rotation period (Figure~\ref{fig1}, panel Ib).

Correspondingly, we find that retrograde flow at the equatorial day side of a WASP-43b-like simulation becomes weaker, when the rotation period is increased. An intermediate rotation period of $P_{\rm orb}\approx 1.5$~days (Figure~\ref{fig2}, panels IIa,b,c) already shows a weaker westward flow at the day side, while a longer rotation period of $P_{\rm orb}\approx 3.5$~days completely removes it. In the latter case,  retrograde flow is replaced at the top ($p<10^{-2}$~bar) by a direct day-to-night-side flow and at deeper layers by a fully superrotating jet stream  (Figure~\ref{fig2}, panels IIIa,b,c). The broad equatorial jet stream of the slowly rotating WASP-43b-like simulations is similar to the zonal wind structure obtained by \citet{Showman2008} for their HD~209458b simulation with simplified thermal forcing, when its rotation is slowed down by a factor of two (their Figure~9, bottom).

Thus, our results imply that predominantly exoplanets with short orbital periods ($P_{\rm orb}\leq 1.5$~days) are prone to develop deep wind jets and thus may have retrograde flow emerging at the day side in the observable horizontal wind flow ($p\leq 1$~bar). This result is in line with previous theoretical work in shallow water models \citep{Penn2017}, which predict strong retrograde flow to occur for $P_{\rm orb}\leq 2$~days. These results also explain why simulations of HD~189733b ($P_{\rm orb}= 2.2$~days) do not appear to exhibit strong deviations from superrotation, even when the lower boundary is placed deeper than 200~bar \citep{Showman2009,Showman2015}. A more thorough analysis of how and why equatorial superrotation may be perturbed in fast rotating, dense hot Jupiters will be performed and analysed in Sections~ \ref{sec: Rhines} and \ref{sec: EP}.

We further note that for the slow ($P_{\rm orb}=3.5$~days) WASP-43b-like simulation, thermally direct day-to-night-side flow emerges in the upper atmosphere, like in the nominal HD~209458b simulation (Figure~\ref{fig:day-to-night}). Conversely, in the fast ($P_{\text orb}=0.8135$~days) HD~209458b-like simulation, the wind-jet dominated regime extends further upwards compared to the nominal simulation, indicating that the dynamical wave response in the upper atmosphere is faster with faster planetary rotation as already noted in Section~\ref{sec: HD20}.

\subsection{Comparison to observations of WASP-43b}
\label{sec: Comparison_obs}
3D GCMs with a simplified treatment of coupling between wind flow and irradiation are less suited for quantitative comparison with observational data than 3D GCMs with full radiative coupling (see e.g. \citet{Amundsen2014}).  However, they are well suited to investigate general flow tendencies and to test underlying assumptions and, as in this case, physical mechanisms that can shape the flow. In this work, we focus predominantly on horizontal flow patterns at the equator, which may be influenced by deep wind jets. One possible outcome is full superrotation if there are no deep wind jets present, leading to efficient day-to-night-side heat transport. Another possible outcome is equatorial retrograde flow at the day side, embedded in at least a part of the planet's observable atmosphere. This would inhibit heat transport from the day towards the night side. Thus, the difference between full superrotation and superrotation with embedded retrograde flow at the equatorial day side could be diagnosed via observed day-to-night-side temperature differences in WASP-43b.

Spitzer observations of WASP-43b \citep{Stevenson2017} reveal predominately the heat flux and wind flow structure at the day-side equator and for $p<0.1$~bar (see \citet{Zhang2017}). These are the same atmosphere pressure levels, where we find an embedded retrograde flow at the day side in our WASP-43b simulations. We use \textit{petitCODE} \citep{Molliere2015, Molliere2017} using equilibrium chemistry to calculate thermal emission spectra averaged over the different planetary phases visible during one orbit from our simulated 3D climate models in post-processing. The disk-integrated spectra are obtained from constructing 400 vertical temperature columns, evenly spaced in longitude and latitude. To obtain the vertical temperature profiles we interpolate the 3D temperature solution of the GCM using radial basis functions. For every column, we then use \textit{petitCODE} to calculate the angle-dependent intensity at the top of the atmosphere. In total, \textit{petitCODE} solves for the radiation field along 40 angles, spaced on a Gaussian grid in $\mu = \cos(\theta)$ space. This results in the intensities at the planetary surface. The planetary coordinate frame is then rotated such that its new pole is the point directly facing the observer. The disk-integrated flux is then calculated from segmenting the hemisphere that faces the observer in tiles along the new latitude and longitude directions, calculating the projected surface area of said tiles, interpolating the local angle-dependent intensity to yield the intensity traveling towards the observer. Finally, we integrate the total flux received by the observer by integrating over the solid angle containing the visible planet hemisphere. The angle between the imaginative detector surface and incoming intensity is also taken into account.

The resulting thermal phase curves are based on our WASP-43b simulations for different lower boundary prescriptions: with lower boundary at depth $p_{\text{boundary}}=700$~bar and magnetic drag (`nominal'), with lower boundary at depth $p_{\text{boundary}}=700$~bar and without deep magnetic drag  (`temp+rad\_stab', see also Table~\ref{table_sims}), and with $p_{\text{boundary}}=200$~bar and no lower boundary stabilization (`no\_stab', see Figure~\ref{fig2m} top panel for this particular version). The first two WASP-43b simulations have embedded retrograde flow at the day side, whereas the third has very strong superrotating flow throughout the atmosphere\footnote{This simulation never reaches a full steady-state (see Figure~\ref{fig2m}, top). The data for the phase curve calculation were taken after 1900~days, corresponding to a fast superrotating jet stream.}.

\begin{figure*}%
	\centering%
     \includegraphics[width=0.95\columnwidth]{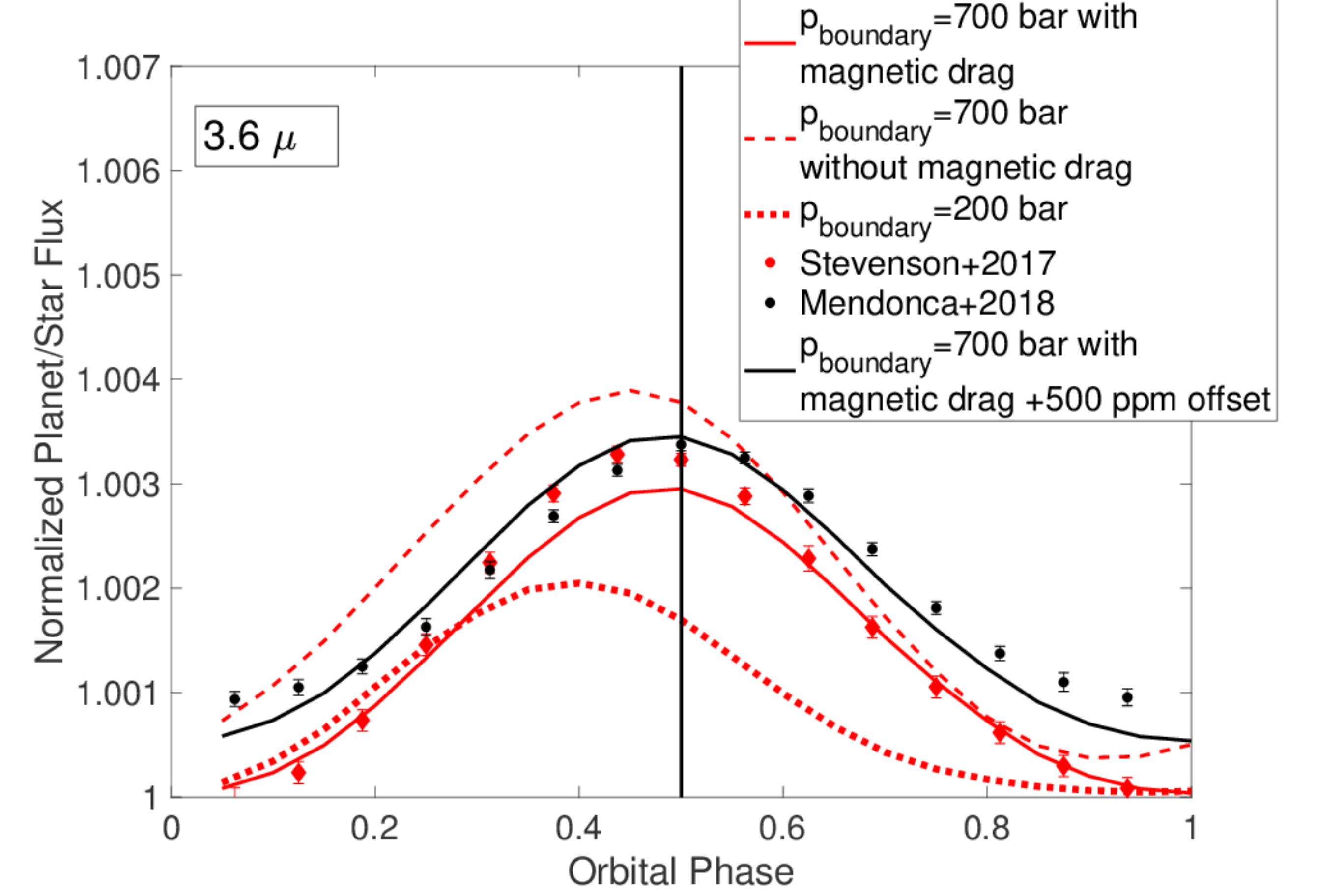}
    \includegraphics[width=0.95\columnwidth]{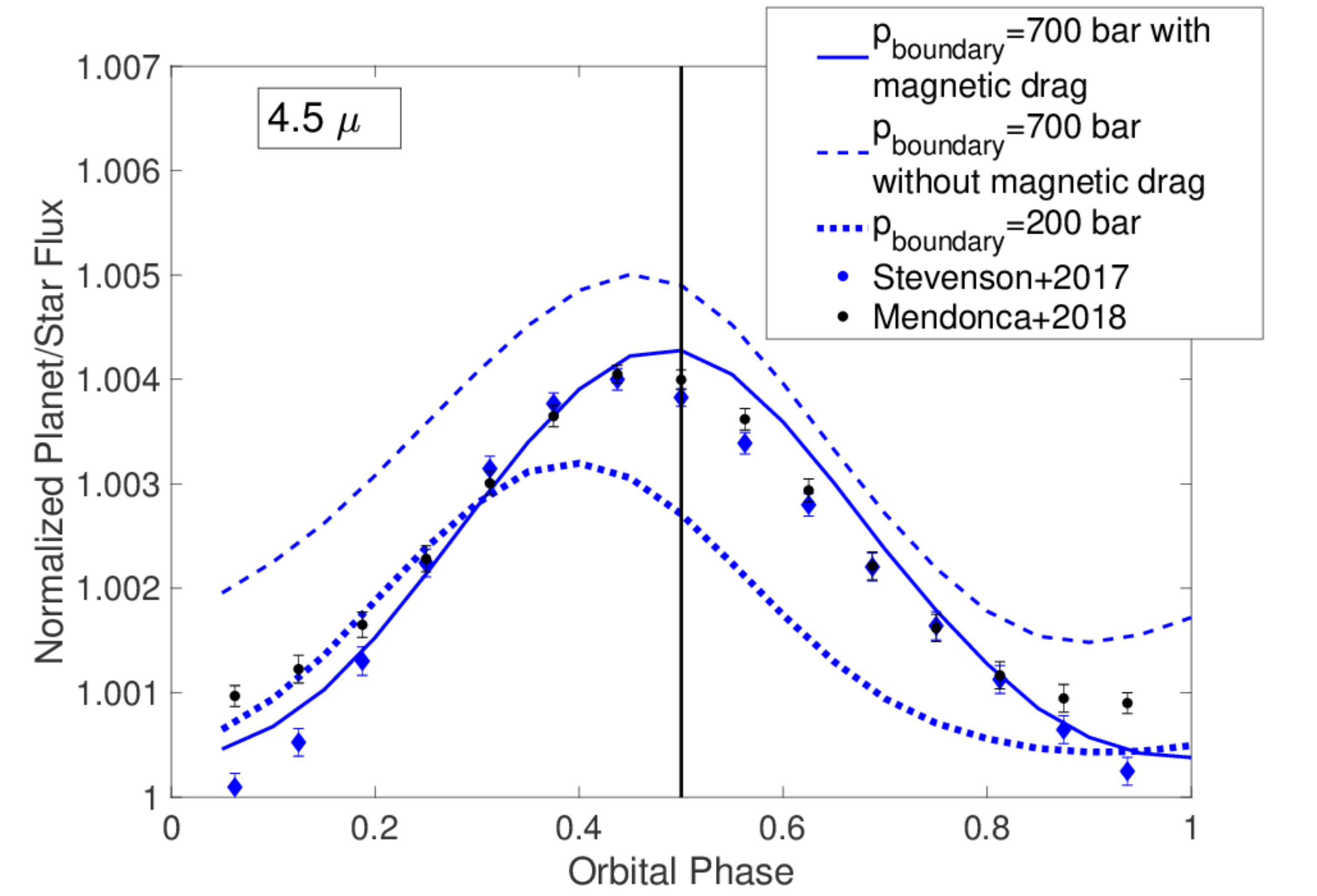}
	\caption{Predictions for orbital phase curve observations of the relative thermal emission (planetary/stellar) normalized to 1 as in \citet{Stevenson2017} of WASP-43b in Spitzer IRAC 1 channel (3.6 micron, right panel) and IRAC 2 channel (4.5 micron, left panel). Predictions are based on different simulations in our deep circulation model framework (solid red lines: simulation with deep zonal wind jet truncation by magnetic drag; solid black line: same as red solid line but with 500 ppm offset, dashed lines: simulation without deep zonal wind jet truncation; dotted lines: shallow model with $p_{\text{boundary}}=200$~bar; diamonds with error bars: WASP-43b Spitzer observations  \citep{Stevenson2017}, where the 3.6 micron data are taken from visit 2). Black dots and errorbars are from the reanalysis by \citet{Mendonca2018}. Solid vertical lines denote the orbital phase 0.5 or the day-side facing observation. }%
    \label{fig4}%
\end{figure*}

From all our models, the nominal WASP-43b simulation with deep magnetic drag agrees best with the large day-to-night-side contrast in Spitzer observed by \citet{Stevenson2017} (Figure~\ref{fig4}, red solid line and diamonds). However, even with this `best' model that exhibits very inefficient horizontal heat transport, both our predicted day-to-night-side contrast and hot spot offset are still smaller in the IRAC 1 channel (Figure~\ref{fig4}, left panel) compared to the data by \citet{Stevenson2017}. Interestingly, in this channel we agree to first order with the re-analysed Spitzer data by \citet{Mendonca2018} (black dots) in terms of day-to-night-side gradient and heat spot shift. This is best seen when adding a 500 ppm offset to the original prediction (black line in the left panel). Apparently, our first order prediction based on a simplified model yields too little flux in the IRAC 1 channel compared to the data reported by \citet{Mendonca2018}. In the IRAC 2 channel (Figure~\ref{fig4}, right panel), our WASP-43b simulation lies in between \citet{Stevenson2017} and \citet{Mendonca2018} for the night side (orbital phases 0-0.25 and 0.75 -1) and yields a slightly too small hot spot offset compared to both \citet{Stevenson2017} and \citet{Mendonca2018}. Again we stress that we are `only' seeking to compare the qualitative not quantitative properties of our simplified GCM simulations with observations to understand if these, and the physical effect that lead to deviations from superrotation, could aid our understanding of WASP-43b.

The model with deep lower boundary and without deep magnetic drag reproduces the eastward shift postulated by \citet{Stevenson2017} best in both channels, but also yields too much thermal flux in those channels (dotted line). The shallow WASP-43b simulation with $p_{\text{boundary}}=200$~bar produces an eastward hot spot shift that is always too large and a much too shallow day-to-night-side thermal contrast due to very efficient horizontal heat transport via superrotation (dashed line). Again, we stress that the simplified climate model that we use here is designed to investigate horizontal heat transport and how basic differences in climate states would influence observations in terms of day-to-night-side thermal flux contrast and hot spot offset. Here, the models clearly highlight how changes in the wind flow pattern in the thermosphere yield very different results - just due to dynamical reasons.

Considering that the deep frictional drag is a parametrized way to include the coupling of the deep atmospheric layers with the magnetic field of the planet, we tentatively argue that weaker or stronger deviations from equatorial superrotation in hot Jupiters (via retrograde wind flow) will shed a light on where the planetary magnetic field couples to the wind flow at depth. `Deep magnetic drag' is presented here as an alternative way to include the effect of magnetic fields in 3D climate simulations compared to previous work  \citep{Rogers2014b,Kataria2015,Parmentier2018}.

\begin{figure*}%
	\centering%
    \includegraphics[width=0.95\columnwidth, height=5cm]{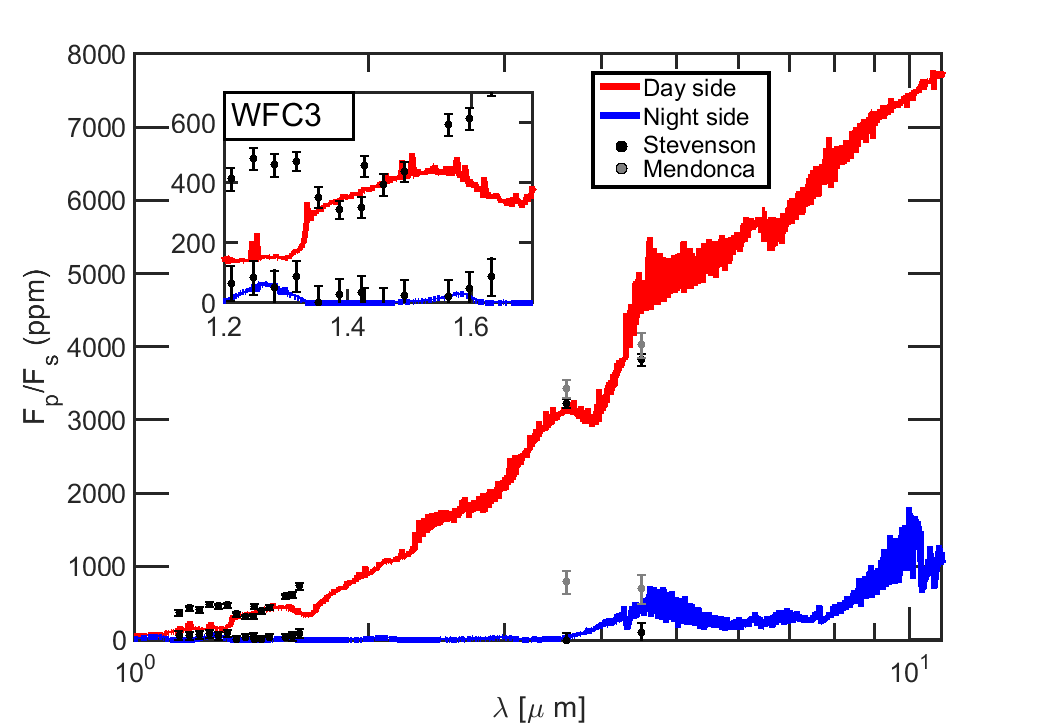}%
    \includegraphics[width=0.95\columnwidth]{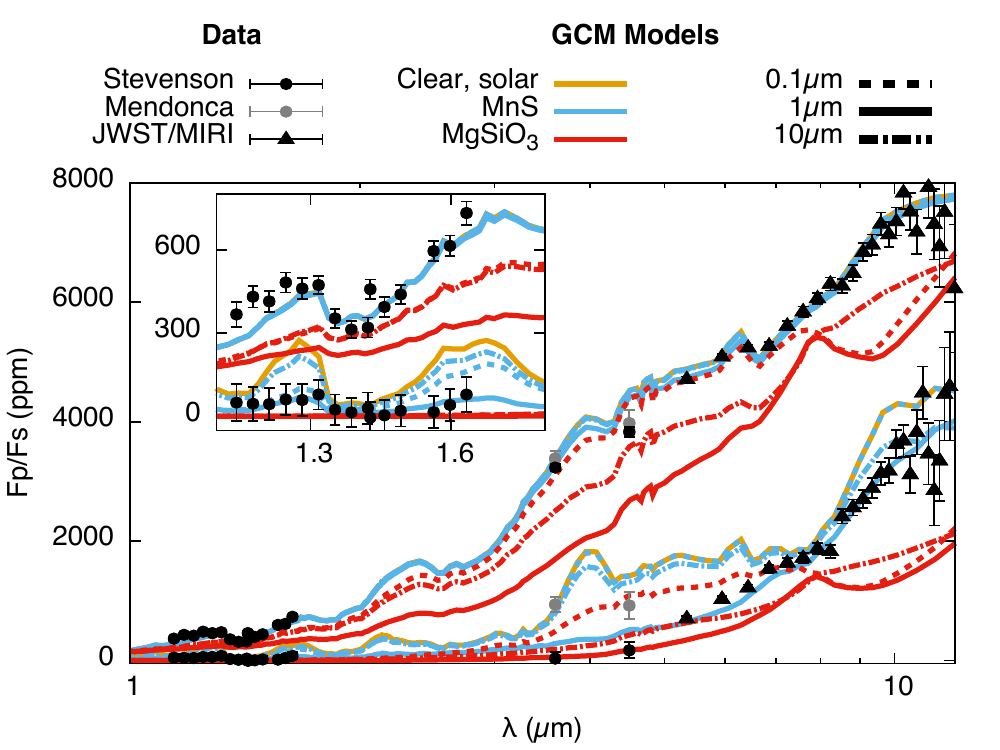}%
	\caption{\textbf{Left:} Integrated day-side (red) and night-side (blue) spectrum of WASP-43b predicted by our 3D cloudless climate model for our nominal simulation with a deep boundary ($p_{\text{boundary}}$=700~bar) and deep magnetic drag. Here, the relative planetary/stellar flux is not normalized and expressed in units of ppm. The inlay shows details of the day-side/night-side spectra for the HST/WFC3 measurements. In our simulation, a water emission band instead of a water absorption band is seen in the spectrum, which is due to a dynamically induced temperature inversion at the day side (Figure~\ref{fig:T_slice}). The black dots denote the observations reported by \citet{Stevenson2017}, the gray dots denote the reanalysis of these data by \citet{Mendonca2018}.\newline
	\textbf{Right}: Day-side and night-side spectrum of WASP-43b predicted by the 3D climate model of \citet{Parmentier2016} with full superrotation (from \citet{Venot2020}, with permission of V. Parmentier). Their day-side thermal emission is very similar to the emission derived from our model with equatorial retrograde flow for $\lambda=5-12$~micron for the cloud-free and MnS cloud case. Their night-side thermal emission is, however, higher for $\lambda=5-12$~micron - even with $\rm MgSiO_3$ clouds.}%
    \label{fig:spec}%
\end{figure*}

The modeled emission spectrum (Figure~\ref{fig:spec}), again using \textit{petitCODE} \citep{Molliere2015, Molliere2017}, shows that the day-to-night-side flux contrast appears to agree to first order with observations of WASP-43b with Spitzer \citep{Stevenson2017,Mendonca2018}. Furthermore, our nominal WASP-43b simulation also recovers the very low night-side flux, which is observed with HST/WFC3 \citep{Stevenson2017}. There is, however, a water emission feature in the predicted spectra, which is instead observed in absorption with HST/WFC3  between 1.1~$\mu$m and 1.7~$\mu$m. This discrepancy can be directly linked to a strong temperature inversion (Figure~\ref{fig:T_slice}) below 1 bar at the cloudless day side in our WASP-43b simulation (see also Section~\ref{sec: shortcomings} for a more detailed discussion).

We stress again that our simplified thermal forcing is well suited to yield general predictions for flow patterns and thus heat transfer tendencies without any strong absorption features, which probe specific parts of the atmosphere for which more complex models like those of \citet{Kataria2015,Mendonca2018,Parmentier2016,Mendonca2018b} are more suitable. The shortcomings of our simplified model clearly shows when comparing with HST/WFC3 data \citep{Stevenson2014}, which covers a strong water feature as already discussed in this section and will be also addressed in Section~\ref{sec: shortcomings}. While there are some disagreements between the predicted spectrum and the observational data for the day-side water absorption feature, the overall day-side and night-side thermal fluxes appear to be qualitatively in agreement with HST and Spitzer data. However, given the ongoing debate of the Spitzer WASP-43b night side measurements, with now three separate analyses of the same measurements \citep{Stevenson2017,Mendonca2018,Morello2019}, and also noting that K-band measurements appear to favor poor heat distribution \citep{Chen2014}, we argue that JWST/MIRI measurements are needed to clarify differences between models \citep{Kataria2015,Mendonca2018b,Parmentier2016} and different interpretations of Spitzer data, as is clearly shown in Figure~\ref{fig:spec}.

The James Webb Space Telescope (JWST) will observe WASP-43b in the near- to mid-infrared range \citep{Bean2018}. JWST observations, in particular in the mid-infrared range (10 -15 micron) should in principle be able to constrain the heat distribution efficiency and thus either confirm or disprove inefficient horizontal heat transport as could be caused by equatorial retrograde wind flow at the day side of WASP-43b.

\section{Discussion}
\label{sec: results}
Using the same Newtonian cooling formalism for WASP-43b and HD~209458b, and implementing the same stabilization measures for the lower boundary of the 3D climate model, we find crucial differences in  model stability between the simulations for HD~209458b and WASP-43b (Section~\ref{sec: disc_deep_boundary}). Furthermore, the wind flow structure in WASP-43b appears to be different from HD~209458b. We link these differences to the very fast rotation of WASP-43b compared to HD~209458b, which apparently gives rise to wave activity in our model set-up.  While we showed that in principle JWST/MIRI could confirm if the results of our WASP-43b simulation are grounded in reality, there are several shortcomings in the predictions based on our simplified 3D GCM, which we will address in more detail in this section.

\subsection{The effects of deep wind jets on the stability of the lower boundary}
\label{sec: disc_deep_boundary}

Generally we find that the HD~209459b simulation is insensitive to how the lower boundary is implemented, whereas the WASP-43b simulation is highly sensitive to the lower boundary set-up. Our simulations are set up with a free slip lower boundary condition as in e.g. \textit{MITgcm/SPARC} \citep{Showman2008}. Thus, the vertical velocity components are set to zero and there is no restriction on the horizontal components. If we then place the lower boundary at $p_{\text{boundary}}=200$~bar, as is customary in many 3D GCMs \citep{Mendonca2018,Kataria2015}, the WASP-43b simulation becomes highly unstable (See Section~\ref{sec: lower_stab}), Figure~\ref{fig2m}).

We postulate in this work that the underlying reason for the greater instability of WASP-43b compared to HD~209458b in our set-up is the tendency of the fast rotator WASP-43b ($P_{\text{orb}}=P_{\text{rot}}=0.8135$~days) to form deep wind jets.  Once the wind flow at depth exceeds 1~km/s, fluctuations of the jet occur at depth, which de-stabilize the model (Figure~\ref{fig3m}). A similar effect was also identified in \citet{Menou2009} their Figures 3 and 6 and \citet{Rauscher2010} their Figure 2, where these authors used a spectral grid and not a finite grid as in this work.

We also find that we cannot avoid shear flow instabilities in our WASP-43b simulation, when we place $p_{\text{boundary}}$ at 200~bar. At least $p_{\text{boundary}}=700$~bar is needed for full stabilization of the simulation. Furthermore, we find that significant vertical transport of zonal momentum can occur at 100~bar in the WASP-43b simulation (Section~\ref{sec: eddy_comp}). We will discuss in Section~\ref{sec: emergence_westward} the physical effect that may lie behind this feature, which in itself may justify to resolve layers down to 700 bar as these layers may be still part of the meteorologically active atmosphere.

Furthermore, we tested several deep stabilization methods described in Section~\ref{sec: low_boundary}: the extension of lower boundary downwards to $p_{\text{boundary}}=700-1500$~bars, a deep convergence time scale, a temperature stabilization and deep drag). We relied here on previous studies that outlined different ways to reach complete steady state down to 100 bar or deeper with stabilization schemes at the lower boundary using different GCMs \citep{Liu2013,Mayne2014,Mayne2017}.

E.g. \citet{Liu2013} investigated the sensitivity of converged flow patterns for different initial conditions. They achieved full convergence after 1000 days simulation time by likewise adopting $\tau_{\rm conv}=10^7$~s for $p>10$~bar at depth ($p>10$~bar). These authors stress the importance of defining the lower boundary appropriately in a 3D GCM for hot Jupiters. Insensitivity to initial conditions is only ensured if the lower boundary is `anchored', i.e.~coupled to the interior, either via frictional drag or by extending the temperature downwards to the common convective adiabat. Otherwise, the globally integrated axial angular momentum of a model may not be conserved over the simulation time. In this work, we choose deep magnetic drag to `anchor' the simulation. In Section~\ref{sec: angular_mom}, we show that angular momentum is indeed adequately conserved within numerical accuracy in our WASP-43b and HD~2093458b simulations, which were fully stabilized at the bottom.

\citet{Mayne2017} also investigated eddy transport in the atmosphere. These authors find that forcing deep atmosphere evolution leads to a deceleration of the superrotating wind jet. We note that in this model, complete steady state was not reached even after 10~000 days simulation time. Interestingly, their deep circulation forcing is not motivated by fast rotation that drives deep jets towards the interior as in our work. Instead, they found a descent of air masses at the poles and ascent over the equators, which could lead to a thermal imbalance between the equator and the polar regions. Their deep circulation  is thus driven by a horizontal temperature gradient at depth ($p>10$~bar), where it was assumed that the polar regions are hotter by several 100~K compared to the equatorial regions \citep{Mayne2017}.

Thus, we confirm with our work that lower boundary conditions can have a significant impact on the hot Jupiter wind flow that is hard to predict. The amount of mass contained in a vertical layer increases with pressure. Therefore, wind flow at depth represents a large zonal momentum reservoir. If even a small fraction of it is transported upwards into the observable atmosphere layers via planetary waves or eddies, it has the potential to modify the photospheric wind flow structure substantially. Therefore, the lower boundary has to be very carefully selected to yield physically consistent, complete and numerically stable results in a 3D GCM for hot Jupiters.

We stress again that the aim of this work is to highlight uncertainties in underlying assumptions that may lead to substantially different simulation results between different 3D climate models for WASP-43b \citep{Mendonca2018, Mendonca2018b} even when the same dynamical core is used, e.g. different results in this work compared to \citet{Kataria2015}. The same is true for different applications of the magnetic drag mechanism used in this work compared to previous work \citep{Kataria2015,Parmentier2018}. We hope that this discussion will serve the community to improve complex 3D climate models for hot Jupiters, that now have to cover a large parameter space in temperature and surface gravity to yield predictions to be verified, for example, with the ARIEL spacecraft \citep{Venot2018,Tinetti2018}. It would be very interesting to use WASP-43b as another benchmark case for dynamics to compare the results of different dynamical solvers, grids and lower boundary prescriptions and to see if and under which they condition they may yield retrograde flow at the equator. Furthermore,\citet{Venot2020} showed that WASP-43b is generally an ideal benchmark case also for retrieval models. So far, only HD~209458b was used for such a benchmark exercise \citep{Heng2011} and only for prograde equatorial flow, that is, superrotation.

\subsection{The emergence of equatorial retrograde flow at the equatorial day side}
\label{sec: emergence_westward}

Both, the superrotating and the retrograde flow at the day side in our hot Jupiter simulations, are the results of interactions between large scale Kelvin and Rossby waves \citep{Showman2010,Showman2011}. One possible result from the interaction between these waves is the dominance by a superrotating flow  \citep{Showman2011}. The other is the `Matsuno-Gill' flow pattern \citep{Matsuno1966,Gill1980} that we either find as dominant wind flow pattern at the day side of WASP-43b, at least for $p\leq 10^{-1}$~bar (Figure~\ref{fig1} panel Ia,b) or in the eddy wind flow (that is after substracting of mean zonal flow) for HD~209458b (Figure~\ref{fig1} panel IIc)  (see also \citet{Showman2010}). Further, it has been demonstrated in shallow water models that the net-flow on the day side can be in either prograde (superrotation) or retrograde direction at the equator \citep{Showman2010, Penn2017}.

Until now, retrograde flow over the equator as one possible climate solution has been demonstrated in fast ($P_{\text{orb}}<3$~days) rotating tidally locked exo-Earths \citep{Carone2015}, and for warm to cool Jupiters with very fast rotation ($P_{\text{orb}}=0.55$~days) \citep{Showman2015}. This work shows that this solution can also appear for dense, hot Jupiters with fast rotations ($P_{\text{orb}}\leq 1.5$~days).). While our WASP-43b simulation does not display the full shift to a climate with off-equatorial jets and retrograde flow over the equator as seen in \citet{Showman2015}, it apparently exhibits a partly on-set of this other climate solution.

A superrotating flow has been linked to horizontal transport of zonal momentum from the poles to the equator and is evident from the tilt of the Rossby wave gyres on the day side (\citet{Showman2011}, see in our simulations e.g. Figure~\ref{fig1}, Ic). However, experiments of \citet{Showman2010} also show that there can be a momentum exchange between the upper, meteorology active atmosphere and the lower atmosphere, which was assumed in their experiments to be `quiescent'. In fact, imposing a dynamically `quiescent' deeper layer was identified by \citet{Showman2010} as a key element to achieve a full superrotating flow in shallow water models. Without such a layer, Matsuno-Gill flow  came to the fore in their simulations. But what if this balance is perturbed because the deeper layers are not dynamically quiescent?  Also \citet{Mayne2017} explicitly point out that vertical angular momentum in balance of horizontal interactions is critical for the generation of a superotating jet.

The Matsuno-Gill flow pattern with its retrograde flow on the day side appears in our simulations to be indeed linked to vertical momentum transport. While the assumption of a dynamically quiescent underlying layer seems to be valid for the inflated hot Jupiter HD~209458b also in this work, we find that it is apparently not true for our simulation of dense, fast-rotating WASP-43b.

The high density of WASP-43b requires a much lower intrinsic temperature ($T_{\rm int}=170$~K) than HD~209458b ($T_{\text{int}}=400$~K), where in the latter case the high $T_{\rm int}$ is used as a proxy for extra energy injected deep into the planet, causing a bloated radius. Consequently, the convective layer starts  deeper in the former, compared to the latter. Already \citet{Thras2011} predicted that fast-rotating hot Jupiters with relatively deep convective layers could exhibit unusually deep wind jets. These deep wind jets are now identified in our simulations to be associated with vertical transport of zonal momentum at depth ($p>10$~bar) that increases with faster rotation.

We also note that WASP-43b is about eight time denser than HD~209458b and thus its radiative timescales (Equation~\ref{eq:tau_rad}) are eight times larger. It has been noted  that, given the same thermal forcing, different radiative time scales lead to differences in the wind flow in shallow-water models \citep{Perez-Becker2013}. Indeed, the 3rd and 4th row of their Figure~3 shows similar flow patterns as the one found on the day side of our 3D WASP-43b climate simulation with equatorial retrograde flow. They also show that simulations with larger radiative time scales have a larger tendency to exhibit equatorial westward flow at day side. The simulations of \citet{Perez-Becker2013}, however, are performed with shallow water models and appear to lack full superrotating tendencies that could counteract retrograde flow. Our model is a full 3D GCM and our simulations exhibit equatorial retrograde flow at the day side,  together with a `head-on collision zone' at the morning terminator where the partly retrograde flow meets equatorial prograde flow at other longitudes.

Additionally, when the rotation period of HD~209458b is decreased such that it has the same rotation period as WASP-43b, HD~209458b, with its longer radiative timescales, also develops westward wind at the day side, albeit to a lesser extent (Figure~\ref{fig1} and \ref{fig2}).

We will show in the next section that the appearance of dynamical deviations from superrotation in hot Jupiters for $P_{\text{orb}} \leq 1.5$~days could be linked to the appearance to another physical effect, which is suppressed for slower rotators. These could potentially be baroclinic instabilities.

\subsection{Dynamical properties for the orbital period regime $P_{\text{orb}} \leq 1.5$~days}
\label{sec: Rhines}

We are not the first to investigate climate changes for different rotation rates and the effect on basic climate dynamics. Also \citet{Showman2015} performed a similar study with the same dynamical core but for the hot Jupiter HD~189733b. HD~189733b (1.14 $M_{Jup}$, 1.14 $R_{Jup}$) is less inflated than HD~209458b and at the same time less dense than WASP-43b. In their work, \citet{Showman2015} investigated hot ($T>1000$~K), warm ($T =1000$~K) and cool  ($T =600$~K) climates for HD~189733b-like mass and radius and for rotation periods of $0.5$, $2.2$ and $8.8$~days. \citet{Kataria2016} performed similarly a study to identify wind flow patterns for ultra-hot ($T\approx 2050$~K), hot and warm ($T\approx 960$~K) Jupiters and for a wide range of orbital periods $0.79 -  4.46$~days. We investigate the hot temperature regime with rotation periods of $0.81$, $1.5$ and $3.5$~days for WASP-43b and HD~209459b. Thus, our work is highly complementary to \citet{Showman2015,Kataria2016}.

To get better insights about the possible physical effects, we examine the Rhines length and the equatorial Rossby radius of deformation over planetary radius for different simulations in our work and that of \citet{Showman2015,Kataria2016}(Table~\ref{table: Rhines_Rossby_comparison}). The Rhines number or Rhines length $L_R$ defines the latitudinal scale at which turbulent flow (e.g. baroclinic eddies) can organize itself into zonal wind jets \citep{Rhines1975} and is calculated by:
\begin{equation}
L_R=\pi \sqrt{\frac{U}{\beta}} \label{eq: Rhines},
\end{equation}
where $U$ is the characteristic zonal wind flow speed and $\beta=2\Omega_P/R_P$ represents the Rossby parameter at the equator, where $\Omega_P$ is the planetary rotation rate and $R_P$ the planetary radius.

The equatorial Rossby radius of deformation $\lambda_R$ \citep{Gill1980} is:
\begin{equation}
\lambda_R= \sqrt{\frac{NH}{\beta}},
\end{equation}
where $H$ is the atmospheric scale height and $N$ is the Brunt-V\"ais\"al\"a frequency. The dimensionless Rossby radius and Rhines length for our simulations and those of \citet{Showman2015} and \citet{Kataria2016} are shown in Table~\ref{table: Rhines_Rossby_comparison}. A Rossby radius of deformation smaller than the planetary radius indicates a climate regime, where  standing Rossby waves can form on the planet, which are necessary to drive superrotation\citep{Showman2011}.

\begin{table*}
	\caption[table]{Rossby radius and Rhines length for our HD 209458b and WASP-43b simulation (in bold) and selected simulations of \citet{Showman2015} and \citet{Kataria2016}. Furthermore, differences in climate are indicated for the different regimes.}
	\label{table: Rhines_Rossby_comparison}
	\centering
	\begin{tabular}{l c c c c c}
    \noalign{\smallskip}
	\hline \hline
	\noalign{\smallskip}
	\textbf{Simulation} & \boldmath{$P_{\textrm{rot}=\textrm{orb}}$} \textbf{[d]} & \textbf{Rossby \boldmath{$\frac{\lambda_P}{R_P}$}} & \textbf{Rhines \boldmath{$\frac{L_R}{R_P}$}} &  \textbf{Crit. \boldmath{$\frac{L_{\textrm{R,crit}}}{R_P}$}} & \textbf{Climate}\\
	\noalign{\smallskip}
	\hline
	\noalign{\smallskip}
    \textbf{HD~209458b (nominal)} & \textbf{3.5} & \textbf{0.78}  & \textbf{2.90} & \textbf{1.5} & 1 strong prograde eq. jet \\
    H$\Omega_{\textrm{intermed.}}$\citep{Showman2015} & 2.2 & 0.62 & 2.3 & 1.3 & 1 strong prograde eq. jet\\
     \citep{Kataria2016} & & & & & \\

    \hline
    \multicolumn{6}{c}{Transition in Rhines length ($L_R/R_P \lesssim 2$) or ($L_{\textrm{R,crit}}/R_P \lesssim 1$)} \\
    \hline
    \textbf{WASP-43b (intermediate)} & \textbf{1.5} & \textbf{0.6}  & \textbf{2.03} & \textbf{1.12} & 1 strong prograde eq. jet \\
     &  &  &  &  & + partly eq. retrograde flow \\
   \textbf{WASP-43b (nominal)} & \textbf{0.81} & \textbf{0.51}  & \textbf{1.77} & \textbf{0.83} & 1 strong prograde eq. jet\\
     &  &  &  &  & + partly eq. retrograde flow \\
   \hline
   \multicolumn{6}{c}{Transition in Rossby radius ($\lambda_P/R_P <0.5$)}\\
   \hline
    \textbf{HD~209458b-like (fast)} & \textbf{0.81} & \textbf{0.45}& \textbf{1.49} & \textbf{0.71} & 1 strong prograde eq. jet\\
      &  &  &  &  & + partly retrograde + off-eq. jets \\
    WASP-19b \citep{Kataria2016} & 0.79 & 0.39 & 1.64 & 0.70 & 1 strong prograde eq. jet\\
     &  &  &  &  & + off-eq. jets \\
    \hline
    \multicolumn{6}{c}{Full transition in Rhines length ($L_R/R_P \lesssim 1$)}\\
    \hline
    H$\Omega_{\textrm{fast}}$ \citep{Showman2015}& 0.55 & - & 0.70 & 0.65 & weak prograde eq. flow \\
     &  &  &  &  & + off-eq. jets \\
    C$\Omega_{\textrm{fast}}$ \citep{Showman2015}& 0.55 & - & 0.55 & 0.65 & retrograde eq. flow \\
     &  &  &  &  & + off-eq. jets \\
    \noalign{\smallskip}
    \hline
	\end{tabular}
\end{table*}%

A further comparison of our simulations with those of \citet{Kataria2016} shows that most of their simulations have a non-dimensional Rossby deformation radius between $0.5<\lambda_R/R_P < 1$ (Table~\ref{table: Rhines_Rossby_comparison}). The only exception is the inflated super-hot Jupiter WASP-19b with $P_{\text{orb}}=0.79$~days and with $\lambda_R/R_P<0.5$, which results in the appearance of a pair of weaker zonal jets at mid-latitudes in addition to the equatorial main jet. A similar wind jet picture can be seen for our fast HD~209458b-like simulation (Figure~\ref{fig2}, panel Ia), which also has $\lambda_R/R_P<0.5$ (Table~\ref{table: Rhines_Rossby_comparison}). This climate transition to more jets in superrotating tidally locked planets for particularly small Rossby radii of deformation ($\lambda_R/R_P<0.5$) was also already pointed out by \citet{Carone2015} for rocky planets. Additionally, \citet{Haqq2018} pointed out that in the same rotation regime another transition can occur: A transition associated with the Rhines length $L_R$ becoming smaller than the planetary radius ($L_R/R_P\lesssim 1$) with smaller orbital periods. In this rotation regime, the climate can switch from one with strong equatorial superrotation to a climate, which instead exhibits a pair of weaker jets at higher latitudes. A similar climate transition was also encountered for the very fast rotating ($P_{\text{rot}=\text{orb}}=0.55$~days) hot Jupiter climate regime H$\Omega_{\textrm{fast}}$ in \citet{Showman2015} (Table~\ref{table: Rhines_Rossby_comparison}).

\citet{Showman2015} noted that these off-equatorial jets are probably driven by baroclinic instabilities and that even retrograde equatorial wind flow can form in this regime, at least for cooler temperatures (C$\Omega_{\textrm{fast}}$). \cite{Showman2015} also found for the hot regime that as long as an equatorial wind jet can form, faster rotation tends to drive this wind jet deeper into the planet (their Figure 3, bottom panel).

However, as already pointed out in \citet{Carone2015}, the switch in climate regimes is not linear. Therefore, we find it illustrative to introduce a critical Rhines length $R_{\textrm{L,crit}}$ that takes the weakening of wind speeds and the shift of the main jets off the equator into account:

\begin{equation}
    L_{\textrm{L,crit}}=\pi\sqrt{\frac{U_{\textrm{crit}}}{\beta_{\textrm{off}}}},
\end{equation}

where we adopt $U_{\textrm{crit}}=800$~m/s and $\beta_{\textrm{off}}=2\Omega_P/R_P \cos(30^{\circ})$ suitable for the appearance of off-equatorial jet in H$\Omega_{\textrm{fast}}$ in \citet{Showman2015} (their Figure 3).

Comparing $L_R$ and $L_{\text{R,crit}}$ (Table~\ref{table: Rhines_Rossby_comparison}), we find that most of our simulations are in the Rhines length regime $L_R/R_P$ between 1 and 2 for climates still dominated by prograde flow over the equator. If the climate shifts from strong equatorial to weaker off-equatorial jets, e.g. by the emergence of baroclinic eddies, this would meet the required Rhines length criterion in $R_{\textrm{L,crit}}/R_P \lesssim 1$.  The Rhines length regime $L_R/R_P$ between 1 and 2, which our simulations exhibit, lie just in between H$\Omega_{intermediate}$ and H$\Omega_{fast}$ investigated by \citet{Showman2015}. The regime lies also just in between HD 189733b (equal to H$\Omega_{\text{intermediate}}$ in \citet{Showman2015}) and WASP-19b investigated by \citet{Kataria2016}.

Interestingly, retrograde flow in the WASP-43b-like simulation for the intermediate rotation period ($P_{\rm orb}= 1.5$~days) is still stronger than in the fast ($P_{\rm orb} = 0.8135$~days) HD~209458b-like simulation. This result indicates that while a rotation period shorter than 1.5~days or $L_{\rm R,crit }\lesssim 1$ is an important factor for the emergence of retrograde equatorial flow, it is not the only factor that has an influence on the strength of retrograde flow.

Apparently, additional important factors are the internal structure and thus density and also the radiative timescales of the planet, as outlined in Section~\ref{sec: HD20}. Also \citet{Showman2015} find different equatorial flow structure on their fast rotating climates (0.55 days) for different temperature regimes. In their work, retrograde equatorial flow becomes more dominant with cooler temperatures. This temperature dependency may also explain why partly retrograde flow at the day side is present on the 1450~K hot, fast rotating (0.81 ~day) hot Jupiter simulations in this work (i.e. the nominal WASP-43b and the fast HD~209458b simulations) but apparently not on WASP-19b, with a temperature of 2050~K \citep{Kataria2016}.

 We now investigate circulation and the Elliassen-Palm flux to get a first understanding about the role of wave activity in our simulations.

 \subsection{Circulation, Elliassen-Palm flux and Potential Vorticity}
 \label{sec: EP}

 There has been one study that coherently linked circulation on tidally locked planets with climate transitions in Rossby radius of deformation for a wide range of orbital periods $P_{\text{orb}}=1-100$~days \citep{Carone2016}. More precisely, they identified four states of climate transition (see \citet{Carone2016}, their Figure 17):
 \begin{itemize}
 \item \textit{State 0} for circulation dominated by one direct circulation cell per hemisphere as on e.g.~Venus for orbital periods of 22 days and slower.
 \item \textit{State 1} for circulation still dominated by one direct circulation cell per hemisphere, but with the appearance of embedded counter-rotating circulation for orbital periods between 12 and 22 days.
 \item \textit{State 2} for two cells per hemisphere, a direct circulation and a fully formed slant-wise counter-rotating cell for orbital periods between 3 and 13 days.
 \item \textit{State 3} for a fragmentation of cells, with three or more per hemisphere, where the direct equatorial cells are strongly diminished for orbital periods shorter than 3~days.
 \end{itemize}

\begin{figure*}%
	\centering%
     \includegraphics[width=0.95\columnwidth]{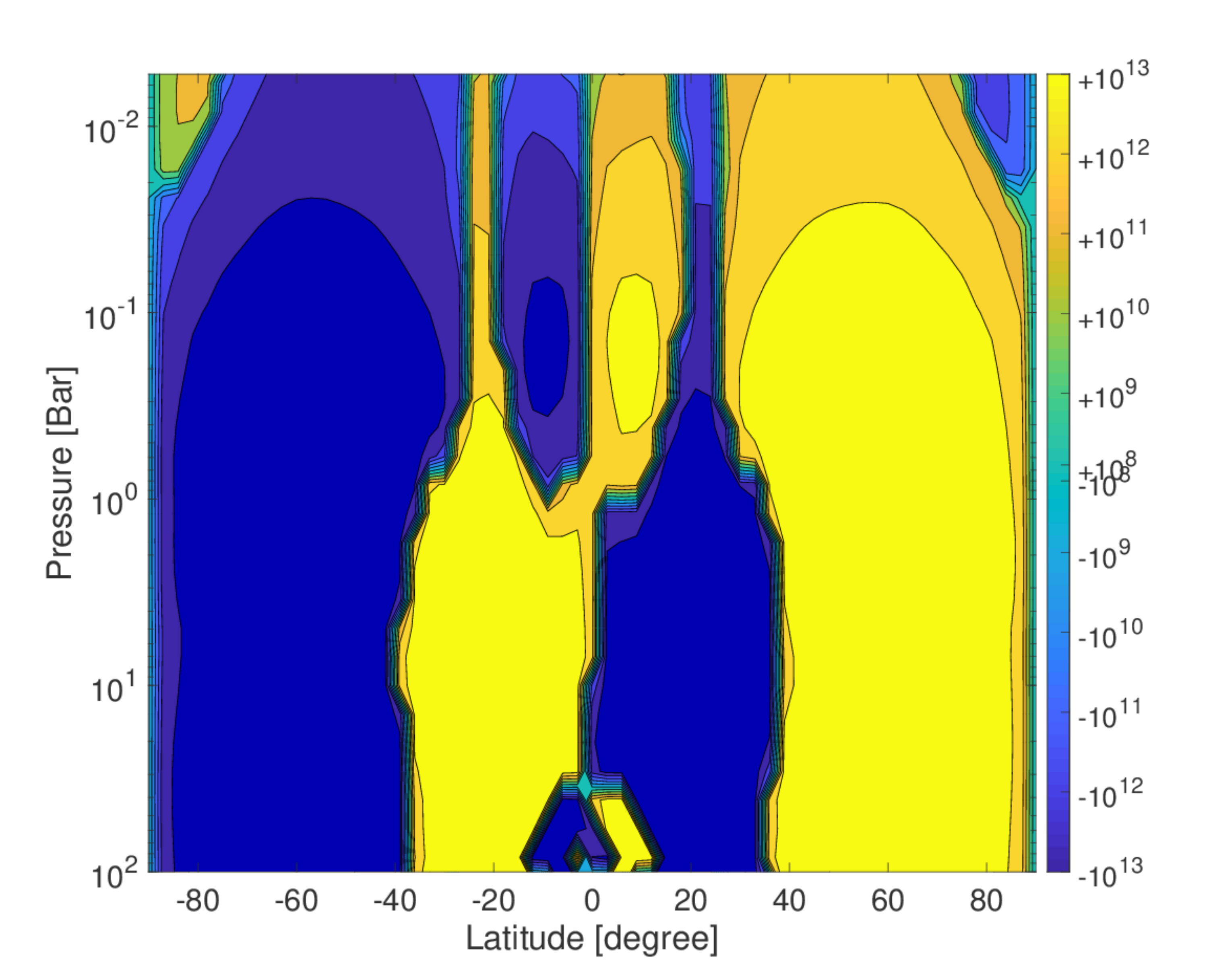}%
    \includegraphics[width=0.95\columnwidth]{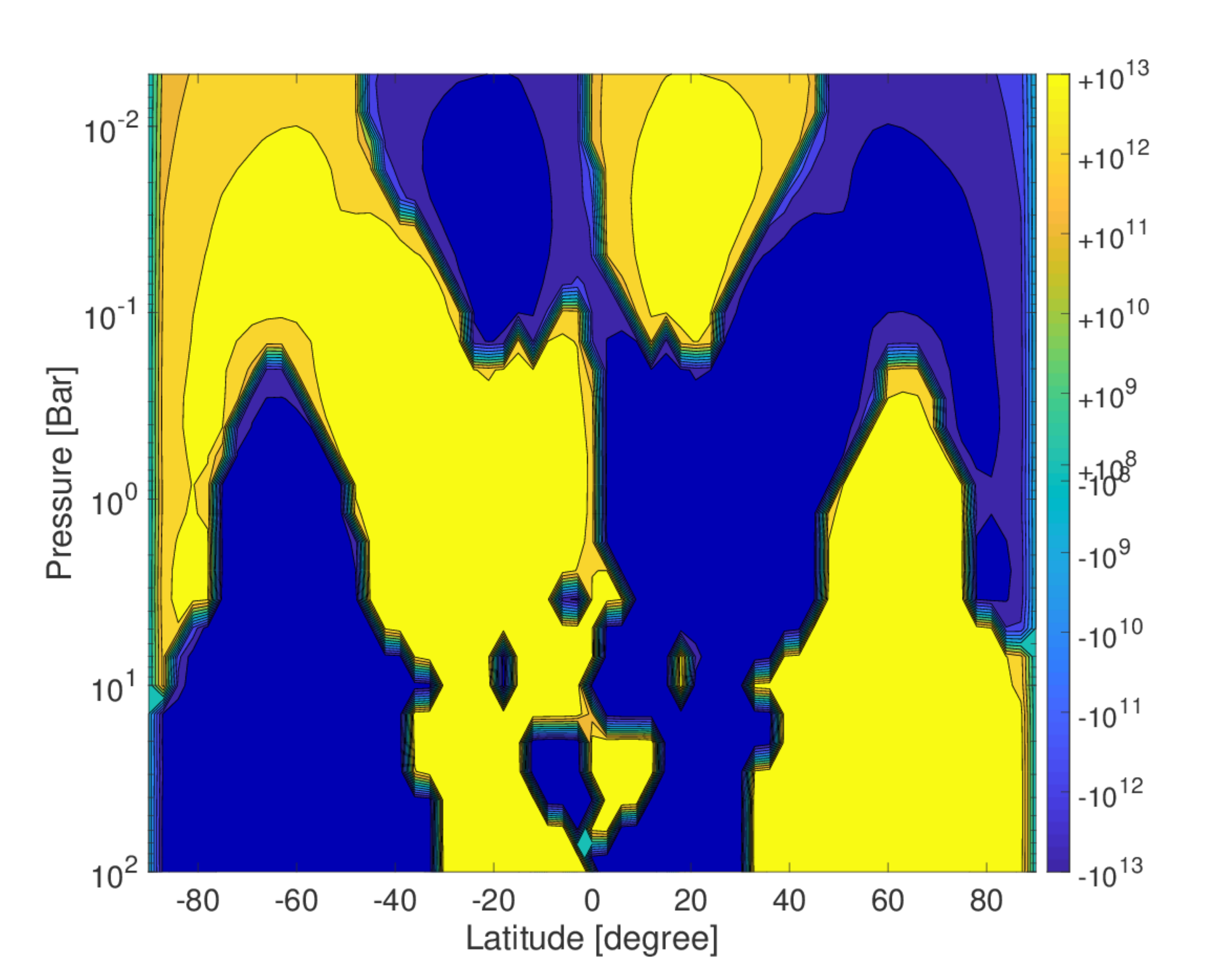}
	\caption{Zonal-mean of overturning stream function in [kg/s] for WASP-43b (left) and HD 209458 b (right). Negative (blue) values are counter-clockwise and positive (yellow) values are clockwise circulation.}%
    \label{figOverturn}%
\end{figure*}

While the study of \citet{Carone2016} has been performed for rocky Exo-Earths, it appears that the circulation states 0 - 3 are also applicable to hot Jupiters. Figure~\ref{figOverturn} shows circulation for WASP-43b (left) and HD~209459b (right). WASP-43b circulation ($P_{\text{orb}}=0.81$~days) is in \textit{state 3} with several fragmented circulation cells per hemisphere, which was identified also in \citet{Carone2016} with a transition in Rossby radius of deformation ($\lambda_R/R_P < 0.5$). \citet{Showman2015} show for their $C\Omega_{\text{fast}}$-scenario also fragmented circulation cells (their Figure~8 f). They identify part of the cells as Ferrell-like circulation cells. On Earth, Ferrell-like circulation is associated with baroclinic eddies.

As a reminder, their $C\Omega_{\text{fast}}$ scenario shows off-equatorial jets, which are driven by baroclinic eddies, and retrograde flow over the equator. More precisely, \citet{Showman2015} attributes deviation from equatorial superrotation in their very fast scenarios to the emergence of baroclinic eddies at mid-latitude that tend to transport angular momentum towards the location where the instability occurs, typically at mid-latitudes. Other work also showed that it is in principle possible to elicit baroclinic instabilities even in hot Jupiters without a surface \citep{Poli2012}. In this work, we also postulate that our WASP-43b simulation shows retrograde equatorial flow at the day side, because also here very fast rotation lead to increased wave activity, which causes deviations from pure equatorial superrotation.

The HD~209458b simulation ($P_{\text{orb}}=0.81$~days), on the other hand, displays a fully formed second circulation cell per hemisphere embedded slant-wise in the direct circulation cell, that is \textit{state 2} circulation. This simulation does not appear to show signatures of baroclinically driven Ferell cells. We further note that \cite{Mendonca2019} display circulation similar to \textit{state 1} identified in \cite{Carone2016} in their hot Jupiter simulation with $P_{\text{orb}}=10$~days (their Figure 14, top): one direct circulation cell with a small embedded counter-circulation cell.

\begin{figure*}%
	\centering%
     \includegraphics[width=1.5\columnwidth]{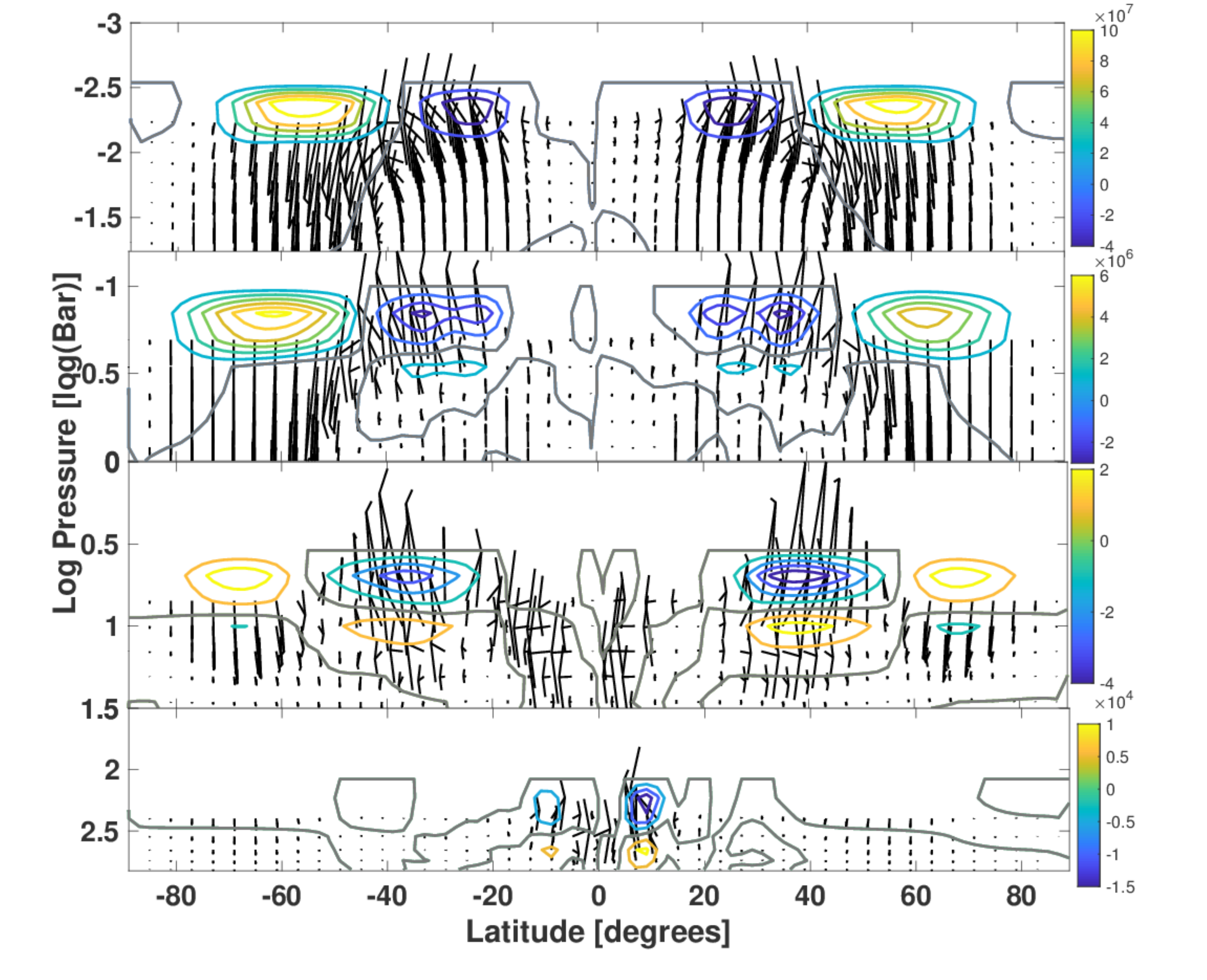}\\%
    \includegraphics[width=1.55\columnwidth]{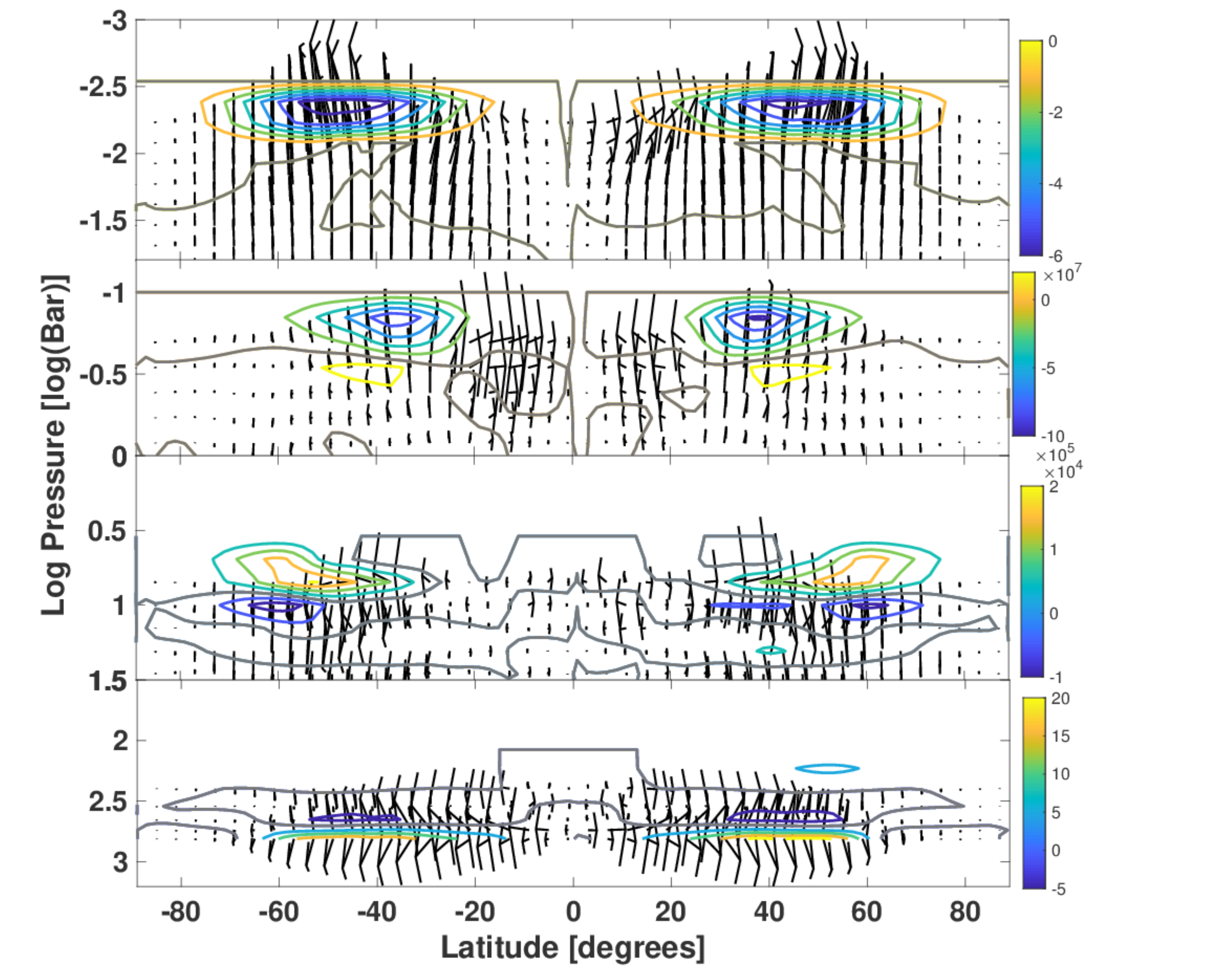}%
	\caption{ Elliassen-Palm flux (black arrows) and Eliassen-Palm flux divergence (colored contours, where grey contours mark zero divergence). Both properties are displayed for WASP-43b (top) and HD~209458b (bottom). Calculation and scaling of these properties are listed in Table~\ref{tab:EP_scale}. Units of  $(\tilde{F}_\phi,\tilde{F}_p)=(m^2,\frac{m^2}{s} Pa)$}%
    \label{figEP}%
\end{figure*}

A look at the Eliassen-Palm (EP) flux and its divergence (Figure~\ref{figEP})  (see also Section~\ref{sec: EP_calc}) supports the view that baroclinic eddies may play a strong role in our WASP-43b simulation (top). The eddy signal is apparent at latitudes $\pm 40^\circ$ (a convergence $\nabla \cdot (EP) <0$ at the top and divergence $\nabla \cdot (EP) >0$ at the bottom). Also the typically upward flow is apparent, but we note that the horizontal direction of the flux is pole-wards and not equator-wards as in the Earth-like climate (see e.g. \citet{Edmon1980}, their Figure~1 and 2).

For HD~209458b (Figure~\ref{figEP}, bottom panel), we see a simpler picture compared to WASP-43b. We do not see the clear vertical pairing of convergence-divergence in the EP flux as in WASP-43b at mid-latitudes. In fact, (positive) divergence is very weak compared to (negative) convergence for most of the atmosphere ($p<1$~bar). Also, the vertical upward component of the flux is missing in the troposphere\footnote{Vertical scaling is here similar for WASP-43b and HD~209458b and thus we can directly compare the vertical flux component, see Appendix~\ref{sec: EP_calc}.} ($p>10^{-1}$~bar).

\begin{figure*}%
	\centering%
     \includegraphics[width=1.05\columnwidth]{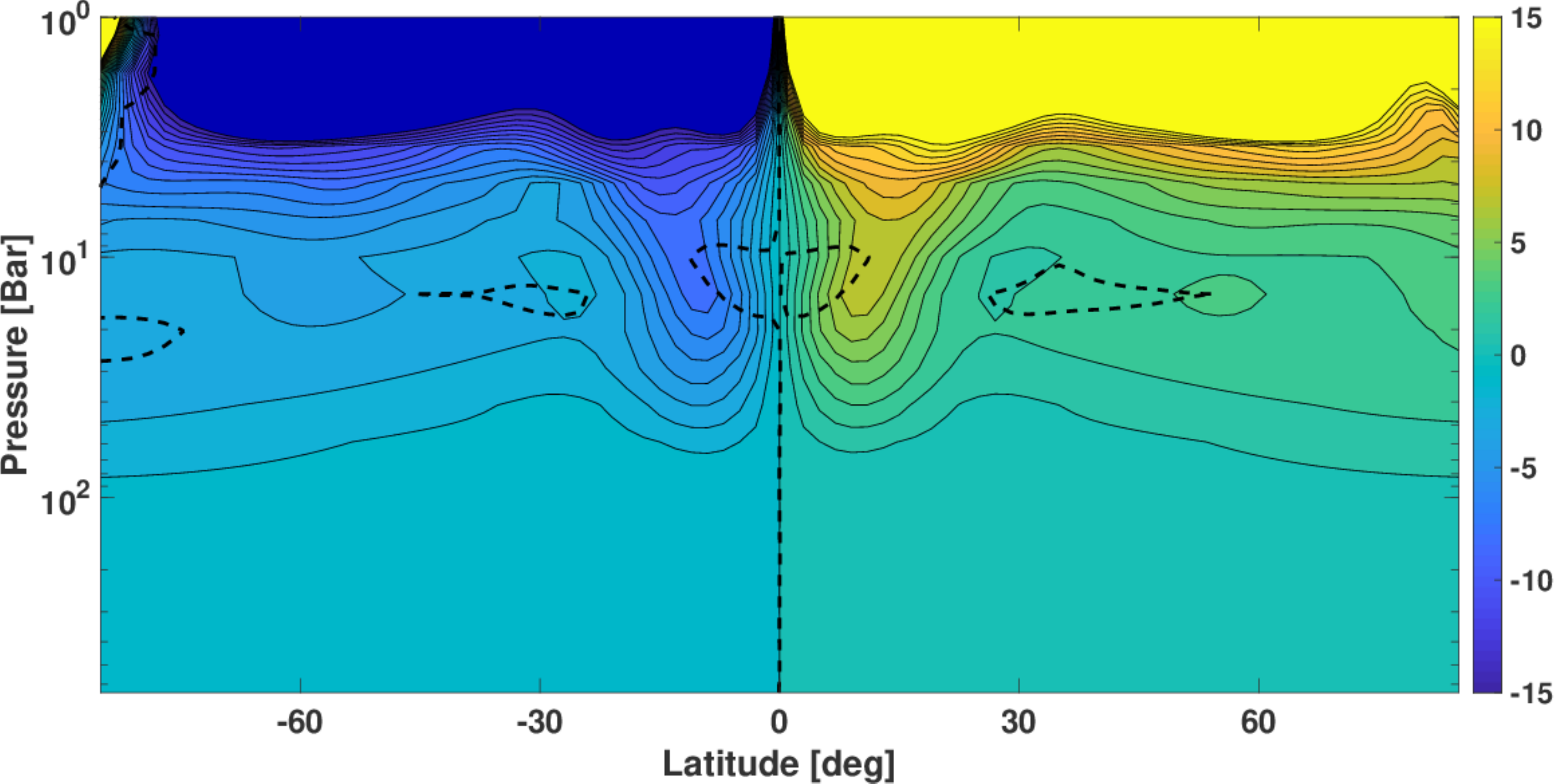}%
     \includegraphics[width=1.05\columnwidth]{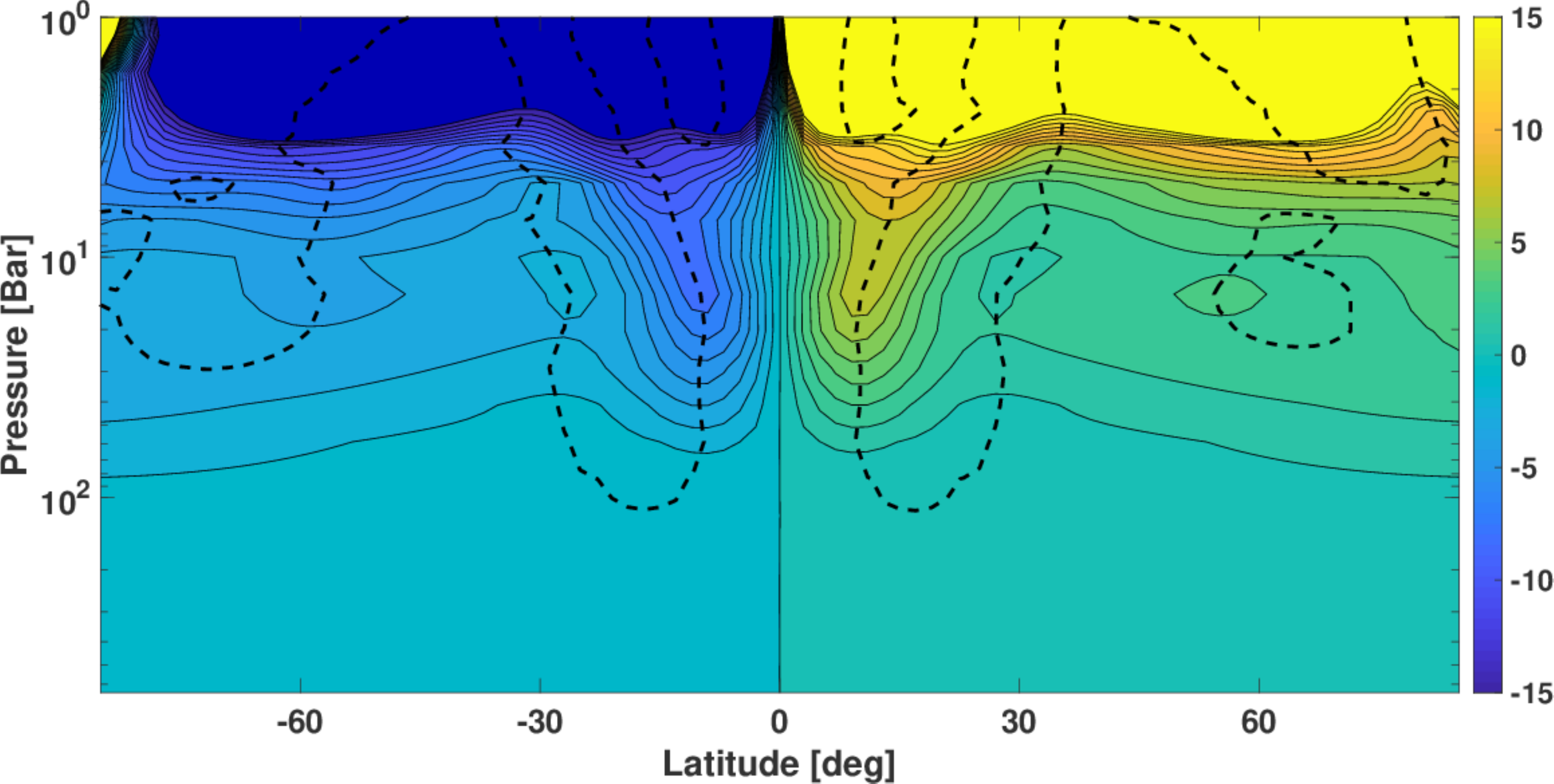}\\%
    \includegraphics[width=1.05\columnwidth]{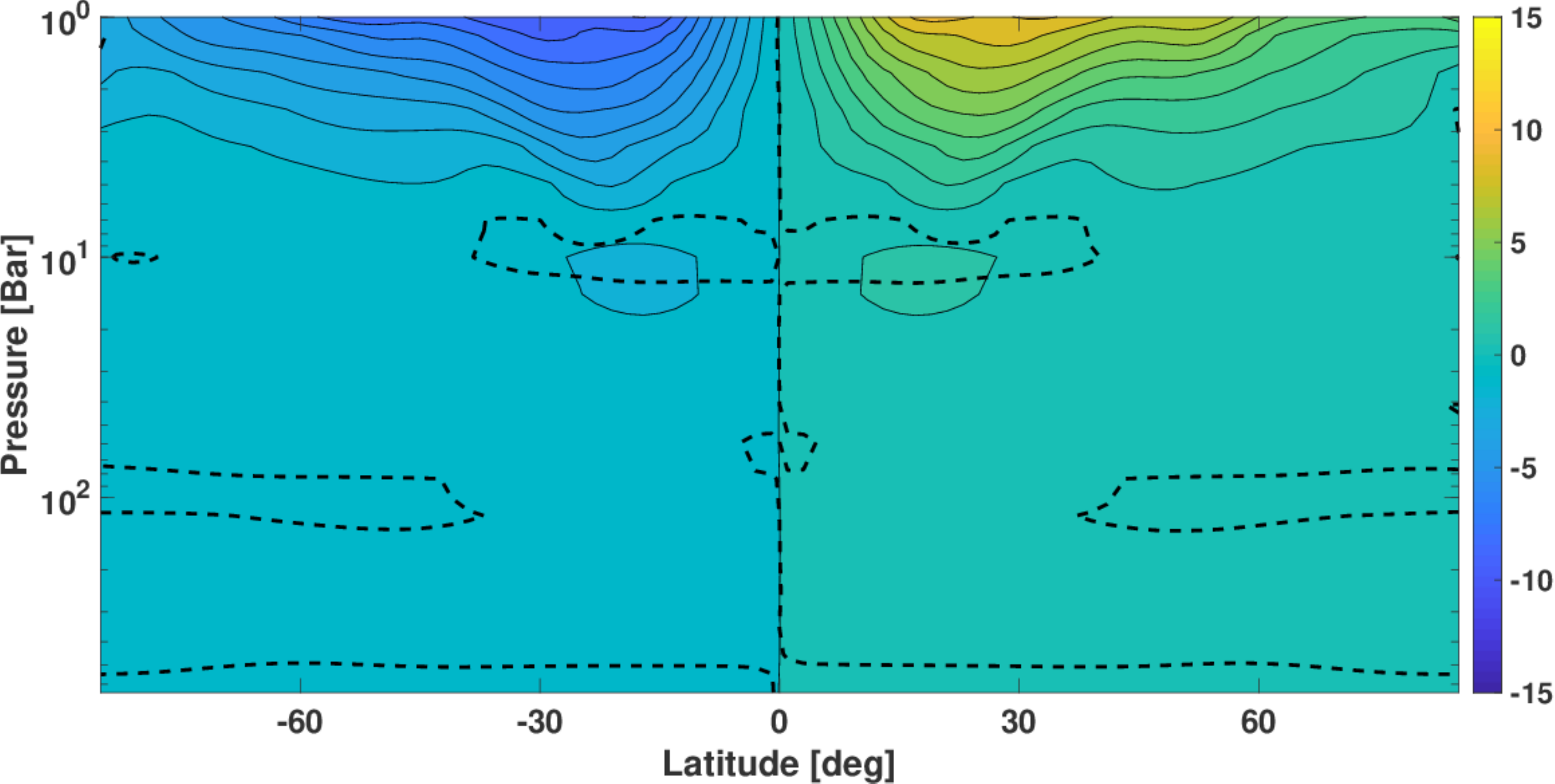}%
    \includegraphics[width=1.05\columnwidth]{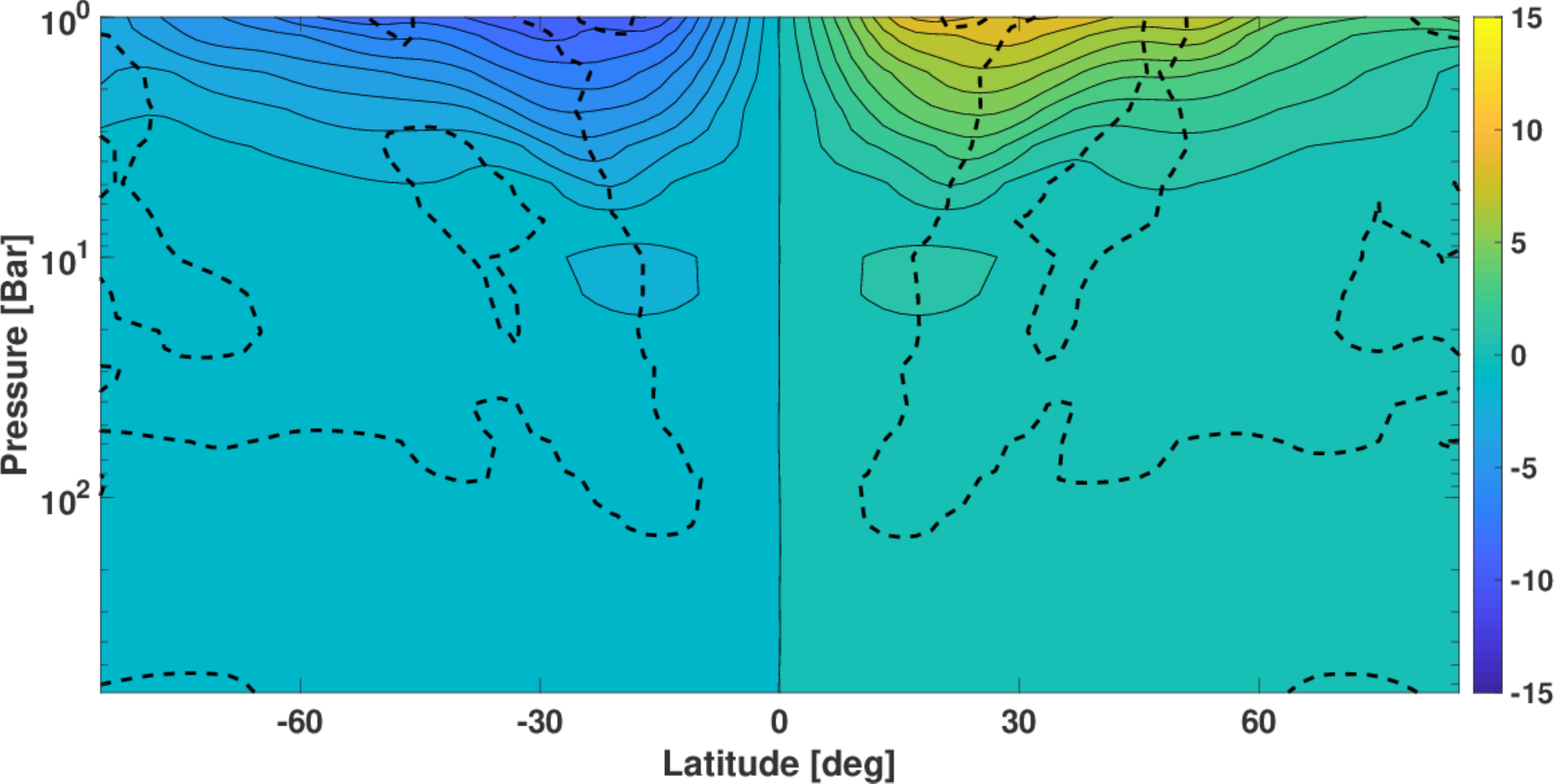}\\
	\caption{Zonal-mean of potential vorticity $\overline{q_E}$ in potential vorticity units (PVU$=10^{-6}$~K m${}^2$kg${}^{-1}$s${}^{-1}$) displayed for WASP-43b (top) and HD~209458b (bottom) for $p>1$~bar. Furthermore, dashed black contours overlaid on the PV map at depth indicate the locations where the vertical gradient ($\partial \overline{q_E} / \partial p $, left) and meridional gradient of the zonal mean of PV ($\overline{q_E} / \partial \theta $ right) become zero, respectively.
	We identify strong anomalies in $q_E$ with sign change in the vertical and horizontal gradient at their inner flanks for the case of WASP-43b, associated with deep atmospheric eddies. HD~209459b on the other hand does not display strong PV anomalies and also does not exhibit such localized sign changes in the vertical and horizontal gradients.}%
    \label{figPV}%
\end{figure*}

We also performed an analysis of the zonal mean of the potential vorticity (PV) on isentropic surfaces $\overline{q_E}$ (see also Section~\ref{sec: EP_calc}), where a sign change in the vertical direction of the meridional gradient is the ``Rayleigh necessary condition'' for baroclinic instabilities \citep{Holton}. More specifically, this condition requires $\partial \overline{q_E} / \partial \theta = 0$ somewhere in the domain.

We find that there is a strong PV anomaly in the equatorial region of the deep atmosphere for WASP-43b associated with vertical sign change in the vertical and meridional gradients of the zonal mean of the PV at their inner flanks (Figure~\ref{figPV} top). That is, the Rayleigh condition for baroclinic instabilities appears to be indeed fulfilled for WASP-43b. A much weaker similar anomaly exists also for HD~209458b (Figure~\ref{figPV} bottom). These PV anomalies appear to be located at similar positions than the regions of upper zonal momentum transport, identified in Figure~\ref{fig1} panels Id and IId for WASP-43b and HD~209458b, respectively.

Strong potential vorticity gradients in the deep $p \gg 1$~bar equatorial regions also fit the picture of the EP flux, where we see for WASP-43b a high upward EP flux originating from deeper regions at latitudes of $\pm 20^\circ$ (see Figure~\ref{figEP}, top panel). These latitudes are exactly where the PV anomalies are located as well. The upwelling EP flux in WASP-43b and its associated divergences at $\pm 60^\circ$ in the higher atmospheric layers seem to correspond to the wind acceleration at the flanks of the Matsuno-Gill flow pattern (see Figure~\ref{fig1}, panels I b and I c). This acceleration appears to contribute to the wave-mean flow interaction that results in maintaining part of the Matsuno-Gill flow pattern, with retrograde wind flow along the equator that is embedded in strong equatorial superrotation at other latitudes.

In contrast to that, HD~209458b lacks this strong upward EP flux and we do not see any strong EP flux divergences in the upper atmosphere. Instead, the EP flux exhibits a wide convergence (negative values) spread across most of the hemisphere (Figure~\ref{figEP}, bottom panel). These convergent EP flux regions are also present in the WASP-43b simulation but only at lower latitudes ($\pm 30^\circ$).

It thus appears that there is a consistent connection between the deep atmosphere with deep jets and the upper atmosphere, that becomes particularly important for very fast rotational speeds. This connection manifests itself in larger vertical momentum transport (see Figure~\ref{fig2} panels I, II an III c) with faster rotation. The EP flux and potential vorticity diagnostics we have used, suggest this connection is associated with wave activity, possibly with upward propagating baroclinic waves.

Similarly, \citet{Showman2015} point out that the occurrence of baroclinic instabilities, at least for Earth-like simulations, depends strongly on the entropy gradient at the lower boundary. This statement of \citet{Showman2015} thus fits our observation that the retrograde flow over the equatorial day side is highly sensitive to conditions imposed at the lower boundary, which may be affected by deep wind jets and also potentially magnetic field interactions. \citet{Mayne2017} also show that deep polar to equatorial temperature gradients can elicit circulation at depth for HD~209458b ($P_{\text{rot}}=P_{\text{orb}}=3.5$~days). Such deep wind flow was likewise found to result in a diminishing of superrotation. In their simulation, however, no retrograde flow along the equator is present. \citet{Mayne2017} also point out that a balance between horizontal and vertical momentum transport is needed for superrotation. We thus postulate that their deep circulation `just' disturbs the balance in the momentum transport. Here, however, an additional physical effect occurs via wave activity that drives wind westwards along the equator.

In summary, we have analysed a series of diagnostics of vertical momentum transport and atmospheric wave activity, and qualitatively compared these for simulations of HD~209458b and WASP-43b. Our figures of the meridional circulation patterns (Figure~\ref{figOverturn}), the (divergence of the) EP flux (Figure~\ref{figEP}) and the potential vorticity (Figure~\ref{figPV}) all support the view that a different wave-mean flow interaction in the deep $p>1$~bar atmosphere of the fast rotating WASP-43b can trigger transport of zonal momentum upwards, resulting in a different circulation regime, namely the occurrence of retrograde flow over the equator. We identified the likely mechanism of transport as associated with wave activity, potentially baroclinic eddies, evidenced by the EP flux and a strong potential vorticity anomaly. Furthermore, the fast rotation rate in our simulations of WASP-43b (and the HD~209458b experiment with faster rotation) is the observable physical parameter that is chiefly responsible for deviations from pure superrotation in the form of retrograde flow along the equatorial day side. In addition to our comparative approach, a full global analysis of these atmospheric wave-mean flow diagnostics in the deep atmosphere would be instructive, especially the distribution of the EP flux, which has never been analysed for hot Jupiters before.

Clearly more work is needed to investigate if and why deviations from equatorial superrotation can occur in some WASP-43b simulations and if these are grounded in physical reality. So far, however, the dynamical picture appears to be consistent with previous work on tidally locked hot Jupiter \citep{Showman2015,Kataria2016} and rocky Earths \citep{Carone2015,Carone2016}: That deviations from equatorial superrotation can occur if $L_{R,crit}/R_P<1$, which appears to be true for hot Jupiters with $P_{\text{orb}}\leq 1.5$~days. We further note that very recently \citet{Wang2020} have independently confirmed how important it is to take into account the deep atmosphere for a complete picture of the observable wind flow.

\subsection{The effects of magnetic field on deep wind jets and the observable atmosphere}

Frictional drag at the lower boundary, which we chose as a first order representation of interaction between wind flow and magnetic fields at depth, appears to have a significant effect on the horizontal heat transport at $p<0.1$~bar. We stress, however, that drag is only a factor that modifies the strength of retrograde flow. It is not the underlying cause for this wind flow as we have shown in previous sections.

The introduction of deep drag at the bottom has several benefits. It allows to reach full stabilization from bottom to top in the WASP-43b simulation. In addition, our nominal WASP-43b simulation with magnetic drag at depth yields the best agreement between our WASP-43b simulation and observational data, in particular with Spitzer phase curves. It reproduces to first order the observed large day-night flux contrast and the small hot spot shift (Figures~\ref{fig4}).

Work of \citet{Hindle2019} (with a shallow-water magnetohydrodynamic model) shows that magnetic fields that couple to wind flow can, in principle, explain the westward wind flow in HAT-P-7b and CoRoT-2b, although it requires very strong magnetic fields for the latter. We note that \citet{Hindle2019} included a realistic coupling between the magnetic field and dynamical atmosphere, but in their shallow-water model did not take into account the possible influence of deep circulation. Interestingly, CoRoT-2b -- with an orbital period of 1.7 days -- appears to lie just at the edge of the rotation period range, for which we find retrograde wind flow driven by deep circulation ($P_{\text{orb}}\leq 1.5$~days). In the future, it would be interesting to investigate for HAT-P-7b and CoRoT-2b, to which degree deep circulation may be a contributing factor for the observed hot spot offset in the westward direction with respect to the substellar point \citep{Dang2018}.

The importance of deep wind jets that couple to magnetic fields has also been explicitly stated by \citet{Rogers2014b}, who note that \textit{``larger flow speeds at depth could increase the value of the Ohmic dissipation there, and therefore deep jets may be required to inflate planets.''} Thus, deep wind jets may also help to solve the `inflated hot Jupiter conundrum', where further evidence appears to favor Ohmic dissipation as the main mechanism for inflation \citep{Thorngren2018}.

The proposed connection between deep wind jets, magnetic fields and inflation via Ohmic dissipation begs the question why WASP-43b, the planet for which we find very deep wind jets driven by its fast rotation, is apparently not inflated. Further work is required to investigate if the deep wind jets in WASP-43b do indeed lead to large Ohmic dissipation.

One noteworthy tentative prediction from our simplified WASP-43b simulation is that the night side will emit overall very little flux in the near- to mid-infrared ranges (Figure~\ref{fig:spec}) -- even compared to predicted spectra derived from cloudy GCMs that display full superrotation in their simulations \citep{Parmentier2016}.  This prediction will be testable with observations by JWST/MIRI \citep{Bean2018,Venot2020}. There are some shortcomings in these tentative predictions as we will show in the next section.

\subsection{Shortcomings of our observational predictions}
\label{sec: shortcomings}

\begin{figure*}%
	\centering%
    \includegraphics[width=2.1\columnwidth]{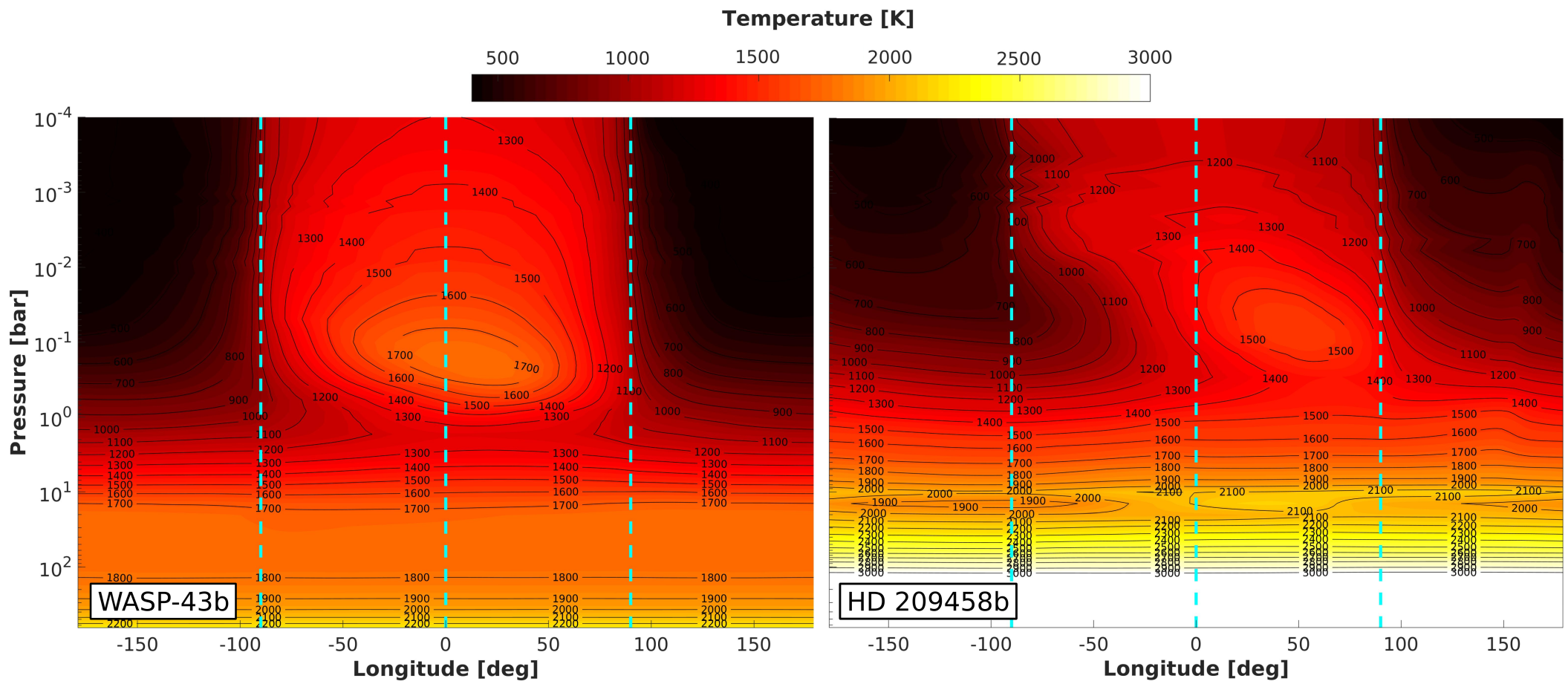}%
	\caption{Maps of the meridionally averaged temperature as a function of longitude and pressure for WASP-43b (left) and HD~209458b (right). The temperature averages shown are weighted with factors of $\cos{\theta}$ with $\theta$ the latitude. The dashed vertical lines from left to right indicate the morning terminator, substellar point and evening terminator, at longitudes of $-90^\circ$, $0^\circ$ and $+90^\circ$ respectively.}%
    \label{fig:T_slice}%
\end{figure*}

Our WASP-43b simulations are simplified to study the general flow and heat circulation regime for different underlying assumptions. They have, however, some apparent shortcomings when specific parts of the atmosphere are probed, which lead to deviations between the predicted spectra and the water feature at the day side, which is captured with the HST/WFC3 data taken by \citet{Stevenson2014}. This deviation is very clear in the day-side averaged spectrum of WASP-43b (Figure~\ref{fig:spec}, left panel inlay). There, apparently, our model yields a water emission instead of absorption feature in the HST/WFC3 wavelength range. The existence of an emission feature can be linked to the temperature inversion that we see at the day side of WASP-43b between 0.5 and 10 bar, where the temperature drops steeply from 1700 K to 1300 K and rises back to 1700 K again (Figure~\ref{fig:T_slice}, left panel).

We attribute the strong temperature inversion in our WASP-43b model to the shortcomings of our Newtonian cooling prescription, where inaccuracies in radiative time scales result in insufficient coupling to the wind flow at the most strongly forced region -- the day side -- in particular. Dynamical temperature inversions at the day side typically change with different treatment of irradiation: This can be seen in the comparison between the day-side temperature structures simulated for HD~209458b using simplified forcing \citep{Showman2008} (their Figure~6) and using full dynamical-radiative coupling \citep{Showman2009} (their Figure 18). Their HD~209458b simulations likewise exhibited a stronger day-side temperature inversion at higher vertical levels in the simplified thermal forcing set-up compared to the fully coupled model.

We note that \citet{Kataria2015} have for their WASP-43b model with 5 times solar metallicity a temperature structure that is qualitatively similar to our WASP-43b model with deep circulation (their Figure 10, bottom panel). However, while their maximum temperature of the temperature inversion occurs at similar pressures  (between  0.1 and 1 bar) compared to our model, the temperature gradient below is in the simulation of \citet{Kataria2015} very small (1800 K to 1700 K and back to 1800 K) and extends over a smaller pressure range (1 - 10 bar). Consequently, their simulations do not yield a `flip' from absorption to water emission (their Figure 11 top).

Despite this shortcoming in the details of the temperature structure which can have an effect on spectral features like the water feature in near-infrared, the overall observed temperatures at the day side and night side of WASP-43b appear to be not so very dissimilar compared to the results yielded by our simulations (Figures~\ref{fig4} and ~\ref{fig:spec}).

More work involving a 3D GCM with deep circulation formalism and with full coupling between irradiation and deep circulation would be beneficial. Only with a more sophisticated 3D climate model that also includes clouds (e.g. \citet{Parmentier2016,Woitke2019,Helling2019,Helling2016,Powell2019}) is an in-depth analysis warranted between model-based predictions and the full set of HST phase curves as performed by \citet{Mendonca2018}. We also note that \citet{Mendonca2018b} invokes disequilbrium chemistry in a further work on WASP-43b, whereas we assume equilibrium chemistry, both in the forcing and when producing spectra and phasecurves with \textit{petitCODE}  \citep{Molliere2015,Molliere2017}.

\subsection{Comparison to other possible mechanisms governing heat redistribution}

Cloud-free GCMs that assume solar metallicity routinely predict larger eastward phase shifts and hotter night sides than are observed (see e.g. \citet{Zellem2014,Kataria2015,Parmentier2016,Mendonca2018} for HD~209458b, WASP-43b, Kepler-76 and HAT-P-7b). As one possible solution, cloud coverage was proposed \citep{Parmentier2016}. We propose another possibility: that the eastward hotspot shift is smaller and the night side is colder for dynamical reasons, because superrotation is diminished via a combination of wind flow at depth and wave activity in very fast rotating hot Jupiters ($P_{\text{orb}}\leq 1.5$~days).

Another mechanism that is invoked to explain the discrepancy between predictions from cloud-free 3D GCMs  for tidally locked Jupiters and observations, is a proposed interaction between the fast wind jets and magnetic fields in hot Jupiters. \citet{Kataria2015} explored this possibility for WASP-43b by imposing frictional drag with constant time scales on the whole atmosphere in their 3D GCM as a representation of magnetic wind drag. Their frictional drag mechanism also achieves retrograde wind flow (or Matsuno-Gill flow) at the day side (see their Figure~5). Thus, by applying friction everywhere in the atmosphere, the emerging general wind flow pattern is similar to our simulated wind flow for WASP-43b (Figure~\ref{fig1}, panel Ib). That is, \citet{Kataria2015} find an equatorial retrograde wind on the day side and a reduced eastward hotspot shift.

The mechanism we propose in this work invokes instead the influence of wave activity for WASP-43b, which does not reduce the overall wind speeds in the observable atmosphere. Only at the very bottom of the simulation, wind speeds are reduced due to drag (see also Section~\ref{sec: complete_stab} and Figure~\ref{fig4m} for models without frictional drag). In contrast, the magnetic drag mechanism imposed in \citet{Kataria2015} reduces the overall strength of the observable wind flow -- indiscriminate  of the wind direction (compare also Figure~4 (top) in \citet{Kataria2015} with Figure~\ref{fig1}, panel Ia and b). Even for moderate magnetic drag, the zonal flow in \citet{Kataria2015} is relatively weak:  $u\approx 3$~km/s. This is significantly lower than the wind speeds in our WASP-43b simulation ($u\approx \pm 5$~km/s) for prograde/retrograde flow, and also smaller than their WASP-43b simulation with full superrotation ($u\approx 5$~km/s) and without magnetic drag (Figure 2 in \citet{Kataria2015}). Apparently, reducing the strength of wind flow in the whole atmosphere as it is done in e.g.~\citet{Kataria2015,Parmentier2018} also changes the force balance between the superrotating (prograde) and retrograde wind flow tendencies. The latter tendency is always `lurking' in the background -- even in a superrotating hot Jupiter simulation without obvious retrograde flow at the equatorial day side (see~Figure~\ref{fig1} panel IIc and \citet{Showman2010}).

\section{Conclusions}
\label{sec: Conclusions}

We applied a deep circulation framework to WASP-43b and HD~209458b. We choose these planets as representative examples of well-studied hot Jupiters. They have roughly similar effective temperature ($T_{\rm eff}\approx 1450$~K) and thus similar radiative time scales within one order of magnitude. However, they have very different rotation periods ($P_{\text{orb}}=0.81$~days) and ($P_{\text{orb}}=3.5$~days), interior temperature ($T_{\text{int}}=170$~K and $400$~K), densities ($\rho=0.25\rho_{\text{jup}}$ and $2 \rho_{\text{jup}}$) and thus also different surface gravity ($g=20.4$~m/s${}^2$ and $10$~m/s${}^2$).

We find crucial differences in  wind flow for HD~209458b and WASP-43b (Section~\ref{sec: disc_deep_boundary}): unperturbed equatorial superrotation with shallow wind jets in the first and general superrotation with embedded retrograde flow at the equatorial day side together with very deep wind flow in the latter. We further find that retrograde equatorial flow can only develop for tidally locked hot Jupiters with relative short orbital periods, $P_{\rm rot}=P_{\rm orb}\leq 1.5$~days.

 The orbital period of 1.5~days coincides with the Rhines length over planetary radius becoming smaller than 2 for a climate with equatorial superrotation in hot Jupiters. In principle, however, also a (critical) Rhines length over planetary radius smaller than 1 is possible for such planets, like WASP-43b. This would lead to a climate dominated by off-equatorial wind jets, possibly driven by baroclinic eddies (Table~\ref{table: Rhines_Rossby_comparison}). Baroclinic eddies have been postulated by \citet{Showman2015} to drive wind flow and circulation for even shorter orbital periods $P_{\text{orb}}=0.55$~days.

Based on a number of dynamical diagnostics, we tentatively conclude that our WASP-43b simulation shows retrograde flow over the equatorial day side due to a combination of the following effects: very fast rotation with $P_{\text{orb}}<1.5$~days that can give rise to wave activity in the form of possible baroclinic eddies, the effect of which is strengthened by deep wind jets ($p>200$~bar), which can be seen by increased vertical momentum transport.

 WASP-43b is thus planet that seems to be particularly prone to develop equatorial retrograde winds, due to its fast rotation and high surface gravity (longer radiative timescales). JWST/MIRI could potentially show this via very low thermal flux at the night side of this planet.

 Furthermore, the deep wind jets may allow to better quantify the coupling between magnetic fields, wind flow and the location of clouds. As \citet{Rogers2014b} explicitly noted \textit{"larger flow speeds at depth could
increase the value of the Ohmic dissipation there, and therefore deep jets may be required to inflate planets"}, where the authors identify 90 bar as the region of depth of interest. A `head-on collision zone' at the morning terminator (mainly at the equator), which is only present with partly retrograde flow at the equatorial day side, can potentially lead to strong vertical mixing compared to climates with pure prograde flow at the equator. It thus may increase local cloud formation compared to simulations with unperturbed superrotation. All these effects may be important for a more complete understanding of the atmospheres of hot Jupiters and their heat redistribution.

Our work further shows that it is worthwhile to re-evaluate the lower boundary treatment of 3D GCMs for hot Jupiters. There is -- as of now -- no uniform agreement on how to treat the lower boundary. What works for one planet (HD~209458b), like setting $p_{\rm boundary}=100$~bar, might not work for a different planet (WASP-43b), which apparently requires a larger $p_{\rm boundary}=700$~bar in our model. WASP-43b appears thus to be a planet that is well suited for comparing different models with each other (see also e.g. \citet{Venot2020}). The significance of the lower-boundary treatment has been recently highlighted again for GJ~1214b, a fast rotator ($P_{\rm rot}=1.5$~days), where it was found that the dynamics in the deep atmosphere significantly impact the observables in the upper atmosphere \citep{Wang2020}.

\section{Outlook}
\label{sec: Outlook}
In future work, we will apply our model framework to also investigate the role of deep wind jets in hotter and colder planets than WASP-43b and HD~209458b. There we will elucidate more on the important role of radiative time scales for the emergence of retrograde equatorial flow \citep{ZhangShowman2017}.

Furthermore, we will investigate, under which conditions a westward shift of the hot spot can be achieved in tidally locked Jupiters like HAT-P-7b and CoRoT-2b. Previous theoretical work in the dynamics of tidally locked planets has shown that such a westward shift of the hottest atmospheric point with respect to the substellar point is in principle possible \citep{Carone2015,Penn2017,Penn2018} and it has been recently proposed to be achievable by including magnetic fields for HAT-P-7b \citep{Hindle2019}.

Although we have not taken into account the effects of cloud coverage or chemistry in our model, we can use basic principles of climate dynamics to hypothesize that these aspects will be strongly affected by retrograde wind flow. In particular, the `head-on collision zone' between the prograde and retrograde equatorial flows, which is present in our WASP-43b simulations at the morning terminator, will be interesting to investigate. Such strong horizontal flow convergence will  be accompanied by strong vertical mixing of chemical species via the `chimney'-effect  \citep{Zhang2018,Parmentier2013}. \citet{Mendonca2018b} and also \citet{Venot2020} investigate the impact of disequilibrium chemistry on WASP-43b but since their simulated 3D atmosphere did not exhibit strong horizontal flow convergence at the morning terminator, the atmospheric flow that they produced, did not show a strong vertical advection chimney. \citet{Mendonca2018b} report that zonal quenching is the dominant chemical disequilibrium mechanism, also \citet{Venot2020} suggest that the chemical composition is homogenized with longitude (via zonal flow) to that of the dayside.  We speculate that the retrograde zonal flow in our model, and the associated vertical transport at the morning limb, can also have an important effect on the disequilibrium chemistry in WASP-43b. Furthermore, there may be dynamical pile-up of clouds in the cool morning regions. In principle, it may be observable as a flattening of absorption features in transmission spectra. However, transmission spectra contain not only information about the morning but also from the warmer evening terminator and the polar regions. Disentangling the effect of a cloudy morning terminator from the evening terminator has been difficult so far \citep{Line2016,VonParis2016}. Also, the observations to date do not yield a consistent picture for cloud coverage over the terminator. One transmission spectrum of WASP-43b appears to be consistent with thick clouds \citep{Chen2014}, other spectra appear to favor a cloud-free atmosphere \citep{Kreidberg2014,Weaver2019}. Our work thus provides incentives for more work on cloud and out-of-equilibrium chemistry models for WASP-43b. We point out that the latter has been performed by \citet{Mendonca2018b,Mendonca2018} for WASP-43b in good agreement with the HST and Spitzer phase curves. It will be, however, interesting to investigate if possible deviations from superrotation can have an additional impact on cloud formation and disequilibrium chemistry in WASP-43b beyond the effects due to unperturbed superrotation that \citet{Mendonca2018b} considered.

We note that very recently \citet{Wang2020} confirmed that the deep atmosphere can indeed change the flow pattern in the observable atmosphere - also in Mini-Neptunes. The authors further state that ``results on the circulation of tidally locked exoplanets with thick atmospheres may need to be revisited''. In addition, Qatar-1b, another temperate hot Jupiter with $T_{\rm eff}\approx 1500$~K, has an intermediate mass between WASP-43b and HD~209458b and a rotation faster than $1.5$~days ($P_{\rm orb}=1.42$~days). It was likewise found to have no discernible hot spot shift \citep{Keating2020}. Qatar-1b could thus exhibit the effect of retrograde equatorial flow at the day side, as proposed here for WASP-43b. Finally, anomalous flow patterns, similar to those produced and analysed in this paper for WASP-43b, have also been found in simulations of irradiated brown dwarfs orbiting white dwarfs \citep{Lee2020}. Thus, the dynamic connection between the deep interior and the observable flow patterns, potentially affecting chemical mixing and clouds, seems to be relevant for a whole range of tidally locked, irradiated substellar atmospheres.

\section*{Acknowledgements}
L.C. acknowledges funding by DLR grant  50OR1804. R.B. is Ph.~D.~fellow of the Research Foundation - Flanders (FWO). L.D. acknowledges support from the ERC consolidator grant 646758 AEROSOL and the FWO Research Project grant G086217N. O.V. thanks the CNRS/INSU Programme National de Plan\'etologie (PNP) and CNES for funding support. P.M. acknowledges support from the European Research Council under the European Union's Horizon 2020 research and innovation programme under grant agreement No. 6945713. P.S. acknowledges the support from the Swiss National Science Foundation under grant BSSGI0$\_$155816 ``PlanetsInTime''. Parts of this work have been carried out within the framework of the NCCR PlanetS
supported by the Swiss National Science Foundation.
We further thank V. Parmentier and N. Mayne for insightful discussions and comments.

The data underlying this article will be shared on reasonable request to the corresponding author.




\bibliographystyle{mnras}
\bibliography{BreakingSuperrotRev4} 



\appendix

\section{Stability and Model Steady-State}
\label{sec: lower_stab}

\begin{table*}
	\caption[table]{Representative simulations, showing the hierarchy of the lower boundary stabilization measures tested in this paper.}
	\label{table_sims}
	\centering
	\begin{tabular}{l c c c}
    \noalign{\smallskip}
	\hline \hline
	\noalign{\smallskip}
	\textbf{Model name} & \textbf{Deep common temperature} & \textbf{\boldmath{$\tau_{\rm conv}=10^6$}~s} &  \textbf{Deep friction} \\
	\noalign{\smallskip}
	\hline
	\noalign{\smallskip}
	\textit{no\_stab} & \ding{55} & \ding{55} & \ding{55} \\
    \textit{temp\_stab} & \ding{51} & \ding{55} & \ding{55} \\
    \textit{rad+fric\_stab} & \ding{55} & \ding{51} & \ding{51} \\
    \textit{temp+rad\_stab} & \ding{51} & \ding{51} & \ding{55} \\
    \textit{temp+fric\_stab} & \ding{51} & \ding{55} & \ding{51} \\
    \textit{full\_stab} & \ding{51} & \ding{51} & \ding{51} \\
    \noalign{\smallskip}
    \hline
	\end{tabular}
\end{table*}%

Here, we present in more details the tests and investigations that we performed to validate our nominal deep lower boundary set-up as presented in Section~\ref{sec: low_boundary}.

\subsection{Shear flow instabilities at the bottom of the WASP-43b model}
\label{sec: shear}
\begin{figure*}%
	\centering%
    \includegraphics[width=1.5\columnwidth]{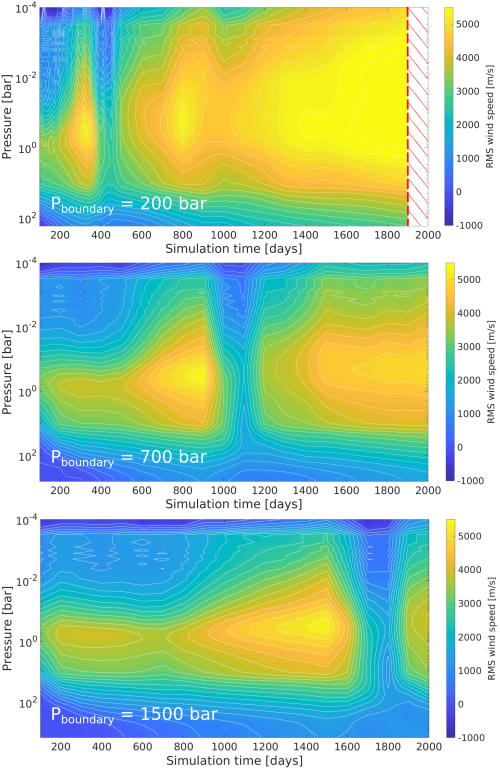}%
	\caption{Evolution of the root mean squared horizontal wind velocity as a function of pressure during three simulations of WASP-43b (using model `temp + rad\_stab', Table~\ref{table_sims}). From top to bottom, the lower simulation boundary was placed at $p_{\text{bottom}}=200$ bar, 700 bar and 1500 bar. The simulations were terminated after 2000 (Earth) days, but the 200 bar-simulation crashed after 1900 days due to a numerical instability. In all simulations abrupt, non-monotonic variations can be seen, after 400, 1000 or 1700 days simulation time, respectively, depending on the depth of the lower boundary. Simultaneously with these events, a shear flow instability at the bottom boundary occurs.}%
    \label{fig2m}%
\end{figure*}

\begin{figure*}%
	\centering%
    \includegraphics[width=1.95\columnwidth]{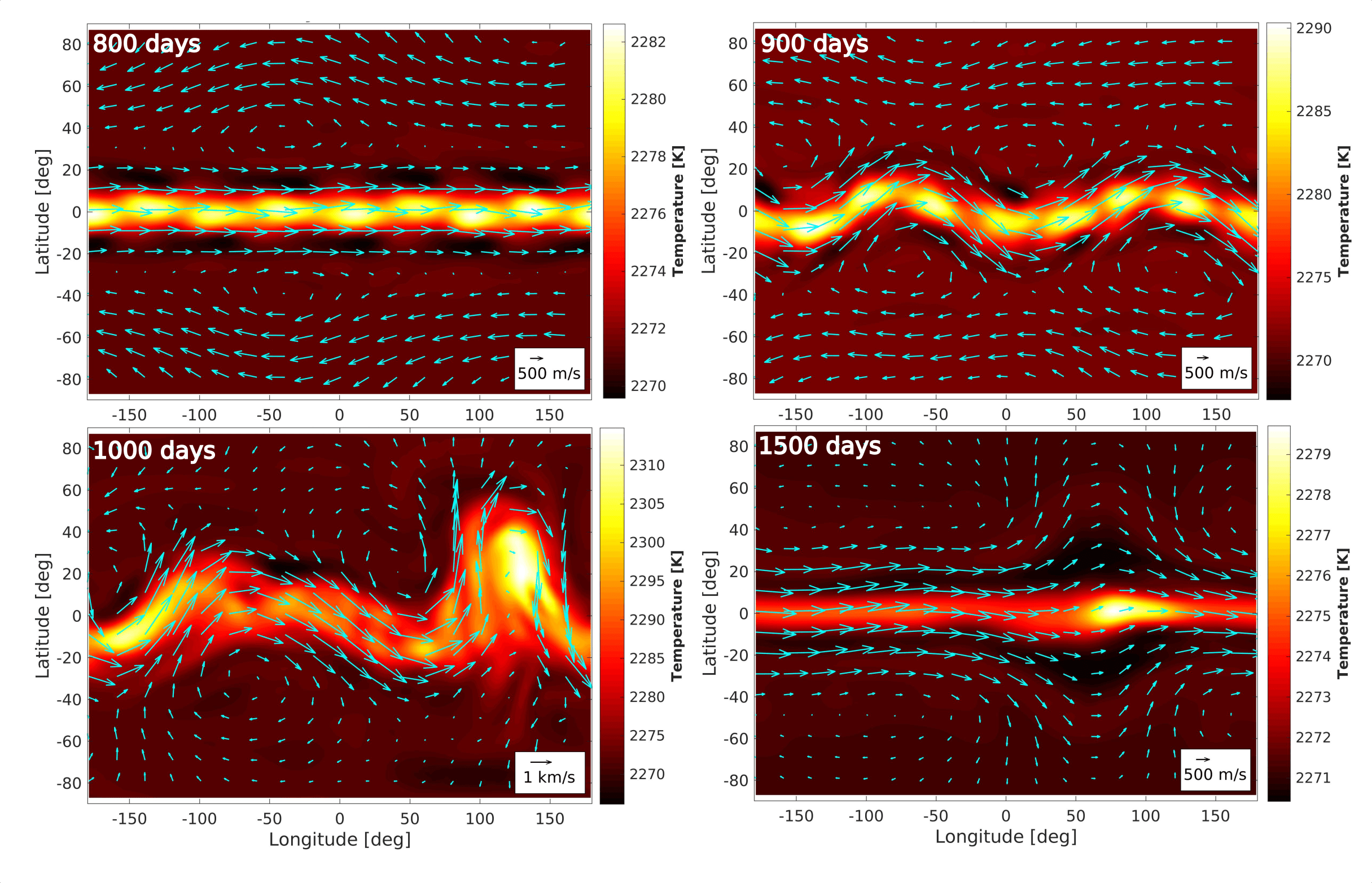}%
	\caption{Snapshots of the shear flow instability occurring at the lower boundary of the WASP-43b simulation. The displays show the temperature and wind maps at $p = 650$~bar after 800, 900, 1000, and 1500 days in the simulation.}%
    \label{fig3m}%
\end{figure*}

We find that every WASP-43b simulation eventually becomes unstable during the spin-up of the model, when the wind at the lowest boundary reaches speeds of about 1~km/s. Winds faster than 1~km/s at the lowest boundary were found to trigger instabilities. These result in an abrupt and discontinuous shift of the momentum budget in the whole atmosphere. Since the jet streams in WASP-43b show a tendency of penetrating comparatively deep inside the planet, this threshold of about 1~km/s will always be reached before the model spin-up has finished, although the instability event can be delayed by placing the lower at higher pressures (Figure~\ref{fig2m}). The development of the bottom-boundary shear flow instability is illustrated in detail in Figure~\ref{fig3m}.

As has been established in \citet{Menou2009,Rauscher2010}, there is no physical reason for deep wind jets to meander at great depth as shown in Figure~\ref{fig3m}. Instead, flow at depth can cause
numerical difficulties even in dynamical cores that use instead of a finite grid the spectral method \citet{Menou2009,Rauscher2010}. It is beyond the scope of this work to discuss in more detail the origin of the instabilities. Instead we provide a phenomenological description of the instability, as well as a documentation of our measures to circumvent it. We note that the shear instability appears to occur due to the lower boundary description, where no restrictions on the tangential velocity are prescribed (`free-slip') as justified by e.g. \citet{Heng2011}. This prescription appears, however, to be invalid for the particular case of WASP-43b, which exhibits fast wind flow at $p > 100$~bar. In fact, the occurrence of numerical shear flow instabilities as a consequence of very deep wind jets in simulations of fast-rotating hot Jupiters was already predicted theoretically in \citet{Thras2011}, and other work also cautioned to be careful to `anchor' a 3D climate  model at depth to improve the numerical stability  \citep{Liu2013}. If the model is not `anchored', problematic behavior can occur in a complex 3D climate model, as we demonstrate here.

\begin{figure*}%
	\centering%
    \includegraphics[width=1.9\columnwidth]{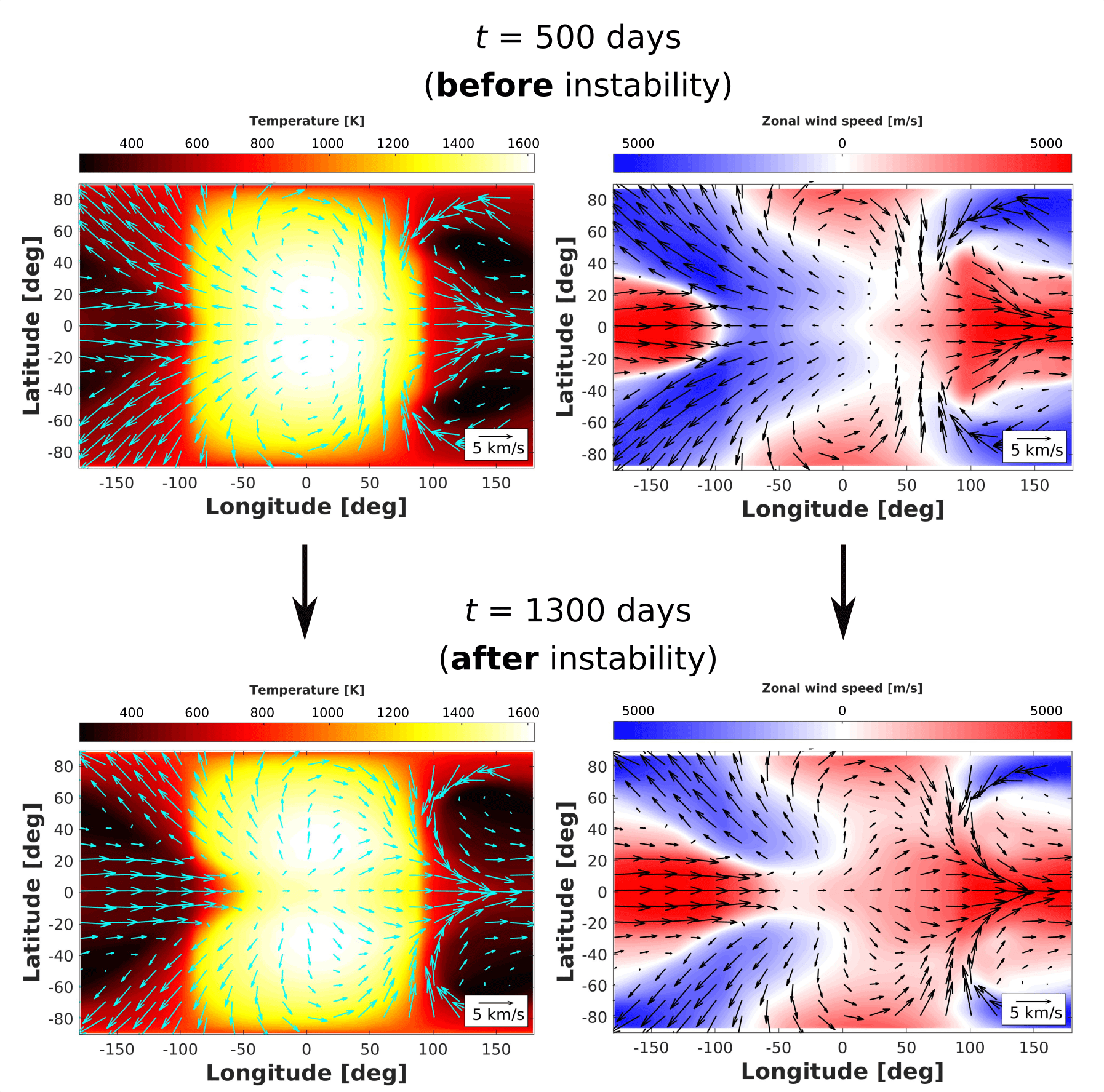}%
	\caption{Temperature and circulation maps of WASP-43b before and after the shear flow instability. This model has $P_{\rm bottom}= 700$~bar, a common deep temperature adiabat and $\tau_{\rm conv}=10^6$~s, corresponding to model setup \textit{temp+rad\_stab} in Table~\ref{table_sims}. For two different times (500 and 1300 days), the temperature map and zonal wind speeds at 12~mbar are shown in color, with the arrows representing the horizontal wind. The selected simulation times are exemplary for the circulation regime before and after the occurrence of the shear instability at the deep atmosphere boundary. The circulation transits from a climate state with a strong retrograde component to a fully superrotating regime.}%
    \label{fig6m}%
\end{figure*}

 In our simulations, the circulation is significantly altered after the occurrence of shear flow instabilities at the lowest boundary: a state of retrograde day-side flow is followed after the instability by a state of equatorial superrotation (Figure~\ref{fig6m}). This occurs even when the GCM `recovers' from the on-set of the instability and settles into a new dynamical state. The occurrence of superrotation in our model is thus dubious and appears to be triggered by numerics and not physics. We conclude that for particular cases, such as WASP-43b, an insufficiently stable model can still give rise to plausible, but ultimately incorrect atmospheric solutions, and one should be very cautious with the interpretation of any unstable or unconverged model.

Simulations of HD~209458b, on the other hand, an inflated slower rotating hot Jupiter, do not develop fast zonal wind jets that extend deeper than 200~bar (within 2000 days of simulation time). Therefore, we do not see the shear flow instabilities that we find in the simulation of the dense, fast-rotating hot Jupiter WASP-43b. Therefore, simulations for a highly inflated, tidally locked planet with relatively long rotation period ($P_{\rm orb}>1.5$~days) like HD~209458b do not require special treatment of the lower boundary.

\subsection{Investigation of complete flow stability for the WASP-43b simulations}
\label{sec: complete_stab}

\begin{figure*}%
	\centering%
    \includegraphics[width=1.45\columnwidth]{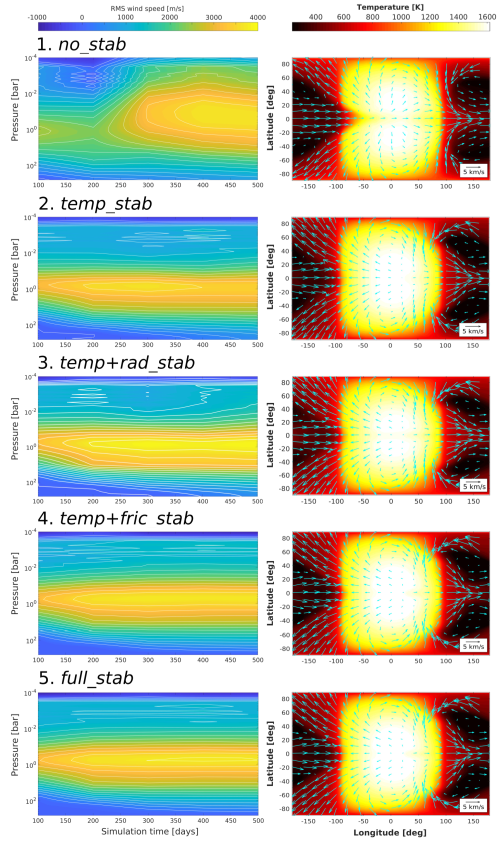}%
	\caption{Evolution of the root mean squared horizontal wind velocities throughout the simulation \textit{(left)} and the corresponding temperature and horizontal flow pattern at $p=12$~mbar, after 500~days \textit{(right)}, for simulations of WASP-43b with different deep atmosphere stabilization measures. The lower boundary pressure is $p_{\rm bottom}=700~$bar everywhere. The corresponding stabilization methods are summarized in Table~\ref{table_sims}. Note that all simulations exhibit qualitatively the same retrograde flow pattern, except for the case in which no stabilization measures were applied (\textit{no\_stab}). This latter case produces superrotation, after the occurrence of shear instabilities.}%
    \label{fig5m}%
\end{figure*}

\begin{figure*}%
	\centering%
    \includegraphics[width=1.8\columnwidth]{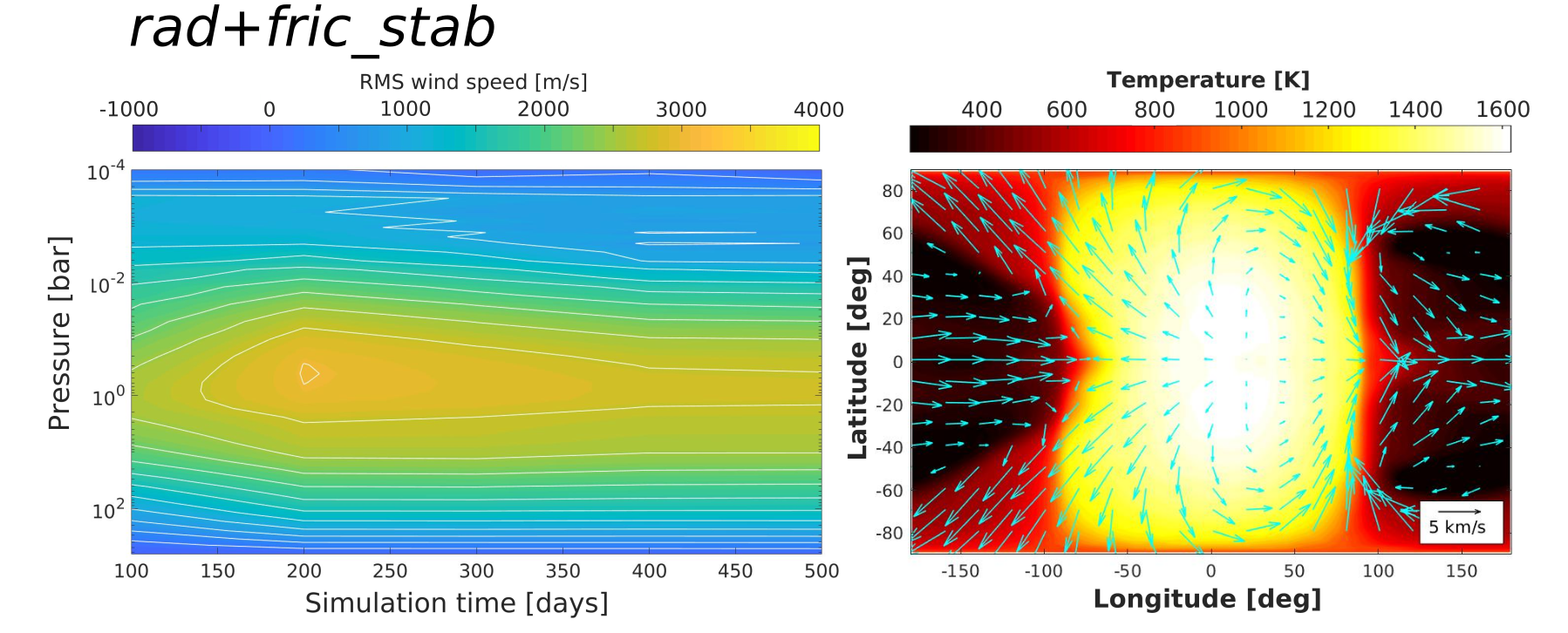}%
	\caption{Evolution of the root mean squared horizontal wind velocities throughout the simulation \textit{(left)} and the corresponding temperature and horizontal flow pattern at $p=12$~mbar, after 500~days \textit{(right)}, for a simulation of WASP-43b with accelerated convergence time scales and bottom boundary friction, but without forcing the model to a planetary averaged temperature in the deep atmosphere (`rad+fric\_stab', see Table~\ref{table_sims}). The simulation is complementary to the `temp\_stab' case in Figure~\ref{fig5m} in its stabilization measures, and shows the same retrograde flow pattern.}%
    \label{fig5mx}%
\end{figure*}

Since it was established in Section~\ref{sec: low_boundary} that the lower boundary treatment and overall model stability are extremely important for WASP-43b, we investigate these elements in a systematic way. Table~\ref{table_sims} and Figures~\ref{fig5m} and \ref{fig5mx} summarize concisely the stabilizing effect of every lower boundary measure employed in the deep circulation framework (\ref{sec: low_boundary}) and their feedback on the observable horizontal wind flow.

For example, we compare the complete steady state results (`full\_stab'), yielded by the simulation with $p_{\rm bottom}=700$~bar, with the quasi-steady state results for atmospheric layers at $p<1$~bar where no accelerated evolution at depth was imposed (`temp\_stab', `temp+fric\_stab').
These tests confirm that the quasi-steady state without artificial lower boundary convergence, and the complete steady-state reached with accelerated convergence time scales, yield similar results for the flow patterns and temperatures in the observable atmosphere ($p\leq 1$~bar) for WASP-43b. Note that this quasi-steady state for $p \leq 1$~bar is reached after 400~days in all our simulations. Only the model without any stabilization treatment (`no\_stab') does not exhibit a retrograde wind flow at the day side. Instead, the simulation results eventually in equatorial superrotation, but only after the shear flow instability has triggered an abrupt change in the atmospheric circulation.

Upon inspection of Figure~\ref{fig5m}, one might conclude that the emergence of retrograde day-side flow in our WASP-43b simulations is caused by the prescription of equilibrium temperatures that are converged onto a common adiabat below 10~bar. In order to check whether it is indeed the forcing to a planetary averaged temperature in the deep atmosphere that is causing the retrograde flow, we have conducted the `rad+fric\_stab' test, which is complementary to the `temp\_stab' test. More specifically, it is a simulation of WASP-43b with accelerated radiative time scale and deep friction (see Table~\ref{table_sims}), but forced to the equilibrium temperature profiles of \textit{petitCODE} without interpolation (i.e.~the dotted lines rather than the full lines in Figure~\ref{fig:rad}, Ia). For such a model, Figure~\ref{fig5mx} shows that the day-side retrograde flow is still present. In summary, our suite of tests show that not any single stabilization measure seems to be responsible for the emergence of retrograde flow in WASP-43b.

\begin{figure*}%
	\centering%
    \includegraphics[width=2.1\columnwidth]{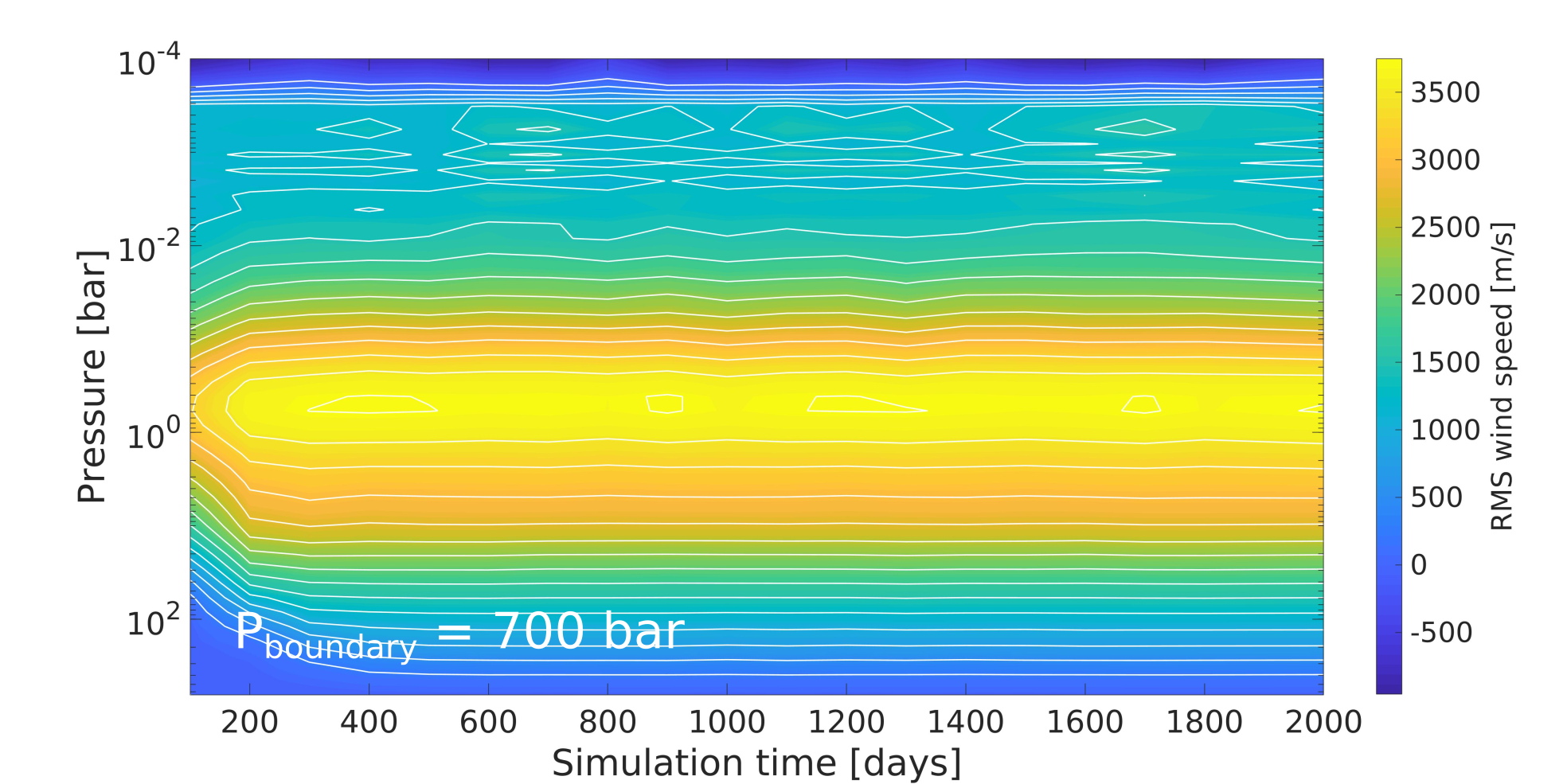}%
	\caption{Evolution of the root mean squared horizontal wind velocity as a function of pressure for WASP-43b, where we have applied `deep magnetic drag' between 490 and 700 bar. This fully stabilized model (full\_stab, Table~\ref{table_sims}) reaches a steady state after $\sim 500$ (Earth) days.}%
    \label{fig4m}%
\end{figure*}

Full atmospheric wind flow stability, from top to bottom, is reached after application of `deep magnetic drag' (`full\_stab'), with which we `anchor' our model to the planet interior \citep{Liu2013}. This situation is depicted in Figure~\ref{fig5m} bottom and in Figure~\ref{fig4m}.  At the same time, the full steady-state simulation has atmospheric flow patterns and temperatures that are qualitatively and quantitatively very similar to  WASP-43b simulations without magnetic drag ('temp+rad\_stab' and 'temp\_stab', see Figure~\ref{fig5m}).
Thus, we have verified that `deep magnetic drag' stabilizes the simulation and still retains the important physical effect that we propose in this work to be important to shape the flow: vertical transport out of the depth that injects westward zonal momentum at the day side into the upper planetary photosphere ($p\leq 0.1$~bar).

Furthermore, we find that retrograde day-side wind at the equator establishes itself during the model spin-up and is also maintained after a complete model steady state is reached (`full\_stab') and also if we choose not to average equilibrium temperature forcing at p=10~bar (Figure~\ref{fig5mx}). Thus, we conclude that equatorial retrograde flow in WASP-43b is a robust result in our model framework that is present even after spin-up of the simulations once instabilities at the lower boundary are avoided. This is in contrast to previous simulations of the hot Jupiters HD~189733b ($P_{\text{orb}}=P_{\text{rot}}=2.2$~days)  \citep{Showman2011} and HD~209458b ($P_{\text{orb}}=P_{\text{rot}}=3.5$~days) \citep{Mayne2017}. In these simulations, retrograde wind flow was exclusively present only during spin-up when the model was initiated from rest. This flow pattern was quickly superseded by full superrotation with longer simulation times.

\begin{figure*}
	\centering
    \includegraphics[width=1.8\columnwidth]{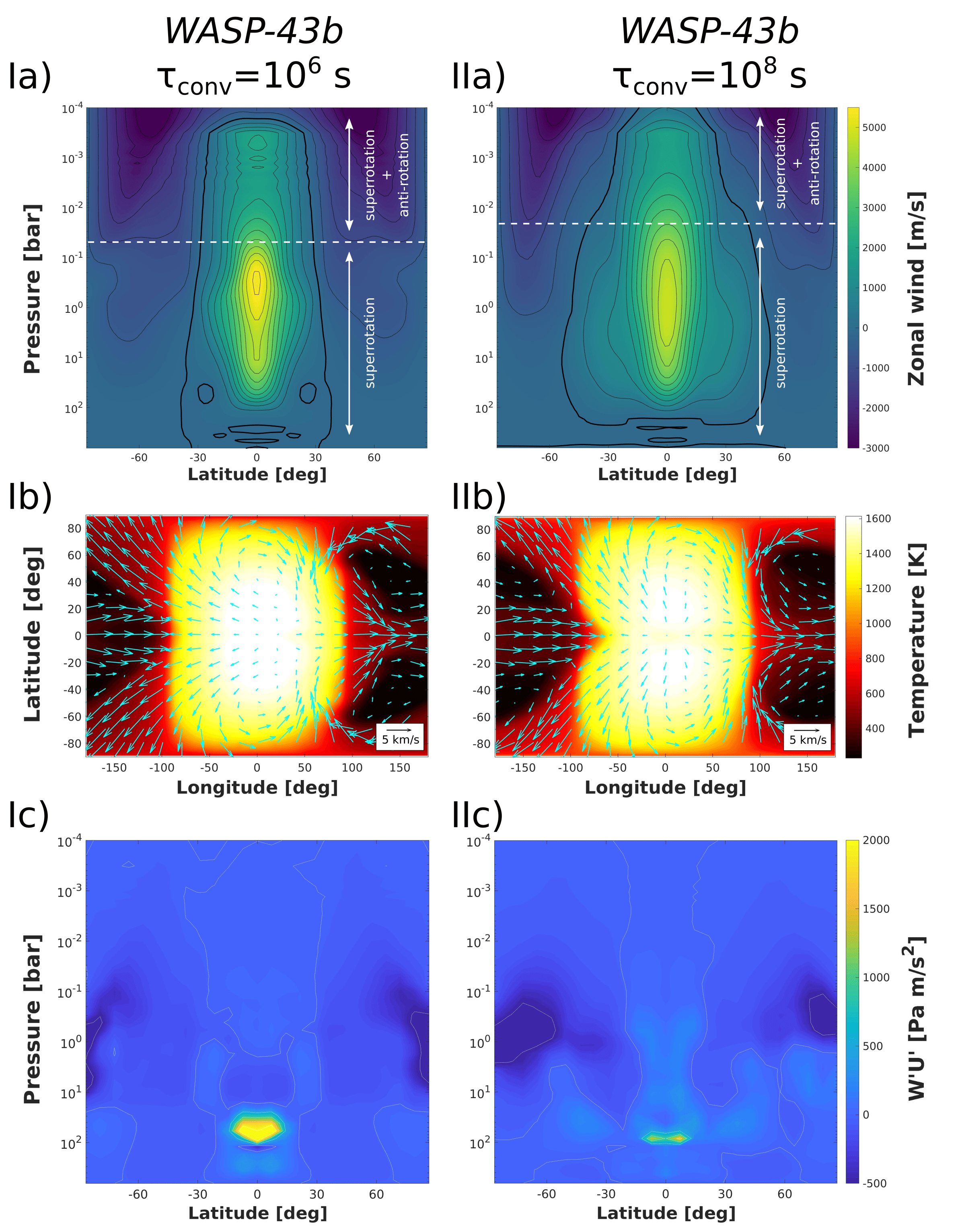}%
	\caption{Comparing WASP-43b simulations (with `deep magnetic drag' in both cases, which truncates the wind jet compared to Figure~2, panel Ia) with different deep radiative convergence timescales: $\tau_{\rm conv} = 10^6$ s (I) and $\tau_{\rm conv} = 10^8$ s (II). The top row (\textit{a}) shows the longitudinally averaged zonal (eastward) wind speed with indication of the dominant wind flow type. Contours are shown every 500 m/s and the contour of 0 m/s is indicated bold. The second row (\textit{b}) shows the temperature and the horizontal wind velocity at $p = 12$ mbar. The third row (\textit{c}) shows the longitudinally averaged vertical transport of zonal momentum $\overline{[w'u']}$.}
    \label{fig3}
\end{figure*}

We further observe that the retrograde wind flow pattern in WASP-43b is quantitatively affected by the radiative convergence time scale $\tau_{\text{conv}}$ adopted for the deepest vertical layers (see Figure~\ref{fig3}). For $\tau_{\rm conv}=10^8$~s, deep wind jets at depth $p>1$~bar have a smaller latitudinal extent and the vertical gradient in the zonal wind strength is smaller than in the nominal simulation with $\tau_{\rm conv}=10^6$~s (see Figure~\ref{fig3} panels Ia and IIa).  Consequently, a slower deep circulation convergence ($\tau_{\rm conv}=10^8$~s) leads to weaker retrograde flow in the horizontal wind flow at higher atmospheric levels (Figure~\ref{fig3} panels Ib and IIb). Checking the zonal momentum reservoir at depth $p\approx 100$~bar enforces this view. It is seen that vertical transport of zonal momentum is weaker for larger $\tau_{\rm conv}$ (Figure~\ref{fig3} panels Ic and IIc). The dependence of retrograde flow on the strength of the deep circulation forcing is another line of evidence which indicates that there is a tight link between deep circulation and retrograde flow. We stress that these sets of experiments need to be confirmed by more physically motivated prescriptions at depth and/or by future JWST observations that may either disprove or confirm the effect of deep wind jets on WASP-43b wind flow. For now, based on these experiments we propose that it is in principle possible to establish a link between deep wind flow and the observable atmosphere in WASP-43b.

\subsection{Sponge Layer Comparison}
\label{sec: sponge layer}

To ascertain whether the sponge layer we have implemented is not affecting the general flow in our simulations in unintended ways, we compare the results of models with different sponge layer descriptions. The aim is to check if the sponge layer affects the angular momentum budget in such a degree that the general flow changes to a qualitatively different regime. Another study examining the effect of the sponge layer on the flow in a hot Jupiter GCM was conducted by \citet{Deitrick2019}.

In addition to the sponge layer used in our nominal setup, which is based on a direct damping of the wind speeds via friction (see Section~\ref{sec:atmo}), we also choose to test a model without any sponge layer, and a model with a `softer' sponge layer that damps only the zonal anomalies, i.e.~the wind speed components when the zonal mean is subtracted \citep{Mendonca2018b, Deitrick2019}. To set up the latter case on the cubed-sphere grid of \textit{MITgcm}, we base ourselves on the approach laid out in \citet{Deitrick2019}. We divide the grid into 20 latitude bins of equal size (9$^\circ$) and compute the average of the zonal and meridional components of the wind speed per bin. We differ from the methodology of \citet{Deitrick2019} in some ways, mostly because of the different grids employed in \textit{MITgcm} and \textit{THOR}. Firstly, we compute the $\eta$ parameter in \citet{Mendonca2018b, Deitrick2019} as
\begin{equation}
    \eta = - a \ln{\left(\frac{p}{p_{\rm bot}}\right)},
\end{equation}
where $p_{\rm bot}$ is the bottom layer pressure of our model and $a$ is a normalization factor to ensure that $\eta$ varies from 0 to 1 in the vertical grid. Further, we do not interpolate between the latitude bins, opting instead to just subtract the zonal mean per bin. Lastly we do not compute the global zonal mean, but only the local zonal mean per cubed-sphere grid tile. Thus, our approach is cruder than the one used in \citet{Deitrick2019}, but we expect it to be an adequate representation of the sponge layer description of \citet{Mendonca2018b, Deitrick2019} nevertheless.

\begin{figure*}
	\centering
    \includegraphics[width=1.78\columnwidth]{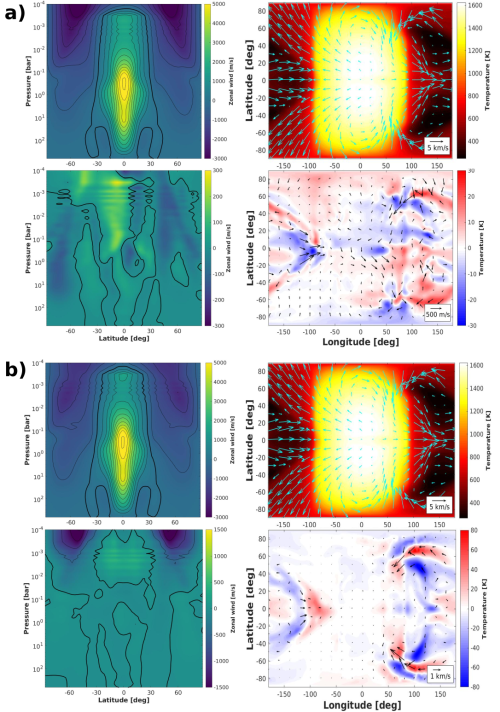}%
	\caption{Comparison between a simulation a) without sponge layer; and b) with a sponge layer description that damps zonal anomalies (cfr. \citet{Mendonca2018b, Deitrick2019}). In both cases the mean zonal wind (\textit{top left}) and an isobaric slice of the tempature at 10~mbar (\textit{top right}) are shown. On the second and fourth row, the differences in these respective diagnostics with our nominal WASP-43b simulation are shown. More specifically, the wind speeds and temperatures of the models without sponge layer (a) and with a zonal anomaly damping sponge layer (b) are subtracted from those of the nominal case. Note the change in color scales.}
    \label{fig_sponges}
\end{figure*}

The results of our sponge layer comparison are shown in Figure~\ref{fig_sponges}. We note once again that the situation without any sponge layer (panels a) may have gravity wave reflection at the top of the model, which could influence deeper atmosphere flow. There is some small difference in zonal wind flow of order 250~m/s compared to the set-up with sponge layer between $p=10^{0}$ and $p=10^{-1}$~bar. As expected, however, the effect of a change in sponge layer implementation is mostly visible in the upper part of the atmosphere ($p < 1$~mbar). There, the zonal wind speeds show deviations of up to 300~m/s in the case without sponge layer, and 1.5~km/s in the case with zonal anomaly-damping sponge layer. In particular, we note that the mid-latitude westwards wind jets in the upper atmospheric layers of our nominal WASP-43b model are much faster (i.e.~more negative) than in the model with a sponge layer that uses zonal anomaly damping. The equatorial region is very similar in this set-up compared to our nominal sponge layer setup.

The temperature maps deeper in the atmosphere (at 10~mbar, see Figure~\ref{fig_sponges}) display the same general picture independent of the sponge layer that was adopted, with local temperature fluctuations of up to 30~K in the case without sponge layer, and 80 K when only the zonal anomalies are damped. The temperatures change the most around the mid-latitudes of the night side and the equatorial morning limb. This is also where the horizontal wind speeds are most affected. We conclude that even though the wind speeds in the upper atmosphere are strongly affected by the sponge layer description, the general regime of the wind flow in our WASP-43b simulations is unaffected by it. In particular, we find retrograde wind flow on the equatorial day side of all simulations. While a more in-depth study of different sponge layers and their effect on hot Jupiter GCMs would be interesting, it is outside the scope of this paper.

\section{Angular momentum conservation}
\label{sec: angular_mom}
\begin{figure*}
	\centering
    \includegraphics[width=1.99\columnwidth]{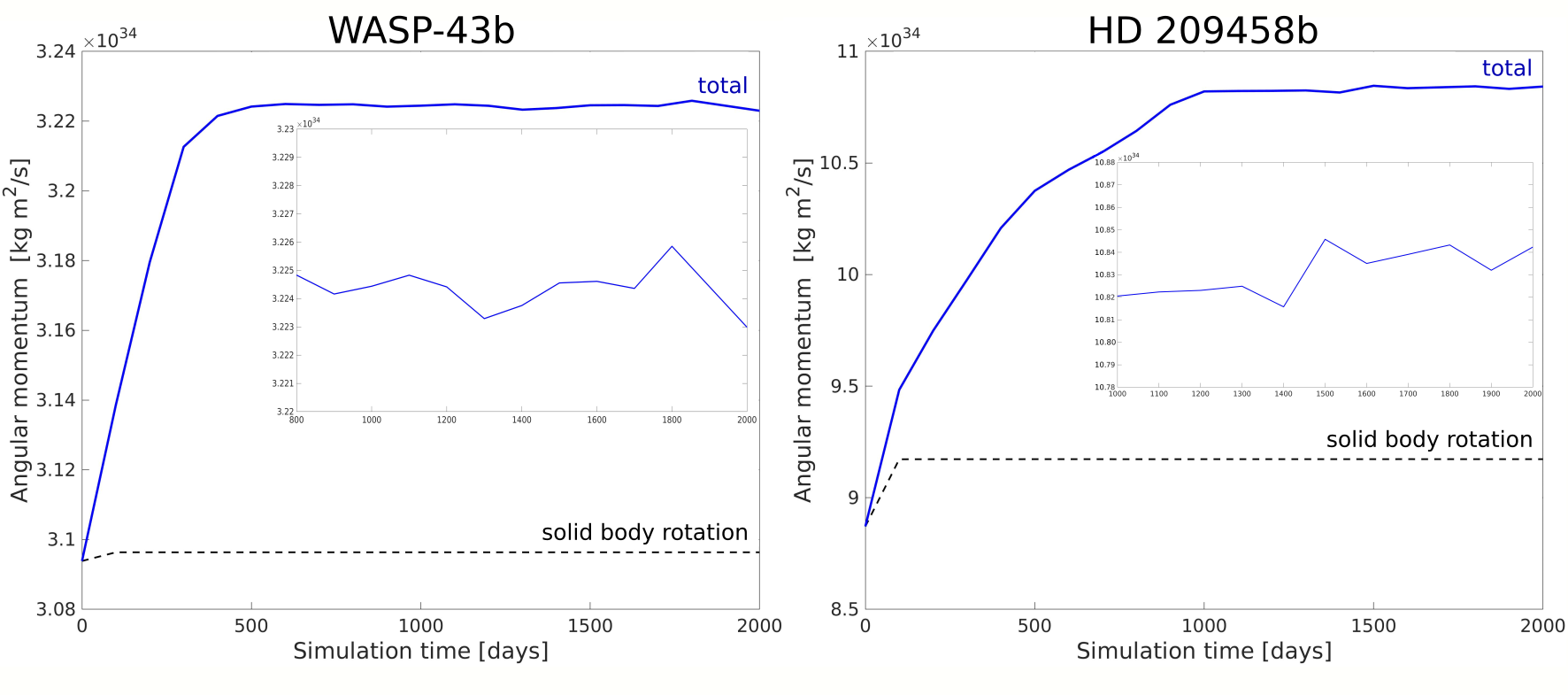}%
	\caption{The angular momentum evolution during the simulation is shown here at 100-day interval for WASP-43b (left) and HD~209458b (right). The angular momentum of the solid-body rotating atmosphere is plotted (black dashed line), as well as the total angular momentum due to solid-body rotation and atmospheric wind flow (blue solid line). The insets are a zoomed in version of the steady-state part: from 800 to 2000 days in the case of WASP-43b and from 1000 to 2000 days in the case of HD~209458b.}
    \label{fig7m}
\end{figure*}

An important diagnostic in the study of superrotation in GCMs is the total axial angular momentum budget of the atmosphere, and whether or not it is conserved within orders of numerical accuracy during a simulation time of 1000 days. It is a diagnostic that can be used to verify whether the simulation has reached a steady-state, and -- after spin-up -- provides global information about the circulation of the atmosphere. We note that angular momentum is not strictly conserved in a GCM simulation. \cite{Deitrick2019}  provide a thorough discussion of the numerical limits of angular momentum conservation for their GCM and \citet{Polichtchouk2014} already provides a numerical analysis of the \textit{MITgcm} dynamical core. Here, we aim to verify whether the total axial angular momentum reaches a steady-state within the context of our simulations (including all stabilizing measurements) and whether this steady-state axial angular momentum budget is maintained over the course of the simulation. The specific angular momentum $l$ is given by \citep{Lebonnois2012, Lee2012}:
\begin{equation}
    l = r \cos{\theta} \left( r \cos{\theta} \Omega + u \right),
\end{equation} where the first term captures the angular momentum due to the solid-body rotation of the atmosphere (the `planetary' angular momentum) and the second term describes the contribution due to the zonal wind flow. The total angular momentum $L$ of the atmosphere can then be computed by integrating over the mass:
\begin{align}
    L &= \int{l \text{d} m} \\
      &= \int{\Omega r^2 \cos^2\theta \text{d} m} + \int{u r \cos \theta \text{d} m}.
\end{align} Here, the mass element $\text{d}m = \rho \text{d}x \text{d}y \text{d}z$ can be expressed in spherical, pressure-based coordinates as $\text{d}m = r^2 \cos{\theta} \text{d}\phi \text{d}\theta \frac{\text{d}p}{g}$, assuming hydrostatic equilibrium. Using this expression, the total angular momentum (solid-body + wind contribution) is computed at 100-day intervals in our nominal simulations of WASP-43b and HD~209458b (Figure~\ref{fig7m}).

In both cases, the total axial angular momentum can be seen to quickly increase during an initial spin-up phase, the duration of which is significantly shorter in the WASP-43b model ($\sim 500$~days) than in the HD~209458b ($\sim 1000$~days). After spin-up, in both cases a steady state of relatively small variation in total angular momentum is reached. From the start of this steady state up to the end of the simulations after 2000 days, the total angular momentum exhibits variations of up to about 0.1\% and 0.4\% for the WASP-43b and HD~209458b simulations respectively (see the insets in Figure~\ref{fig7m}). We thus conclude that the axial angular momentum of our steady-state model is conserved in our simulations within the accepted numerical error. This a very satisfactory result, especially considering that the dynamical core of \textit{MITgcm} in the cube-sphere grid configuration is known to have relatively poor angular momentum and kinetic energy stability \citep{Polichtchouk2014, Cho2015}. However, due to the addition of lower boundary drag, torques can act on the atmospheric system during spin-up, resulting in a much better conservation of the total angular momentum and removing the model's sensitivity to initial conditions \citep{Liu2013, Cho2015, Mayne2017}. The stabilizing effect of adding bottom-boundary drag can indeed clearly be seen in Figure~\ref{fig5m}.

We note that the total angular momentum in both simulations is well above the value of solid-body rotation. Nevertheless, a quantitative difference is present in the degree of superrotation between both planets. The total axial angular momentum of WASP-43b is only 4\% higher than its angular momentum associated with solid-body rotation, whereas for HD~209458b this is 15\% higher. Also the \textit{no\_stab} model of WASP-43b (see Table~\ref{table_sims} and Figure~\ref{fig5m}), which displays an equatorial superrotating jet stream, has a total angular momentum budget that is more than 10\% higher than its solid-body rotation.

Hence, it seems that the partially retrograde wind flow in WASP-43b is reflected in its total angular momentum budget after spin-up. However, WASP-43b is still showing an excess in total angular momentum compared to solid body rotation because deep layers ($p>1$~bar) still exhibit an unperturbed equatorial eastward jet stream. Due to the amount of mass in them, these  deep layers tend to dominate the total angular momentum budget \citep{Mayne2014}.

\section{Eddy-mean flow analysis}
\label{sec: perturb}

\subsection{Eddy wind and momentum}
To diagnose the wind flow tendencies and momentum transport throughout the atmosphere, we analyse the interaction between eddies and the mean flow. This method is used in theoretical meteorology as a diagnostic tool to interpret general circulation patterns \citep{Holton}, and it involves studying deviations from the zonally or longitudinally averaged flow, temperature and other quantities in a 3D atmosphere model. It has been further demonstrated in e.g. \citet{Mayne2017} that diagnosing the horizontal and vertical momentum transport is highly informative for understanding emergent flow patterns in hot Jupiter 3D climate simulations.

Key parameters are divided into a basic state, which is averaged in time and longitude, and an eddy component, which contains the local deviation from the basic mean state. Since we use pressure as the vertical coordinate, we focus in this work on deviations from the horizontal velocity $\mathbf{v}=(u,v)$, where $u$ is the zonal component of the velocity (east-west oriented wind) and $v$ is the meridional component (north-south oriented wind). We further examine the vertical velocity $w$ (in units [Pa/s]) and the geopotential height $Z_g(p,\theta, \phi)$. The geopotential height is a measure of the vertical extent of the atmosphere (in units [m]) based on the local temperature and pressure. For example, a hot atmosphere column will have a higher geopotential height than a colder one at a given pressure level. We average over 100~days of simulation time, once the simulation has reached a steady state after the initial spin-up.

The decomposition in mean-flow and eddy components yields
\begin{align}
    &[\mathbf{v}] = \overline{[\mathbf{v}]}+[\mathbf{v}]', \\
    &[Z_g] = [\overline{Z_g}]+[Z_g]', \\
    &[wu] = \overline{[wu]}+[wu]',
\end{align}
where we used the traditional notation in which the brackets denote time averages, the bar denotes longitudinal averages and the prime denotes the eddy components.

We use the eddy horizontal wind $[\mathbf{v}]'$ and eddy geopotential height  $[Z_g]'$ to reveal the interaction of two large scale waves, the Kelvin and Rossby waves, which together form a `Matsuno-Gill' flow pattern \citep{Matsuno1966,Gill1980} (see Figure~\ref{fig: WASP43b_eddy} in the main body of the paper). The Matsuno-Gill flow is characterized by two pairs of Rossby vortices in the eddy wind flow on each hemisphere that coincide with positive and negative geopotential height (or temperature) anomalies. In addition, the eddy wind following the equator trace the Kelvin wave (see Figure~\ref{fig1}, panels Ic and IIc). Superrotation is thought to arise due to shear between the Rossby vortices and the equatorial Kelvin wave, visible by the tilt of the vortices with respect to the North-South axis  \citep{Showman2010,Showman2011}. The wave-mean-flow analysis also facilitates the comparison of the horizontal wind flow tendencies in 3D climate models to results of shallow water models  \citep{Showman2010,Showman2011,Penn2017}.

It has also been shown  \citep{Holton} that for products of zonally averaged atmospheric properties, the following relation holds:
\begin{equation}
[\overline{wu}] =[\overline{w}~\overline{u}] + [\overline{w'u'}].
\end{equation}
The last term in this expression, the eddy vertical transport of zonal (eastward/westward) momentum $[\overline{w'u'}]$, is used to diagnose the direct link between deep wind jets in the interior ($p=20 -500$~bar) and retrograde, that is, westward flow in the upper photosphere ($p<0.1$~bar). We find that the vertical zonal momentum transport $[\overline{w'u'}]$ is a good diagnostic for the interaction between the observable atmosphere and deeper layers in our GCM (see e.g. Figure~\ref{fig1}, panels Id and IId).  Also in shallow water models  \citep{Showman2010}, vertical transport of zonal momentum at the equator was found to be closely linked to the part of the Matsuno-Gill flow that dominates on the equatorial day side: retrograde or prograde flow. It is, however, beyond the scope of this work to identify the exact analogue between the vertical transport of zonal momentum $[\overline{w'u'}]$ in our GCM and the related quantity in a shallow water model.

\subsection{Eliassen-Palm flux and Potential vorticity}
\label{sec: EP_calc}

In the Earth climate, the Eliassen Palm flux is commonly used to diagnose the effect of eddies, in particular those triggered by baroclinic instabilities \citep{Edmon1980}. The Eliassen-Palm flux is defined as (see also \citet{Holton}):

\begin{align}
& F_\phi = -R_P \cos \phi \overline{u'v'},\\
& F_p =  R_P f \cos \phi \frac{\overline{v'\theta'}}{\theta_p},
\end{align}
where $R_p$ is the planetary radius, $\phi$ is the latitude, $f$ is the Coriolis parameter $f=2\Omega_P \sin \phi $, $\theta$ is the potential temperature and $\theta_p= \partial \theta / \partial p$.

We follow in this work the approach by \citet{Edmon1980} to use scaling factors $s_\phi$ and $s_p$ for the individual components of the Eliassen-Palm Flux and scale with latitude by:
\begin{align}
& \tilde{F}_\phi = \cos \phi \frac{1}{R_P} \frac{F_\phi}{s_\phi} r_{\text{Fac}},\\
& \tilde{F}_p =  \cos \phi  \frac{F_p}{s_p} r_{\text{Fac}},
\end{align}
We further adopt the approach of \citet{Taguchi2006} to scale the vertical component of the vectors differently for different pressure ranges by setting $r_{\text{Fac}}=\sqrt{10^5 Pa/p}$. We have however expanded the approach of \citet{Edmon1980,Taguchi2006} by using different scaling factors $s_p$ and $s_\phi$ for four different pressure ranges (Figure~\ref{figEP}) to allow for a better analysis of eddies, as we are not only discussing atmospheric layers between 1~bar and 0.01~bar but from $10^3$ to $10^{-4}$~bar. The scaling factors are listed in Table~\ref{tab:EP_scale}.

\begin{table}
	\caption[table]{Lists of the scaling factors $s_p$ and $s_{\phi}$ used for Figure~\ref{figEP}.}
	\label{tab:EP_scale}
	\centering
	\begin{tabular}{|l|cc|cc|}
    \noalign{\smallskip}
	\hline \hline
	\noalign{\smallskip}
	\textbf{log(pressure)} & \multicolumn{2}{c}{\textbf{WASP-43b}}  & \multicolumn{2}{c}{\textbf{HD~209458b}} \\
	   $[\rm{log(bar)}]$ & $s_p$ & $s_{\phi}$ &  $s_p$ & $s_{\phi}$\\
	\noalign{\smallskip}
	\hline
	\noalign{\smallskip}
	$-3$ to $-1.5$&$5.83 \times 10^{5}$& $\pi\times 2\cdot 10^{-5}$&$3.6 \times 10^{8}$ & $\pi\times 2\cdot 10^{-5}$\\
	$-1.5$ to $0$& $4.01 \times 10^{10}$& $\pi\times 2\cdot 10^{-4}$ &$2.2 \times 10^{10}$   & $\pi\times 10^{-4}$\\
    $0$ to $1.5$&$6.6 \times 10^{10}$& $\pi\times 0.034$ &$3.1 \times 10^{10}$  & $\pi\times 0.034$\\
    $1.5$ to $2.7$& $5.9 \times 10^{10}$& $\pi\times 0.1$  &$3.4 \times 10^{10}$  & $\pi$\\
    \noalign{\smallskip}
    \hline
	\end{tabular}
\end{table}%

Another measure to diagnose the possible presence of baroclinic instabilities is potential vorticity and here, specifically, a sign change in the horizontal gradient of the potential vorticity \citep{Holton}. In this work, we use the approximation to Ertels potential vorticity $q_E$ employed by \citet{Read2009} for Saturn, following their equation~(4):
\begin{equation}
q_E \approx - g \left(f + \zeta_p \right)\frac{\partial \theta}{\partial p},
\end{equation}
where $g$ is surface gravity, $f$ is again the Coriolis parameter and $\zeta_p$ is the vertical component of relative vorticity $\zeta$ evaluated on a pressure surface. The relative vorticity is here defined as $\zeta_p= \vec{k} \cdot (\nabla_p \times \vec{u}) $, where $\vec{u}$ is the horizontal flow velocity and $\vec{k}$ is the vertical unit vector.

\bsp	
\label{lastpage}
\end{document}